\documentclass[fleqn,usenatbib,letters]{mnras}
 
\usepackage{newtxtext,newtxmath}
\usepackage[T1]{fontenc}
\usepackage{graphicx}
\usepackage{amsmath}	
 
\DeclareRobustCommand{\VAN}[3]{#2}
\let\VANthebibliography\thebibliography
\def\thebibliography{\DeclareRobustCommand{\VAN}[3]{##3}\VANthebibliography}

\newcommand{\Msun}{\mathrm{M}_{\odot}}
\newcommand{\Hz}{\mathrm{Hz}}
\newcommand{\Msmbh}{M_{\rm SMBH}}
\newcommand{\AU}{{\rm AU}}
\newcommand{\Gyr}{{\rm Gyr}}

\newcommand{\Mpc}{\mathrm{Mpc}}
\newcommand{\Gpc}{{\rm Gpc}}

\renewcommand{\theenumi}{(\arabic{enumi}}

\title[Characterizing LK-induced BBH Mergers]{Parameter Distributions of Binary Black Hole Mergers Near Supermassive Black Holes as Seen by Advanced Gravitational Wave Detectors}

\author[L. Gond\'an]{
L\'aszl\'o Gond\'an\thanks{E-mail: laszlo.gondan@ttk.elte.hu}
\\
$^{1}$ ELTE E\"otv\"os Lor\'and University, P\'azm\'any P. s. 1/A, Budapest 1117, HU
}

\date{Accepted XXX. Received YYY; in original form ZZZ}

\pubyear{2022}

\begin{document}
 
 \label{firstpage}
 
 \pagerange{\pageref{firstpage}--\pageref{lastpage}}
 
 \maketitle

\begin{abstract}
{
 The environment surrounding supermassive black holes (SMBHs) in galactic nuclei (GNs) is expected to harbour stellar-mass binary black hole (BBH) populations. These binaries were suggested to form a hierarchical triple system with the SMBH, and gravitational perturbations from the SMBH can enhance the mergers of BBHs through Lidov-Kozai (LK) oscillations. Previous studies determined the expected binary parameter distribution for this merger channel in single GNs. Here we account for the different spatial distribution and mass distribution models of BBHs around SMBHs and perform direct high-precision regularized N-body simulations, including Post-Newtonian (PN) terms up to order PN2.5, to model merging BBH populations in single GNs. We use a full inspiral-merger-ringdown waveform model of BBHs with nonzero eccentricities and take into account the observational selection effect to determine the parameter distributions of LK-induced BBHs detected with a single advanced gravitational-wave (GW) detector from all GNs in the Universe. We find that the detected mergers' total binary mass distribution is tilted towards lower masses, and the mass ratio distribution is roughly uniform. The redshift distribution peaks between $\sim 0.15 - 0.55$, and the vast majority of binaries merge within redshift $\sim 1.1$. The fraction of binaries entering the LIGO/Virgo/KAGRA band with residual eccentricities $>0.1$ is below $\sim 10 \%$. We identify a negative correlation between residual eccentricity and mass parameters and a negative correlation between residual eccentricity and source distance. Our results for the parameter distributions and correlations among binary parameters may make it possible to disentangle this merger channel from other BBH merger channels statistically.
 }
\end{abstract}



 
\section{Introduction} 
\label{sec:Intro}
 
 The recent detection of GWs from merging stellar-mass BBHs and neutron star binaries by the Advanced Laser Interferometer Gravitational-Wave Observatory\footnote{\url{http://www.ligo.caltech.edu/}} (aLIGO; \citealt{Aasietal2015}) and Advanced VIRGO\footnote{\url{http://www.ego-gw.it/}} (AdV; \citealt{Acerneseetal2015}) has opened the field of GW astronomy \citep{Abbottetal2016,Abbottetal2017}. To date, nearly 100 BBH mergers have been detected with the aLIGO-AdV-KAGRA\footnote{\url{https://gwcenter.icrr.u-tokyo.ac.jp/en/}} \citep{KagraColletal2018} detector network \citep{Abbotetal2019b,Abbotetal2021d,Abbotetal2021c}, and several additional BBH mergers were identified in the publicly available data \citep{Venumadhavetal2019a,Venumadhavetal2019b,Zackayetal2019a,Zackayetal2019b,Nitzetal2021}. Based on the detected BBHs, the BBH merger rate density was observationally constrained by the LIGO--Virgo--KAGRA Collaboration to the range $17.9 - 44 \, {\rm Gpc}^{-3} {\rm yr}^{-1}$ at a fiducial redshift \citep{Abbotetal2021b}. With ongoing improvements to the operating advanced GW detectors and by expanding the system to five stations with the involvement of LIGO India\footnote{\url{http://www.gw.iucaa.in/ligo-india/}} \citep{Unnikrishnan2013}, the number of GW detections is expected to grow at an unprecedented rate in the upcoming years (e.g. \citealt{Abbottetal2018,Baibhavetal2019}).
 
 Various formation scenarios have been proposed to constrain the possible astrophysical origin of the detected BBH mergers (e.g. \citealt{Abbottetal2019,Baracketal2019,Abbotetal2021a}, and references therein). In most isolated binary channels, the resulting merging BBHs are circular within measurement errors within the sensitive frequency band of advanced GW detectors. However, several merger channels involving dynamical environments may result in BBH merger events with non-negligible eccentricity when reaching the aLIGO/AdV/KAGRA band, thereby distinguishing themselves from other astrophysical merger channels. So far, some potential candidates have been proposed to have a dynamical origin \citep{RomeroShawetal2020,RomeroShawetal2021,Gayathrietal2022,RomeroShawetal2022}, but no compelling evidence has yet been found for non-zero eccentricity \citep{Abbottetal2019c,Wuetal2020}.
 
 The avenues to produce eccentric BBH mergers through dynamical interactions include different formation scenarios and host environments. In the late-inspiral phase, the highest eccentricities are expected following strong gravitational scatterings in GNs (e.g. \citealt{OLearyetal2009,KocsisLevin2012,HongLee2015,Gondanetal2018b,GondanKocsis2021}). This effect may also lead to systems with significantly nonzero eccentricity in globular clusters (e.g. \citealt{Samsingetal2014,SamsingRamirezRuiz2017,Rodriguezetal2018b,Rodriguezetal2018,Samsing2018,Samsingetal2018,Zevinetal2018,Samsingetal2020}) and the gas disks of active galactic nuclei \citep{Tagawaetal2020,Samsingetal2022}. In hierarchical triples, the LK oscillation \citep{Kozai1962,Lidov1962,LidovZiglin1976} may also lead to eccentric BBH mergers in the aLIGO/AdV/KAGRA band, where the tertiary companion may be a stellar-mass object in the galactic field (e.g. \citealt{Antoninietal2017,LiuLai2017,SilsbeeTremaine2017,LiuLai2018,RodriguezAntonini2018,FragioneKocsis2020,Liuetal2020,MichaelyPerets2020}), in globular clusters (e.g. \citealt{Wen2003,Aarseth2012,Antoninietal2014,Antoninietal2016,Breiviketal2016,Martinezetal2020}), or in young massive and open star clusters \citep{Kimpsonetal2016,Tranietal2022}. The tertiary can also be an intermediate-mass BH in globular clusters \citep{FragioneBromberg2019} or an SMBH in GNs (e.g. \citealt{AntoniniPerets2012,Hamersetal2018,Hoangetal2018,RandallXianyu2018a,RandallXianyu2018b,Fragioneetal2019,Zhangetal2019,ArcaSedda2020}). Eccentric BBH mergers are also produced in the aLIGO/AdV/KAGRA band in quadruples \citep{FragioneKocsis2019,LiuLai2019,HamersSafarzadeh2020,Hamersetal2021} and in triples in the vicinity of an SMBH \citep{Fragioneetal2019b,ArcaSeddaetal2021}.
 
 Recent studies have shown that eccentricity can be measured for stellar-mass BBH mergers at the $10 \, \Hz$ frequency band of advanced GW detectors for $e \gtrsim 0.02 - 0.1$  \citep{BrownZimmerman2010,Huertaetal2017,Huertaetal2018,Gondanetal2018a,Loweretal2018,GondanKocsis2019,RomeroShaw2019,Lenonetal2020,Wuetal2020}. Together with other suggested parameters such as spins and mass-dependent parameters, their distributions, and correlations among them may be used to uniquely or statistically disentangle among different BBH merger channels (e.g.  \citealt{Baracketal2019,GerosaFishbach2021,GondanKocsis2021,Speraetal2022}, and references therein). Note that the formation environment may also be constrained with advanced GW detectors by identifying variations in the measured GW signal due to Doppler effects related to a possible movement of the BBH's centre of mass \citep{Inayoshietal2017,Meironetal2017,Chamberlainetal2019} or by measuring the magnification and time delay of secondary GW signals \citep{Kocsis2013,GondanKocsis2022}.
 
 In this study, we focus on the distributions of measurable  binary parameters for BBHs merging in GNs due to the LK mechanism. In this merger channel, BBHs undergo large-amplitude eccentricity and inclination oscillations due to the LK mechanism in the presence of an SMBH (e.g. \citealt{AntoniniRasiol2016,Leighetal2016}), and GW emission drives the binaries to merge if their eccentricities reach sufficiently high values.\footnote{See \citealt{Naoz2016} for a review of the LK mechanism.} This effect accelerates the merger times of BBHs, leading to enhanced merger rate densities, which are competitive with other proposed merger channels \citep{MandelBroekgaarden2022}. This merger channel was first investigated in the pioneering work by \citet{AntoniniPerets2012} for a Milky Way-size nucleus. They predicted merger rates together with the total fraction of BBHs (and binary neutron stars) merging due to LK resonances and also investigated the distribution of eccentricity when BBHs enter the $10 \, \Hz$ frequency band of aLIGO. Subsequent studies have refined the merger rate estimates, predicted merger rate density estimates, and determined the distributions of several orbital and mass-dependent parameters in single GNs, taking into account variations in the underlying BBH and stellar population models, GN models accounting for spherical and non-spherical nuclear star clusters, torques from the stellar cluster, and variant integration techniques of the equations of motion (e.g. \citealt{VanLandinghametal2016,PetrovichAntonini2017,Hamersetal2018,Hoangetal2018,RandallXianyu2018a,RandallXianyu2018b,Fragioneetal2019,Takatsyetal2019,Zhangetal2019,ArcaSedda2020,BubPetrovich2020,Yuetal2020}).
 
 We extend previous analyses to predict the distributions of binary parameters as observed by an advanced GW detector at design sensitivity. In our investigations, we account for the different spatial distribution and mass distribution models of BBH populations around SMBHs and take into consideration the observational selection effect. We determine the distribution of eccentricity at $10 \, \Hz$, the distributions of mass-dependent parameters (e.g. total mass, chirp mass, and mass ratio), and the redshift distribution for aLIGO/AdV/KAGRA detections. Finally, we identify possible correlations among various binary parameters using their observable distributions. The measured correlations and characteristics of parameter distributions may be useful to statistically disentangle this merger channel from other compact object merger channels.
 
 To examine the distributions of binary parameters, we generate mock Monte Carlo (MC) catalogues of LK-induced BBHs as observed by aLIGO/AdV/KAGRA. For this purpose, we first generate mock MC samples of LK-induced BBH merger populations in single GNs using high-precision, fully regularized N-body simulations, including PN terms up to order PN2.5. We use a full inspiral-merger-ringdown waveform model of BBHs with nonzero eccentricities to calculate the signal--to--noise ratio (SNR) values, taking into account the source direction, inclination, and polarization angles. We generate mock samples of GNs with the observed distribution of SMBHs in the detection volume, then sample LK-induced BBH mergers for each GN host from the corresponding redshift-dependent merger rate distribution, and finally discard the sources that do not reach the SNR limit of detection.
 
 The paper is organized as follows. We summarize the characteristics of GN hosts relevant to generating mock MC samples of GNs in Section \ref{sec:Props_GNs}. Similarly, we summarize the main quantities of BBH populations in GNs relevant to generating mock MC samples of LK-induced BBH mergers in single GNs in Section \ref{sec:InitDist_BBHs}. We give an overview of the relevant timescales related to the dynamical processes governing the evolution of binaries around SMBHs in Section \ref{sec:Timescales}. We introduce the methods we applied in the performed MC simulations to generate mock samples of LK-induced BBH mergers in single GN hosts in Section \ref{sec:MCsetup_SingleGN}. We describe the setup of MC simulations resulting in mock catalogues of LK-induced BBH mergers detectable by an advanced GW detector in Section \ref{sec:MC_Setup}. We present our main results in Section \ref{sec:Results}. Finally, we summarize the results of the paper and conclude in Section \ref{sec:SummAndConc}.
 
 We use geometric units ($G = 1 = c$) throughout the paper, where mass $M$ and distance $r$ have units of time: $GM/c^3$ and $r/c$.

\section{Galactic Nuclei} 
\label{sec:Props_GNs}
 
 We start by briefly summarizing the characteristics of GN hosts relevant in generating mock MC samples of GNs and LK-induced BBH mergers in single GN hosts.

\subsection{Galactic nucleus}
\label{subsec:GN}
 
 GNs are dense and massive assemblies of stars and compact objects gravitationally bound to a central SMBH and are found at the centres of most galaxies (e.g. \citealt{Neumayeretal2020}). The orbital evolution of the components of stellar populations is dominated by the SMBH within the GN's radius of influence, which is commonly defined in the literature as
\begin{equation}   \label{eq:rmax}
  r_{\rm max} = \frac{ \Msmbh }{ \sigma_*^2 } 
\end{equation}
 \citep{Peebles1972}. Here, $\Msmbh$ is the mass of the SMBH, and $\sigma_*$ is the velocity dispersion of the underlying stellar populations in the nucleus near the SMBH. In order to estimate $r_{\rm max}$, we use the $\Msmbh-\sigma_*$ fit of \citet{KormendyHo2013},
\begin{equation}   \label{eq:Msigma}
  \Msmbh \simeq 3.097 \times 10^8 \,\Msun \left( \frac{ \sigma_* }{ 200\ \mathrm{km}\ \mathrm{s}^{-1}} \right)^{4.384} \, .
\end{equation}
 
 Note that the radius of influence may be defined alternatively as the distance from the central SMBH at which the enclosed mass in stars equals twice $\Msmbh$ \citep{Merritt2004}. We investigate in Section \ref{subsec:DistaLIGOdet} how this choice for $r_{\rm max}$ influences the distributions of measurable binary parameters for detections with single advanced GW detectors.

\subsection{Spatial distribution of galactic nuclei} 
\label{subsec:SMBH_SpatDist}
 
 The astrophysical reach of advanced GW detectors at design sensitivity ranges between $\sim 4.1 - 5.5 \, \Gpc$ in comoving distances for quasi-circular non-spinning BBHs\footnote{We used the luminosity distance - redshift relation \citep{Hogg1999} and the adopted spatially-flat $\Lambda$CDM cosmology model to set a relation between redshift $z$ and luminosity distance $D_{\rm L}$ and thereby calculate the comoving distance $D_{\rm C}$ as $D_{\rm C} = D_{\rm L}/(1 + z)$ \citep{Hogg1999}. In the case of a single aLIGO detector, the maximum redshift of detection is $z \sim 2$ \citep{HallEvans2019}, which translates into a luminosity distance of $\sim 16 \, \Gpc$ and a comoving distance of $\sim 5.5 \, \Gpc$. The maximum luminosity distance of detection for KAGRA and AdV are $\sim 9.3 \, \Gpc$ \citep{Michimuraetal2020} and $\sim 9.7 \, \Gpc$ \citep{Abbotetal2020a}, respectively, which correspond to comoving distances of $\sim 4.1 \, \Gpc$ and $\sim 4.2 \, \Gpc$, respectively.}, and it is expected to be somewhat larger for initially highly eccentric BBHs with low orbital separations (e.g. \citealt{Eastetal2013}). This detection ranges significantly exceeds the largest scale of inhomogeneity in the mass distribution of the Universe, which arises from cosmic voids \citep{GregoryThompson1978} of maximum radii in the range of $20 - 100 \, h^{-1} \, \Mpc$ \citep{Mao2017} with Hubble parameter $h = H_0 / 100$. Accordingly, we assume that anisotropies and inhomogeneities average out over the corresponding detection volumes and thereby consider a homogeneous and isotropic spatial distribution of GN hosts in comoving coordinates. To obtain the radial distribution of GNs in terms of the luminosity distances $D_{\rm L}$ instead of the comoving distance, we adopt a spatially-flat $\Lambda$CDM cosmology model with parameters given in \citet{Planck2018} and follow the methodology introduced in \citet{GondanKocsis2021}.

\subsection{Mass range of supermassive black holes} 
\label{subsec:SMBH_MassRange} 
 
 Recent observations showed evidence for SMBHs with masses down to $\sim 5 \times 10^4 \Msun$ \citep{Baldassareetal2015}, while the highest SMBH mass ever reported is $\sim 6.6 \times 10^{10} \, \Msun$ \citep{Shemmeretal2004}. As SMBH mass distribution estimates\footnote{\citet{Shankaretal2004} and recent SMBH mass distribution estimates in the literature (e.g. \citealt{Grahametal2007,Shankaretal2009,Shankar2013,Uedaetal2014,Sijackietal2015,Thanjavuretal2016}) which are consistent with it within uncertainties.} can be generally extended to masses as low as of order $\sim 10^5 \, \Msun$ \citep{Barthetal2005,GreeneHo2006}, we set the lower bound of the SMBH mass range of interest to be $10^5 \, \Msun$. Regarding the upper bound, the differential merger rate of LK-induced BBHs in the local Universe $d \Gamma_{\rm locUniv} / d \Msmbh$ was investigated up to $\Msmbh = 10^9 \, \Msun$ \citep{Fragioneetal2019}, so we set the upper bound accordingly.\footnote{Note that the differential merger rate was investigated for SMBH masses down to $10^4 \, \Msun$ \citep{Hamersetal2018}.}

\subsection{Mass distribution of supermassive black holes} 
\label{subsec:SMBH_MassFunc} 
 
 Numerical simulations (e.g. \citealt{Sijackietal2015}) pointed out that  the SMBH mass distribution weakly evolves out to $z \simeq 2$ over the SMBH mass range $\Msmbh \in \left[ 10^5 \, \Msun, 10^9 \, \Msun \right]$. Since the majority of BBHs in the LK channel in GNs are expected to enter the aLIGO/AdV/KAGRA band with negligible eccentricities (e.g. \citealt{AntoniniPerets2012,Hamersetal2018,RandallXianyu2018a,Fragioneetal2019,Zhangetal2019,ArcaSedda2020}), they may be detected by single advanced GW detectors at design sensitivity to a maximum redshift of $z \simeq 2$ \citep{HallEvans2019}. Note that due to the strong dependence of the maximum distance of detection on binary parameters, the vast majority of sources are expected to be detected well below $z \sim 2$ that we check in Section \ref{subsec:DistaLIGOdet}.\footnote{For instance, see the results of \citet{GondanKocsis2021} for the characteristics of the redshift distribution for the GW capture channel in GNs.} Consequently, we neglect the redshift dependence of the SMBH mass distribution in our further investigations.
 
 Note that we sample SMBH masses from the SMBH mass range of interest with equal probability when generating mock MC samples of GN hosts as the target sample size per GN based on $d \Gamma_{\rm locUniv} / d \Msmbh$ that already accounts for the SMBH mass distribution (e.g. \citealt{AntoniniPerets2012,Hamersetal2018,Hoangetal2018,Fragioneetal2019}).

\subsection{Relaxed stellar populations around supermassive black holes} 
\label{subsec:GNs_Relax} 
 
 GNs are assumed to be relaxed systems of spherically symmetric, multi-mass, stellar populations gravitationally bound to a central SMBH. Studies showed that stellar objects undergo dynamical mass segregation and form an approximately power-law number density profile, $n(r,m) \propto r^{-\alpha(m)}$ within the SMBH's radius of influence (e.g. \citealt{BahcallWolf1977,AmaroSeoaneetal2004,Baumgardtetal2004,Freitagetal2006,HopmanAlexander2006,AlexanderHopman2009,Keshetetal2009,OLearyetal2009,PretoAmaroSeoane2010,AharonPerets2016,Alexander2017,Vasiliev2017,Baumgardtetal2018,FragioneSari2018,Panamarevetal2019,EmamiLoeb2020}). Observations of the stellar distribution in our Galactic Centre of old main-sequence stars justified the existence of a cusp with a slope consistent within uncertainties with theoretical expectations (e.g. \citealt{Schodeletal2007,Trippeetal2008,Gillessenetal2009,YusefZadehetal2012,Feldmeieretal2014,GallegoCanoetal2018,Schodeletal2018}).
 
 Recent studies have shown that heavy objects may form a disk-like structure within GNs due to vector resonant relaxation \citep{SzolgyeneKocsis2018,Gruzinovetal2020,Matheetal2022}. Such distribution was observed for young massive stars within the centre of the Milky Way (e.g. \citealt{Paumardetal2006,Bartkoetal2009,Doetal2013,Yeldaetal2014}) and other galaxies (e.g. \citealt{Sethetal2008,Lockhartetal2018}). Since the number density profiles of objects in disks have been uncertain, we follow previous studies (e.g. \citealt{AntoniniPerets2012,Hamersetal2018,Hoangetal2018,Fragioneetal2019}) and restrict our investigations to spherically symmetric stellar populations and BBH populations in the central regions of GNs with an SMBH in their centre in this study.
 
 The equilibrium state in GNs is reached within a few $\Gyr$ in stellar populations around SMBHs with either massive components or components in the outer regions of the GN \citep{Mastrobuono-Battisti2014,Panamarevetal2019,EmamiLoeb2020}. Specifically for our Galactic Centre, the spatial distribution of BHs can be characterized by the predicted power-law density profile after $\sim 5 \, \Gyr$ ($z \sim 1.2$) and a quasi-steady state is established for the outer regions after $\sim 3 \, \Gyr$ ($z \sim 2.2$) \citep{Panamarevetal2019}.\footnote{We use the inverse cosmic time - redshift relation \citep{Hogg1999} and the adopted spatially-flat $\Lambda$CDM cosmology model to associate redshift values to cosmic times.} Note that the majority of LK-induced BBHs are expected to merge relatively far from the SMBH (e.g. \citealt{Hamersetal2018,Fragioneetal2019}). Furthermore, the vast majority of these binaries are thought to be detected well below $z \sim 2$ with single advanced GW detectors at design sensitivity owing to the difference in detection distances for binaries with variant parameters that we also check in Section \ref{subsec:DistaLIGOdet}. Considering these arguments, we conclude that the assumption of an equilibrium state for stellar populations and the BBH population in GNs at relatively high redshift close to the astrophysical reach of advanced GW detectors may have a negligible effect on the parameter distributions of detectable systems. As a consequence of these arguments, we assume relaxed stellar populations and BBH populations around SMBHs and parameterize the components of the stellar populations as prescribed in \citet{GondanKocsis2021}.

\section{Binary Black Hole Populations Around Supermassive Black Holes}
\label{sec:InitDist_BBHs}
 
 In this section, we summarize the quantities of BBH populations in the central regions of GNs relevant to generating mock MC samples of LK-induced BBH mergers in single GN hosts.
 
 We start by introducing the notations we use to describe BBHs around SMBHs. We assume an inner binary orbiting an SMBH of mass $\Msmbh$ with the outer semi-major axis and eccentricity $a_{\rm out}$ and $e_{\rm out}$, respectively. The component masses of the inner binary are denoted by $m_A$ and $m_B$, where $m_B \leqslant m_A$, the total mass of the binary by \mbox{$M_{\rm tot} = m_A + m_B$}, and the mass ratio by $q = m_B / m_A$. The symmetric mass ratio and reduced mass satisfy $\eta = q / (1 + q)^2$ and $\mu = \eta M_{\rm tot}$, respectively, and the chirp mass is calculated as $\mathcal{M} = \mu^{3/5} M_{\rm tot}^{2/5}$. Finally, the inner semi-major axis and eccentricity are respectively denoted by $e_{\rm in}$ and $a_{\rm in}$, and $i$ defines the mutual inclination between the inner and outer orbit.

\subsection{Outer eccentricity and semi-major axis}
\label{subsec:ParamDist_Outer}
 
 Following previous studies (e.g. \citealt{PetrovichAntonini2017,Hamersetal2018,Hoangetal2018,Fragioneetal2019,Zhangetal2019,ArcaSedda2020,BubPetrovich2020}), we draw $e_{\rm out}$ from a thermal distribution \citep{Jeans1919}, i.e. \mbox{$f(e_{\rm out}) \propto e_{\rm out}$}. Note, however, that observations of young stars in the Galactic Centre indicate a much steeper profile, $f(e_{\rm out}) \propto e_{\rm out}^{2.6}$ \citep{Gillessenetal2009}. Therefore, we run additional MC simulations with $f(e_{\rm out}) \propto e_{\rm out}^{2.6}$ to assess the impact of $f(e_{\rm out})$ on the distributions of measurable binary parameters for detections in Section \ref{subsec:DistaLIGOdet}.
 
 Similar to the stellar population around SMBHs in relaxed GNs (Section \ref{subsec:GNs_Relax}), the BBH population is also expected to form an approximately power-law density cusp $n(r) \propto r^{-\alpha(m)}$ within $r_{\rm max}$ due to dynamical mass segregation, where lighter and heavier objects respectively develop shallower and steeper cusps. We adopt the parametrization $\alpha(m) = 3/2 + p(m)$ presented in \citet{OLearyetal2009}, where $p(m) = p_0 m /{\rm max}(m)$ with $p_0 \simeq 0.5 - 0.6$, and we set $p_0 = 0.5$ in our investigations assuming standard mass segregation.\footnote{Note that the 3D number density profile could be steeper due to star formation \citep{AharonPerets2016}, binary star disruption by the SMBH's tidal field \citep{FragioneSari2018}, or when the heavy objects are rare \citep{AlexanderHopman2009}.} For spherically symmetric GNs, the 3D number density profile of mass $m$ objects $n(r) \propto r^{-\alpha(m)}$ can be translated into a distribution of outer semi-major axis as $f(a_{\rm out}) \propto a_{\rm out}^{2 - \alpha(m)} \propto a_{\rm out}^{1/2 - p(m)}$ \citep{Schodeletal2003}. For possible comparison with the results of previous studies on the distributions of binary parameters \citep{AntoniniPerets2012,Hamersetal2018,Hoangetal2018,Fragioneetal2019}, we also consider cusp models with $n(r) \propto r^{-2}$ and $\propto r^{-3}$, where the corresponding $a_{\rm out}$ distributions are $f(a_{\rm out}) \propto a_{\rm out}^{0}$ and $\propto a_{\rm out}^{1}$, respectively. We keep the $\alpha = 3$ case to investigate the impact of extreme mass segregation \citep{Keshetetal2009} on the distributions of binary parameters. We sample $a_{\rm out}$ values from the $f(a_{\rm out})$ distribution between $10 \, \AU$ and $r_{\rm max}$. Recent studies found that $a_{\rm out}$ yields a steep distribution with a cut-off below a few tens of $\AU$ depending on $\Msmbh$ and assumptions on the parameter distributions of the BBH population (e.g. \citealt{Hoangetal2018,Fragioneetal2019}), which motivated our choice for the lower bound. The upper bound corresponds to the radius out to which the SMBH dominates the orbital evolution of stellar objects (Section \ref{sec:Props_GNs}).

\subsection{Inner eccentricity and semi-major axis}
\label{subsec:ParamDist_Inner}
 
 $e_{\rm in}$ is sampled from a uniform distribution. This choice is motivated by observations \citep{Raghavanetal2010} and that dynamical encounters may fail to thermalize the eccentricity distribution of the inner binary even in dense star clusters within the star cluster's lifetime \citep{Gelleretal2019}.
 
 According to Öpik's law \citep{Opik1924} and observations of massive stellar binaries in our Galactic Centre \citep{Sanaetal2012}, we sample $a_{\rm in}$ from a log-uniform distribution, i.e. $f(a_{\rm in}) \propto a_{\rm in}^{-1}$. Following \citet{Hoangetal2018}, we consider $a_{\rm in}$ between $0.1 \,\AU - 50 \, \AU$. Here, the lower bound is chosen such as to avoid common-envelope and mass transfer phases between stars before supernovae take place in which the orbital separation of surviving binaries significantly shrinks (e.g. \citealt{Dominiketal2012,EldridgeStanway2016,Woosley2017}). Regarding the upper limit, results of studies on the parameter distributions of BBHs forming in the LK channel in GNs pointed out that $a_{\rm in}$ yields a steep distribution with a cut-off beyond a few tens of $\AU$ depending on the SMBH mass and assumptions on the parameter distributions of BBH populations (e.g. \citealt{Hamersetal2018,Hoangetal2018,Fragioneetal2019,ArcaSedda2020}). Thus, $50 \, \AU$ is a reasonable upper limit for $a_{\rm in}$.

\subsection{Inclination and relevant angles}
\label{subsec:RelevantAngles}
 
 Following previous studies  (e.g. \citealt{AntoniniPerets2012,PetrovichAntonini2017,Hamersetal2018,Hoangetal2018,Fragioneetal2019,Zhangetal2019,ArcaSedda2020,BubPetrovich2020}), we draw $\cos{i}$ from a uniform distribution between $[-1,1]$. The other relevant angles, such as nodes and mean anomalies, together with the arguments of pericentre, are drawn from a uniform distribution between $[0, 2 \pi]$.

\subsection{Black hole masses}
\label{subsec:BHmasses}
  
 The lower bound of the BH mass range of interest is set to be \mbox{$m_{\rm min} = 5 \, \Msun$}, which is supported by recent GW observations (e.g. \citealt{Abbottetal2019,Abbotetal2021a,Abbotetal2021b}), and results of X-ray observations \citep{Bailynetal1998,Ozeletal2010,Farretal2011} and population synthesis studies (e.g. \citealt{Belczynskietal2012,Belczynskietal2016}). Both pair-instability supernova and pair-instability pulsation may limit the maximum BH mass to $\sim 50 \, \Msun$ in the stellar-mass range ($\gtrsim 130 \, \Msun$ in the intermediate-mass range; see e.g. \citealt{Belczynskietal2016,FishbachHolz2017,Marchantetal2018,Farmeretal2019,Woosley2019}), so we set the upper bound of the BH mass range of interest to be $m_{\rm max} = 50 \, \Msun$.\footnote{BHs with masses above $\sim 50 \, \Msun$ may form under special circumstances from metal-poor  \citep{Woosley2017,Giacobboetal2018,Mapellietal2020}, intermediate-metallicity \citep{Limongietal2018}, and high-metallicity stars \citep{Belczynskietal2020} such as by multiple generations of mergers (e.g. \citealt{MillerLauburg2009,Fishbachetal2017,McKernanetal2018,Rodriguezetal2018,GerosaBerti2019}).}
 
 The mass distribution of BHs in BBHs around SMBHs is still not well understood. Therefore, we follow previous studies (e.g. \citealt{Hoangetal2018,Fragioneetal2019}) and consider a power-law multi-mass distribution $f(m) \propto m^{-\beta}$ with $\beta \in \{ 1, 2, 3\}$. This model is also motivated by the recently inferred parameters of the mass distribution for the observed BBH population \citep{Abbottetal2019,Abbotetal2021a,Abbotetal2021b}.

\section{Timescales in galactic nuclei}
\label{sec:Timescales}
 
 In the dense stellar environment of a GN, various dynamical processes are relevant to the evolution of binaries near the centre, and even otherwise rare dynamical processes can take place and affect the evolution of binaries. This section gives a concise overview of the timescales associated with such processes and relevant to our further investigation.\footnote{Note that similar timescale calculations can be found in recent studies (e.g. \citealt{AntoniniPerets2012,Hamersetal2018}), and we refer the reader to \citet{Merritt2013} for a general overview of galactic nuclei dynamics.}
 
 The timescales of interest are as follows.
\begin{itemize}
   \item Regarding the systems investigated in this study, the LK timescale is associated with the secular evolution of binaries due to the perturbations by a central SMBH. It can be written at the quadrupole level of approximation in terms of the inner and outer orbital parameters and masses of the BBH and SMBH as
\begin{equation}   \label{eq:t_LK}
  t_{\rm LK} = \frac{8}{15 \pi} \frac{M_{\rm tot} + \Msmbh}{\Msmbh} \frac{P_{\rm out}^2}{P_{\rm in}} \left( 1 - e_{\rm out}^2 \right)^{3/2}
\end{equation}
   \citep{Antognini2015MNRAS}, where $P_{\rm in}$ and $P_{\rm out}$ are the periods of the inner and outer orbits, respectively.
   
   \item General relativistic (GR) precession (e.g.  \citealt{Weinberg1972}) in the inner binary is the most relevant effect that can potentially suppress the LK mechanism \citep{LithwickNaoz2011}. The timescale of this effect can be written in terms of the parameters of the inner binary as
\begin{equation}   \label{eq:t_GR}
  t_{\rm GR,inner} = \frac{ 2 \pi a_{\rm in}^{5/2} ( 1-e_{\rm in}^2 ) }{ 3 M_{\rm tot}^{3/2} } \, .
\end{equation}
   Note that relativistic precession in the outer orbit is typically negligible for the investigated triple systems of this study.
   
   \item Binaries may evaporate due to dynamical interactions with field stars and other compact objects in the dense environment of a GN; see e.g. \citet{Leighetal2018} for detailed investigations. The typical timescale of this process is given by 
\begin{equation}   \label{eq:t_evap}
  t_{\rm evap} = \frac{ \sqrt{3} \sigma(r) M_{\rm tot} }{ 32 \sqrt{\pi} a_{\rm in} n_{\rm dens,tot}(r) m_*(r) {\rm ln}\Lambda }
\end{equation}
   \citep{BinneyTremaine1987}, where the 1D velocity dispersion $\sigma(r)$ is determined using the Jeans equation as in \citet{AntoniniPerets2012}, and ${\rm ln}\Lambda$ is the Coulomb logarithm. The combined number density and the average mass for the mixture of relaxed stellar populations are respectively $n_{\rm dens,tot}(r)$ and $m_*(r)$, and these quantities are calculated as in \citet{Gondanetal2018b}.
   
   \item The timescale at which stellar populations in a GN approach a steady state with no memory of the initial conditions is the two-body relaxation timescale, which is given by
\begin{equation}   \label{eq:t_rlx}
  t_{\rm rlx} = \frac{ 0.34 \, \sigma^3(r) }{ n_{\rm dens,tot}(r) \, m^2_*(r) {\rm ln}\Lambda }
\end{equation}
   \citep{Spitzer1987}, where the second moment of the mass function for the mixture of relaxed stellar populations $m^2_*$ is computed as prescribed in \citet{Gondanetal2018b}.
   
   \item Vector resonant relaxation (VRR; \citealt{RauchTremaine1996}) randomizes non-coherently the direction of the external orbital plane that can possibly produce variations in the mutual inclination of the binary orbit with respect to its orbit around the SMBH. This effect can bring a low-inclination binary into an active LK regime in which binaries can merge via LK oscillations \citep{Hamersetal2018}. The VRR timescale can be given as
\begin{equation}   \label{eq:t_VRR}
  t_{\rm VRR} = \frac{ P_{\rm out} }{ 2 } \frac{ M_{\rm SMBH} }{ m_* \sqrt{ N_{\rm tot}(<r) } } 
\end{equation}
   \citep{Hamersetal2018}. Here, $N_{\rm tot}(<r)$ is the enclosed number of the mixture of stellar populations within radius $r$ and is computed as
\begin{equation}   \label{eq:Ntot}
  N_{\rm tot}(<r) = \int_0^r 4 l \pi^2 n_{\rm dens,tot}(l) dl \, ,
\end{equation}
   where $l$ also accounts for the radial distance from the central SMBH. Besides VRR, scalar resonant relaxation \citep{RauchTremaine1996} and two-body relaxation can also occur, changing the outer eccentricity and all outer orbital elements, respectively, but on timescales much larger than $t_{\rm VRR}$ (e.g. \citealt{KocsisTremaine2011,Hamersetal2018}). Accordingly, we can safely neglect these effects besides VRR when selecting the systems in which the BBH merges due to LK oscillations (Section \ref{subsec:MockSamples_GNs}).
   
   \item The merger timescale of an isolated binary with component masses and orbital parameters $\{ m_A, m_B, a, e \}$ can be given as 
\begin{equation}   \label{eq:t_merg}
  t_{\rm merg} = \frac{5 \, a^4 }{256 m_A m_B (m_A + m_B) } \frac{\left(1 - e^2 \right)^{7/2}}{\left(1 +  \frac{73}{24} e^2 + \frac{37}{96} e^4 \right)}
\end{equation}
   \citep{Peters1964}. We use $t_{\rm merg}$ to estimate the GW-driven infall of the inner binary into the SMBH. Regarding the systems investigated in this study, the merger timescale in terms of outer orbital parameters and masses of the BBH and the SMBH is
\begin{equation}   \label{eq:t_merg}
  t_{\rm merg,outer} = \frac{5 \, a_{\rm out}^4 }{256 M_{\rm tot} \Msmbh^2 } \frac{\left(1 - e_{\rm out}^2 \right)^{7/2}}{\left(1 +  \frac{73}{24} e_{\rm out}^2 + \frac{37}{96} e_{\rm out}^4 \right)} \, .
\end{equation}

\end{itemize} 
 
 Note that the listed timescales can be used to pinpoint the region inside the GN where BBHs may merge owing to the LK mechanism (e.g. \citealt{AntoniniPerets2012,PetrovichAntonini2017,Hamersetal2018}).

\section{Generation of Mock Binary Samples in Single Galactic Nuclei Hosts} 
\label{sec:MCsetup_SingleGN}
 
 We introduce a set of criteria to identify triple systems that are likely to produce BBHs merging due to the LK mechanism in Section \ref{subsec:SelectCrit}. In Section \ref{subsec:MockSamples_GNs}, we give an overview of the methodology that we use in generating mock MC samples of LK-induced BBH merger populations in single GNs. Finally, we compare or results with previous papers in Section \ref{subsec:CompPrevRes}.

\subsection{Selection criteria}
\label{subsec:SelectCrit}

 Following previous studies (e.g. \citealt{AntoniniPerets2012,Hamersetal2018,Hoangetal2018,Fragioneetal2019,Zhangetal2019,ArcaSedda2020}), we use a set of criteria (A-D) and consider new ones (E-F) in order to select the triple systems from MC samples in which BBHs are likely to merge due to LK oscillations.
\begin{enumerate}
  \item[(A)] The relative strengths of the octupole and quadrupole level of approximations, i.e. the octupole parameter $\epsilon$, should be below 0.1 in order to ensure dynamical stability for a triple system \citep{Naoz2016}, otherwise, the system can turn from integrable to chaotic \citep{Lieetal2011} which potentially induce orbital flips and even direct collisions (e.g. \citealt{Katzetal2011,LithwickNaoz2011}). Thus, we require for stable systems that
\begin{equation}   \label{eq:Crit_DynStab}
  \epsilon \equiv \frac{ a_{\rm in} }{ a_{\rm out} } \frac{ e_{\rm out} }{ 1 - e_{\rm out}^2 } \frac{ \arrowvert m_A - m_B \arrowvert }{ M_{\rm tot} } < 0.1 \, .
\end{equation}
   
  \item[(B)] An inner binary can be tidally disrupted by an SMBH if it crosses the Roche limit of the SMBH at its orbital pericentre distance (e.g. \citealt{Hills1988,AntoniniPerets2012,NaozSilk2014}). Accordingly, we require the Hill stability criterion \citep{Hill1878} when checking for stable systems, i.e.
\begin{equation}   \label{eq:Crit_Disrupt}
  \frac{ a_{\rm out} }{ a_{\rm in} } > \left( \frac{ 3 M_{\rm SMBH} }{ M_{\rm tot} } \right)^{1/3} \frac{ 1 + e_{\rm in} }{ 1 - e_{\rm out} } \, .
\end{equation}
 
  \item[(C)] LK cycles can be suppressed by relativistic precession in the inner orbit if the LK timescale is much longer than the GR precession timescale \citep{LithwickNaoz2011}. In this case, the inner binary evolves due to close encounters with field stars and compact objects in the GN, and the inspiral is caused by GW emission \citep{Naozetal2013}. To exclude such systems from the initial MC sample, we conservatively set the condition
\begin{equation}   \label{eq:Crit_Quench_GRprec}
  t_{\rm LK} < t_{\rm GR,inner} \, .
\end{equation}
  
  \item[(D)] An inner binary may evaporate due to close encounters with other stellar objects, such as field stars and compact objects in the dense environment of a GN. To take this effect into account, we follow \citet{Hamersetal2018} and allow $10$ LK cycles for a binary before evaporation,
\begin{equation}   \label{eq:Crit_Evap}
   10 \times t_{\rm LK} < t_{\rm evap} (r) \, .
\end{equation} 
  Regarding the initial MC sample, we conservatively set $r := a_{\rm out}$ to discard systems that are potentially evaporated by stellar objects. Note that the number of LK cycles allowed before evaporation is somewhat arbitrary. Therefore, we investigate its impact on the distributions of measurable binary parameters for detections in Sections \ref{subsec:DistaLIGOdet}.
  
  \item[(E)] We pointed out in Section \ref{subsec:GNs_Relax} that GNs can be considered relaxed within reach of advanced GW detectors. In this case, the merger timescale of the inner binary into the SMBH due to GW emission should be longer than the two-body relaxation timescale, which we conservatively estimate by setting $r := a_{\rm out}$. Accordingly, we require that
\begin{equation}   \label{eq:relax_tmergouter}
  t_{\rm merg,outer} > t_{\rm rlx}(a_{\rm out}) \, . 
\end{equation}
  
  \item[(F)] The impact of VRR on the dynamical behaviour of the system is negligible when the adiabatic parameter $\mathcal{R} = t_{\rm LK} / t_{\rm VRR}$ is negligible ($\mathcal{R} << 1$), and it gradually increases with $\mathcal{R}$. Based on the results of \citet{Hamersetal2018}, a significant eccentricity excitation can occur even when $\mathcal{R} \simeq 0.1$. Since we neglect VRR effects in our investigations, we conservatively keep systems from the initial MC sample that satisfy the condition
\begin{equation}   \label{eq:Crit_VRR}
  t_{\rm LK} < 0.1 \times t_{\rm VRR} \, .
\end{equation}
\end{enumerate}
 
 Note that the fraction of systems satisfying the conditions (A)-(F) slightly increases for larger $\mathcal{R}$ limits.

\subsection{Mock samples of binary black hole mergers in single galactic nuclei}
\label{subsec:MockSamples_GNs}
 
 In this section, we describe how we obtain an MC sample of binaries merging through the LK mechanism in a single GN host.
 
 We first choose the free parameters related to the SMBH $\{\Msmbh\}$ (Section \ref{subsec:SMBH_MassRange}) and the BBH population $\{ \alpha, \beta \}$ (Sections \ref{subsec:ParamDist_Outer} and \ref{subsec:BHmasses}) from their appropriate domain. Then, we sample $\mathcal{N}_{\rm init}$ inner and outer orbital parameters $\{ a_{\rm out}, e_{\rm out}, a_{\rm in}, e_{\rm in} \}$, component masses $\{ m_A, m_B \}$, and relevant angles from their distributions introduced in Section \ref{sec:InitDist_BBHs} to generate an initial MC sample of triple systems.
 
 Next, using the criteria introduced in Section \ref{subsec:SelectCrit}, we select the systems in which BBHs can survive long enough around the SMBH to merge due to LK oscillations. The introduced selection criteria significantly alter the initial parameter distributions of BBHs related to inner and outer orbital parameters and masses as the considered criteria incorporate these parameters (Sections \ref{sec:Timescales} and \ref{subsec:SelectCrit}). Accordingly, the distributions of relevant angles remain unchanged after the selection process. We use the set of remaining triple systems to initialize MC runs.
 
 For the successfully selected triples, we use the \emph{ARCHAIN} code \citep{MikkolaMerritt2006,MikkolaMerritt2008} to integrate the system of differential equations of motion of the three bodies. \emph{ARCHAIN} is a fully regularized code that includes PN corrections up to order PN2.5, and is able to model the evolution of systems of arbitrary mass ratios and eccentricities with high accuracy, even over long periods of time.\footnote{Note that \emph{ARCHAIN} has been updated in \citet{ArcaSeddaCapuzzo2019} most recently by taking into account both the gravitational field of the galaxy and the dynamical friction effect and was utilized to investigate the overall merger probability of BBHs around an SMBH \citep{ArcaSeddaGualandris2018,ArcaSeddaCapuzzo2019}.} Note that \emph{ARCHAIN} does not incorporate spin effects. We, therefore, consider non-spinning BBHs in this study. During the evolution of a triple system, we check for tidal disruption (Equation \ref{eq:Crit_Disrupt}) and evaporation (Equation \ref{eq:Crit_Quench_GRprec}) using root findings within the numerical integration of the ordinary differential equations \citep{Hamersetal2018} and stop the integration, whichever happens. Otherwise, we integrate the system either until they merge or for a duration of $10 \, \Gyr$, whichever is the shortest. We assume the merger of a BBH if the Schwarzschild radii of the two BHs overlap directly.
 
 For the merging BBH population, we use the peak GW frequency defined by \citet{Wen2003} as
\begin{equation}   \label{eq:fGW}
  f_{\rm GW} = \frac{ \sqrt{ M_{\rm tot} } }{ \pi }  (1 + e_{\rm in})^{1.1954}  \left( a_{\rm in} (1 - e_{\rm in}^2) \right)^{-3/2} 
\end{equation}
 to determine the residual eccentricity $e_{\rm 10 Hz}$ with which binaries enter the $10 \, \Hz$ frequency band of aLIGO/AdV/KAGRA \citep{Abbottetal2018} since eccentric binaries emit GWs with a broad frequency spectrum. For binaries at redshift $z$, the GW frequency scales as \mbox{$f_{\rm GW} \rightarrow f_{\rm GW} / (1 + z)$}, where $1/(1+z)$ is the cosmological scale factor. Taking into account the evolution of isolated binaries at redshift $z$, e.g. using the evolution equations of \citet{Peters1964}, one may find numerically that more massive binaries with larger initial orbital separations at higher redshifts enter the aLIGO/AdV/KAGRA band with systematically lower residual eccentricities. Finally, the output sample of $\mathcal{N}_{\rm merg}$ merging binaries contains the parameters $\{ a_{\rm out}, e_{\rm out}, a_{\rm in}, e_{\rm in}, e_{\rm 10Hz}, M_{\rm tot}, q, i \}$.
 
 We used the output distributions to fix $\mathcal{N}_{\rm init}$ based on their convergence and to validate the presented MC method as follows. First, we verified the convergence of the binary parameter distributions as a function of sample size by evaluating the Kolmogorov--Smirnov test with respect to the final distributions and set $\mathcal{N}_{\rm init}$ accordingly. Note that we selected a low $\mathcal{N}_{\rm init}$ value leading to convergence due to the high computational cost of simulations. Furthermore, we validated the presented MC method by comparing our results with that presented in \citet{Fragioneetal2019} for the merger fraction $f_{\rm merg} = \mathcal{N}_{\rm merg} / \mathcal{N}_{\rm init}$ and the distributions of the parameters $\{ a_{\rm out}, a_{\rm in}, i, M_{\rm tot}, e_{\rm 10 Hz} \}$ since we used the same software package for the evolution of triple systems. In the performed validation tests, we set the free parameters $\{ \Msmbh, \beta, \alpha \}$, binary parameter distributions, selection criteria for triple systems, assumptions for the stopping criteria, and the classification of outputs identical to that presented in \citet{Fragioneetal2019}.

\subsection{Comparison with previous results}
\label{subsec:CompPrevRes}
 
 We developed an MC method in Section \ref{subsec:MockSamples_GNs} to investigate the distributions of key binary parameters including their orbital parameters, mass-dependent parameters, and mutual inclination. There are some previous studies that also investigated the distributions of several key parameters \citep{AntoniniPerets2012,Hamersetal2018,Hoangetal2018,Fragioneetal2019,Zhangetal2019}. However, they typically used different methods to solve the equations of motion, applied somewhat different selection criteria, or considered different initial parameter distributions for orbital parameters and component masses. Consequently, only a qualitative comparison is possible with those results, which we discuss in this section. Regarding previous results, we set the free parameters $\{ \Msmbh, \beta, \alpha \}$ according to previous studies for possible comparison and restrict our interest to those results where the impact of VRR was excluded.
 
 We start by comparing our results for the inclination distribution with \citet{Hamersetal2018,Hoangetal2018,Fragioneetal2019}.  Consistent with these studies, most of the BBHs merge with initial inclinations $\sim 90^{\circ}$, and the obtained distribution is approximately symmetric to the peak.
 
 We find that the $a_{\rm out}$ distribution $P(a_{\rm out})$ is shifted toward higher and lower values compared to that in \citet{Hoangetal2018,Fragioneetal2019} and \citet{Hamersetal2018}, respectively. The former mainly comes from the fact that we did not cut off the initial $a_{\rm out}$ distribution at $0.1 {\rm pc}$ (Section \ref{subsec:ParamDist_Outer}), and the latter mainly arises due to the exclusion of binaries with $\mathcal{R} > 0.1$ (Section \ref{subsec:SelectCrit}). Note, further, that $P(a_{\rm out})$ yields a steep distribution with a cut-off well below $r_{\rm max}$ for all models considered in this study. Similarly, the $a_{\rm in}$ distribution $P(a_{\rm in})$ is also between those presented in \citet{Hamersetal2018} and \citet{Fragioneetal2019}. Furthermore, $\Msmbh$ (and $\alpha$) significantly affects both $P(a_{\rm out})$ and $P(a_{\rm in})$, resulting in lower semi-major axes for more massive SMBHs (and steeper number density profiles). However, these distributions are slightly shaped by $\beta$, and a steeper BH mass function leads to somewhat lower semi-major axes. These findings are consistent with that in \citet{Fragioneetal2019}.
 
 The obtained $M_{\rm tot}$ distribution $P(M_{\rm tot})$ is shifted toward lower masses compared to those in \citet{Fragioneetal2019}, mainly due to the considered lower $m_{\rm max}$ limit (Section \ref{subsec:BHmasses}). We find that $P(M_{\rm tot})$ is mainly governed by $\beta$, does not depend significantly on $\Msmbh$, and is slightly shaped by $\alpha$. Furthermore, a larger $\beta$ leads to lower binary masses, while a lighter SMBH results in more massive binaries. The obtained results and trends are in agreement with that presented in \citet{Fragioneetal2019}.
 
 Consistent with \citet{Hamersetal2018}, we find a roughly uniform mass ratio distribution $P(q)$. Note that $P(q)$ is closer to a uniform distribution for $\beta = 2$, and it is slightly tilted toward lower and higher values for $\beta = 1$ and $\beta = 3$, respectively. Furthermore, it is affected somewhat by both $\alpha$ and $\Msmbh$.
 
 Similar to the findings of \citet{AntoniniPerets2012} and \citet{Fragioneetal2019}, the $e_{\rm 10Hz}$ distribution $P(e_{\rm 10Hz})$ has a double peak at of order $e_{\rm 10Hz} \sim 10^{-2}$ and at $e_{\rm 10Hz} \sim 1$. By tracking back the evolution of systems with high eccentricities, we find consistent with \citet{AntoniniPerets2012} that the dynamics of these binaries are dominated by GW radiation within one LK cycle. Furthermore, for the representative case of a Milky Way - size nucleus with an SMBH mass of $\sim 4 \times 10^6 \, \Msun$ \citep{Gillessenetal2017}, $\sim 5 - 17\%$ of binaries enter the aLIGO/AdV/KAGRA band with $e_{\rm 10Hz} > 0.1$ depending on both $\beta$ and $\alpha$. This fraction is similar to that presented in \citet{AntoniniPerets2012} but much larger than that obtained in \citet{Hamersetal2018}. The latter may come from the fact that we get lower $a_{\rm in}$ values than \citet{Hamersetal2018} as discussed above, which results in larger residual eccentricities (Section \ref{subsec:MockSamples_GNs}). The presented fractions are somewhat lower than those in \citet{Fragioneetal2019}, which may be explained by the facts that we get larger $a_{\rm in}$ and lower $M_{\rm tot}$ values in our simulations, but $f_{\rm GW}$ has a stronger dependence on $a_{\rm in}$ (Equation \ref{eq:fGW}). Finally, we find that the fraction of binaries entering the aLIGO/AdV/KAGRA band mildly decreases with $\Msmbh$, which is consistent with the trend obtained in \citet{Zhangetal2019}.

\section{Generation of Mock Binary Catalogues} 
\label{sec:MC_Setup}
 
 We give an overview of the developed MC framework we use in generating mock catalogues of LK-induced BBH mergers in the local Universe in Sections \ref{subsec:MockSamples_locUniv}. Then, we extend the introduced framework to generate mock catalogues of merging binaries detectable by a single GW detector in Section \ref{subsec:MockSamples_DetUniv}.

\subsection{Mock catalogues of binary black hole mergers in the local Universe}
\label{subsec:MockSamples_locUniv}
 
 In this section, we discuss the generation of merging populations in the local Universe.
 
 We start by generating a mock MC sample of GNs. We first draw $N_{\rm GN} = 10^3$ SMBH masses with equal probability between $[10^5 \, \Msun, \, 10^9 \, \Msun]$ (Section \ref{subsec:SMBH_MassRange}) and select the remaining free parameters $\beta$ and $\alpha$ from their appropriate domain (Sections \ref{subsec:ParamDist_Outer} and \ref{subsec:BHmasses}). We assume identical GNs in terms of $\beta$ and $\alpha$ to generate model-related catalogues.
 
 In order to accurately model the merging BBH population in the local Universe, we generate a mock merger sample for each GN host with fixed parameters $\{ \Msmbh, \beta, \alpha \}$ in the mock GN sample as prescribed in Section \ref{subsec:MockSamples_GNs}: (i) we sample $\mathcal{N}_{\rm init}$ inner and outer orbital parameters, BH component masses, and relevant angles from their specific distributions, (ii) select triple systems that are likely to produce LK-induced BBH mergers, (iii) evolve the selected triples with the \emph{ARCHAIN} code in order to obtain a mock merger sample, (iv) and determine $e_{\rm 10Hz}$ for the merging binaries. For each GN host, the output parameters for the generated mock merger sample are $\{ a_{\rm out}, e_{\rm out}, a_{\rm in}, e_{\rm in}, e_{\rm 10Hz}, M_{\rm tot}, q, i \}$.
 
 An estimate for the target sample size per GN is also necessary to accurately model the merging BBH population in the local Universe. Recent studies have pointed out that the differential merger rate of LK-induced BBH (and other compact objects) mergers in GNs in the local Universe $d \Gamma_{\rm locUniv} / d \Msmbh$ decreases weakly with $\Msmbh$ for SMBH masses above $\sim 10^5 \, \Msun$ without VVR \citep{Hamersetal2018,Fragioneetal2019}. In this case, it can be parameterized in terms of $\Msmbh$ as $d \Gamma_{\rm locUniv} / d \Msmbh \propto f_{\rm merg} \Msmbh^{-1/4}$ \citep{Fragioneetal2019}. The differential merger rate is shaped by various SMBH mass distribution and stellar population-dependent quantities and dynamical processes\footnote{Such quantities and dynamical processes include the galaxy density, the fraction of galaxies containing an SMBH, the compact object supply rate, the fraction of stars forming compact object binaries, and the fraction of BBH mergers in simulations (e.g. \citealt{AntoniniPerets2012,Hamersetal2018,Fragioneetal2019}).}. However, its redshift dependence can be neglected in further investigations as the redshift dependence of the SMBH mass distribution is also negligible (Section \ref{subsec:SMBH_MassFunc}) and relaxed GNs can be assumed (Section \ref{subsec:GNs_Relax}). Note that $f_{\rm merg}$ somewhat depends on $\{ \Msmbh, \beta, \alpha \}$, but we conservatively neglect these dependencies as they are not significant \citep{Fragioneetal2019}. Taking it all together, we choose an MC sample size of a mock merger sample in a single GN host $\mathcal{N}_{\rm 1GN}$ by setting a fiducial MC sample size $\mathcal{N}_{\rm 1GN,fid}$ for a fiducial GN host with $M_{\rm SMBH} = 10^5 \, \Msun$ and then assigning $\mathcal{N}_{\rm 1GN}$ to a GN of mass $\Msmbh$ according to $d \Gamma_{\rm locUniv} / d \Msmbh$ as
\begin{equation}   \label{eq:SampSize_LocUniv}
   \mathcal{N}_{\rm 1GN} = \Bigg\{ \mathcal{N}_{\rm 1GN,fid} \times \left( \frac{ 10^5 \, \Msun }{ \Msmbh } \right)^{1/4} \Bigg\} \, .
\end{equation}
 Here, the bracket $\{ \, \}$ denotes the floor function. Finally, we randomly select $\mathcal{N}_{\rm 1GN}$ merger events from each mock BBH merger sample in order to construct the catalogue of LK-induced BBHs in the local Universe. The selected binaries define the mock catalogue of BBH mergers for a given $(\beta, \alpha)$ model, and the catalogue contains the parameters $\{ a_{\rm out}, e_{\rm out}, a_{\rm in}, e_{\rm in}, e_{\rm 10Hz}, M_{\rm tot}, q, i \}$.
 
 We fixed the $\{ \mathcal{N}_{\rm init}, \mathcal{N}_{\rm 1GN,fid} \}$ pair in preliminary MC runs such as to ensure the convergence of binary parameter distributions as a function of sample size in the Kolmogorov–Smirnov test. Accordingly, the binary catalogue contains $\sim 5 - 7 \times 10^4$ BBH mergers depending on $\beta$ and $\alpha$. Note that the generated catalogues contain enough merger events to ensure the convergence of correlations among various source parameters; see Section \ref{subsec:CorrBBHparams} for details.

\subsection{Mock catalogues of binary black hole mergers detectable by a single GW detector}
\label{subsec:MockSamples_DetUniv}
 
 In this section, we introduce the MC method with which we generate mock catalogues of LK-induced BBH mergers detectable by a single GW detector.
 
 We start by adopting a Cartesian coordinate system to the selected L-shaped interferometer, as discussed in \citet{GondanKocsis2021}, in order to be able to assign sky positions to GNs in the detection volume and to define angular momentum unit vectors to binaries with respect to the interferometer.
 
 Next, we generate a mock MC sample of GN hosts in the detection volume. We first draw $N_{\rm GN} = 10^4$ SMBH masses with an equal probability between $[10^5 \, \Msun, \, 10^9 \, \Msun]$ (Section \ref{subsec:SMBH_MassRange}). Then, we assign luminosity distances to each SMBH by sampling $D_{\rm L}$ from the detection range of the selected advanced GW detector (Section \ref{subsec:SMBH_SpatDist}) using the probability distribution function derived in \citet{GondanKocsis2021}. Redshift values are also assigned to each SMBH by using the inverse luminosity distance--redshift relation \citep{Hogg1999} and the adopted spatially-flat $\Lambda$CDM cosmology model. Furthermore, we generate an isotropic random sample of their sky position angles $\theta_N$ and $\phi_N$ (Section \ref{subsec:SMBH_SpatDist}) relative to the detector by drawing $\cos{\theta_N}$ and $\phi_N$ from a uniform distribution between $[-1, 1]$ and $[0, 2 \pi]$, respectively. Finally, we select $\beta$ and $\alpha$ from their appropriate domain (Sections \ref{subsec:ParamDist_Outer} and \ref{subsec:BHmasses}) and assign them to each GN host to generate model-related catalogues.
 
 For each GN host in the mock GN sample, we generate a mock MC sample of BBH mergers as prescribed in Section \ref{subsec:MockSamples_GNs}. In order to accurately model the merging populations in the detection volume, we combine $\mathcal{N}_{\rm 1GN}$ (Equation \ref{eq:SampSize_LocUniv}) with the cosmological scale factor $1/(1+z)$ to get the target number of sampled binaries per GN with $\{ \Msmbh, z\}$ as
\begin{equation}   \label{eq:SampSize_DetVol}
   \mathcal{N}_{\rm 1GN,z} = \Bigg\{ \frac{ N_{\rm 1GN,fid,z} }{ 1 + z} \times \left( \frac{ 10^5 \, \Msun }{ \Msmbh } \right)^{1/4} \Bigg\} \, .
\end{equation}
 Then, for each GN host, we randomly select $\mathcal{N}_{\rm 1GN,z}$ merger events from the mock BBH merger sample, and assign the $\{ z, D_{\rm L}, \theta_N, \phi_N \}$ parameters of the GN host to the binaries in the sample. Besides $\{ \theta_N, \phi_N \}$, the angular momentum unit vector angles $\{ \theta_L, \phi_L \}$ and the polarization angle $\psi$ are also necessary to determine the SNR through the antenna pattern functions $F_+ $ and $F_{\times}$ of a GW detector. Accordingly, for each selected binary, we extract $\{ \theta_L, \phi_L \}$ from their spatial coordinates before the merger, and we measure these angles in the GW detector frame due to the consideration of spherically symmetric GNs (Section \ref{subsec:GNs_Relax}). Furthermore, we assign a $\psi$ value to each binary by randomly sampling it between $[0, 2 \pi]$.\footnote{Note that $\psi$ is practically unknown because the GW's polarization itself is typically unknown \citep{Thorne1987}. Therefore, we assume a uniform distribution of $\psi$ between $[0, 2 \, \pi]$.}
 
 In order to select the detectable binaries from the obtained merger sample, we calculate the SNR values. To this end, we determine the antenna pattern functions $F_+$ and $F_{\times}$ for binaries with $\{ \theta_N, \phi_N, \theta_L, \phi_L, \psi \}$ and the selected L-shaped GW detector as introduced in \citet{Thorne1987}. Furthermore, we use the waveform generator described in \citet{Eastetal2013} to simulate full inspiral-merger-ringdown waveforms of BBHs with nonzero eccentricities. Note that the adopted waveform generator includes generic spin configurations, but we simulate BBHs with zero spins as we neglect spin effects in this study (Section \ref{subsec:MockSamples_GNs}). We start generating the $h_+$ and $h_{\times}$ waveform components when the peak GW frequency enters the aLIGO/AdV/KAGRA band and extract the necessary parameters from simulations to initialize the waveform generator accordingly. The measured strain is given by $h = h_+ F_+ + h_{\times} F_{\times}$, and the SNR using a perfectly matched filter is computed as
\begin{equation}   \label{eq:SNR_onedet} 
  \mathrm{SNR} = \left( 4 \int_0 ^{+\infty } \frac{|\tilde{h}(f)|^2 }{S_h(f)} df \right)^{1/2} \, .
\end{equation} 
 Here, $f$ is the GW frequency in the observer frame, $|\tilde{h}(f)|$ is the amplitude of the eccentric waveform model in the frequency domain, and $S_h(f)$ is the one-sided power spectral density of an advanced GW detector at design sensitivity \citep{Abbottetal2018}. A simplifying, while still satisfactorily accurate, detection criterion commonly used in the astrophysical literature involves computing the ${\rm SNR}$ of events and assuming detection if ${\rm SNR} > {\rm SNR}_{\rm lim}=8$ for a single GW detector (e.g. \citealt{Abadieetal2010,OShaughnessyetal2010,Dominiketal2015,Belczynskietal2016b}). Thereby, we keep only those binaries in the sample that satisfy the detection criterion. Finally, the remaining set of binaries defines the mock catalog of detectable BBH mergers for a given $(\beta, \alpha)$ model, and the output parameters are $\{ a_{\rm out}, e_{\rm out}, a_{\rm in}, e_{\rm in}, e_{\rm 10Hz}, M_{\rm tot}, q, i, z, D_{\rm L} \}$.
 
 Similar to the case of mock catalogues in the local Universe (Section \ref{subsec:MockSamples_locUniv}), we also fixed the $\{ \mathcal{N}_{\rm init}, \mathcal{N}_{\rm 1GN,fid,z} \}$ pair in preliminary MC runs to ensure the convergence of final binary parameter distributions as a function of sample size in the Kolmogorov–Smirnov test. Accordingly, the mock binary catalogue contains $\sim 7 - 9 \times 10^4$ BBHs depending on $\beta$ and $\alpha$. Note that the generated mock catalogues contain enough mergers to ensure the convergence of correlations among various source parameters (Section \ref{subsec:CorrBBHparams}).

\section{Results} 
\label{sec:Results}
 
 We present the distributions of orbital parameters and mass-dependent parameters for binaries detectable by aLIGO in Section \ref{subsec:DistaLIGOdet}, and possible correlations among various source parameters for aLIGO detections are investigated in Section \ref{subsec:CorrBBHparams}.

\subsection{Distributions of binary parameters for Advanced LIGO detections}
\label{subsec:DistaLIGOdet}
 
 In this section, we present results for the parameter distributions of LK-induced BBHs as seen by aLIGO, focusing on the parameter set measurable by advanced GW detectors.
 
 GWs of compact binaries encode masses, mass ratios, spins, and luminosity distances (e.g. \citealt{CutlerFlanagan1994,PoissonWill1995}). Accordingly, we present results for the commonly investigated mass-dependent parameters $\{ M_{\rm tot}, q, \mathcal{M} \}$ but neglect spins as we consider non-spinning binaries in this study (Section \ref{subsec:MockSamples_GNs}). Furthermore, we present results for redshift $z$ instead of $D_{\rm L}$ in order to present the cosmological distribution of LK-induced BBH mergers. Besides the listed parameters, residual eccentricity at $10 \, \Hz$ (e.g. \citealt{Gondanetal2018a,Loweretal2018,GondanKocsis2019,RomeroShaw2019}) and initial orbital parameters \citep{Gondanetal2018a,GondanKocsis2019}, such as the initial dimensionless pericenter distance and initial orbital eccentricity, can also be extracted from the GWs of compact binaries. Consequently, we present results for $e_{\rm 10Hz}$. However, we neglect initial orbital parameters since an LK-induced binary decouples gradually from the SMBH as GW radiation starts to dominate the dynamics over the LK mechanism that prevents the unambiguous definition of initial orbital parameters for the inspiral phase.

\begin{table}
\centering  
   \begin{tabular}{@{}cc|cccc}
     \hline
      $\beta$ & $\alpha$ & aLIGO det. & AdV det. & KAGRA det. & local Univ.  \\
     \hline\hline
     $1$ & $\alpha(m)$ & $2.6 \%$ & $2.8 \%$ & $3.2 \%$ & $6.7 \%$  \\
     $2$ & $\alpha(m)$ & $4.3 \%$ & $5.2 \%$ & $4.8 \%$ & $9.3 \%$  \\
     $3$ & $\alpha(m)$ & $6.8 \%$ & $7.4 \%$ & $7.1 \%$ & $12.3\%$  \\
     $1$ &     $2$     & $4.5 \%$ & $6.1 \%$ & $5.9 \%$ & $11.3\%$  \\
     $2$ &     $2$     & $7.1 \%$ & $8.3 \%$ & $8.2 \%$ & $16.3\%$  \\
     $3$ &     $2$     & $11.5\%$ & $12.1\%$ & $11.2\%$ & $19.7\%$  \\
     $1$ &     $3$     & $3.7 \%$ & $6.2 \%$ & $4.2 \%$ & $8.8 \%$  \\
     $2$ &     $3$     & $5.9 \%$ & $7.0 \%$ & $6.4 \%$ & $13.3\%$  \\
     $3$ &     $3$     & $9.8 \%$ & $10.5\%$ & $9.6 \%$ & $17.1\%$  \\
   \end{tabular} 
   \caption{The fraction of LK-induced BBH sources having residual eccentricities $e_{\rm 10Hz}$ larger than $0.1$ when their GW signals enter the $10 \, {\rm Hz}$ frequency band of advanced GW detectors. Results are presented for BBH mergers with $\rm S/N > 8$ for detections with a single aLIGO (AdV, KAGRA) detector at design sensitivity and for BBH mergers per unit volume in the local Universe. Isotropic GNs with different BH mass functions ($m^{-\beta}$) and 3D number density profiles ($r^{-\alpha}$) of the BBH population are considered as labelled in the first two columns, and a BH mass range of $[5 \, \Msun, 50 \, \Msun]$ is assumed for the BBH component masses.
   \label{tab:Frac_e10Hz_0p1} }
\end{table}

\begin{table}
\centering  
   \begin{tabular}{@{}cc|cccc}
     \hline
      $\beta$ & $\alpha$ & aLIGO det. & AdV det. & KAGRA det. & local Univ.  \\
     \hline\hline
     $1$ & $\alpha(m)$ & $1.1 \%$ & $1.2 \%$ & $1.1 \%$ & $1.9 \%$  \\
     $2$ & $\alpha(m)$ & $1.7 \%$ & $1.8 \%$ & $1.6 \%$ & $2.7 \%$  \\
     $3$ & $\alpha(m)$ & $2.3 \%$ & $2.4 \%$ & $2.1 \%$ & $3.3 \%$  \\
     $1$ &     $2$     & $0.2 \%$ & $0.3 \%$ & $0.2 \%$ & $0.8 \%$  \\
     $2$ &     $2$     & $0.4 \%$ & $0.5 \%$ & $0.4 \%$ & $1.7 \%$  \\
     $3$ &     $2$     & $1.1 \%$ & $1.2 \%$ & $0.9 \%$ & $2.6 \%$  \\
     $1$ &     $3$     & $1.7 \%$ & $2.1 \%$ & $1.7 \%$ & $3.0 \%$  \\
     $2$ &     $3$     & $2.5 \%$ & $3.1 \%$ & $2.4 \%$ & $4.6 \%$  \\
     $3$ &     $3$     & $3.9 \%$ & $4.3 \%$ & $3.5 \%$ & $6.0 \%$  \\
   \end{tabular} 
   \caption{Same as Table \ref{tab:Frac_e10Hz_0p1}, but for the fraction of sources having $e_{\rm 10Hz} > 0.9$.  \label{tab:Frac_e10Hz_0p9} }
\end{table}

\begin{figure*}
    \centering
    \includegraphics[width=80mm]{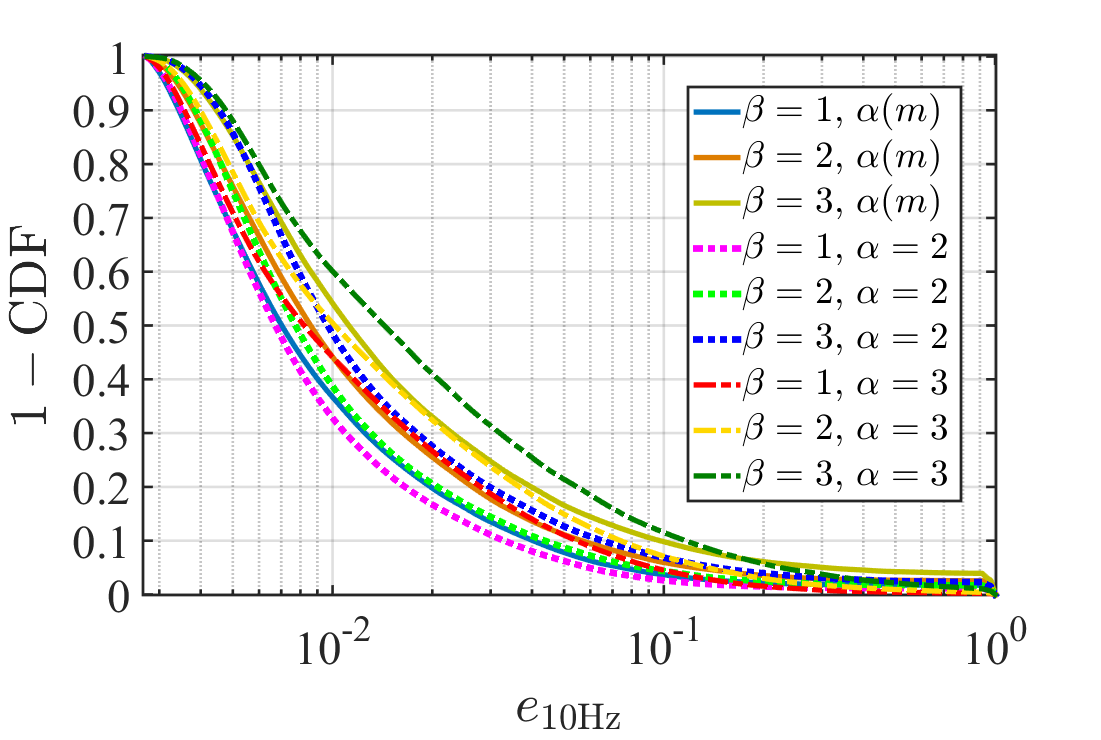}
    \includegraphics[width=80mm]{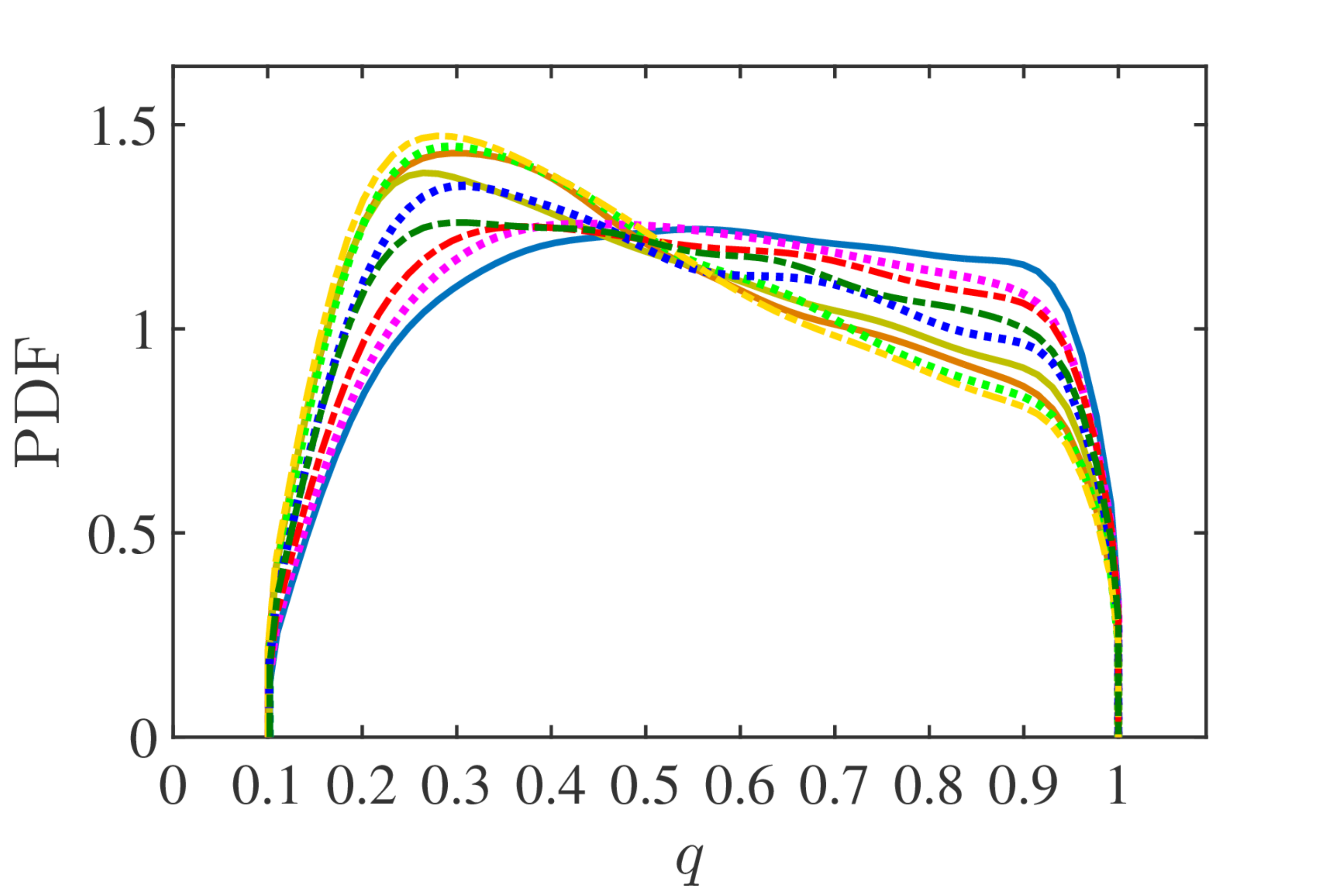}
    \\
    \includegraphics[width=80mm]{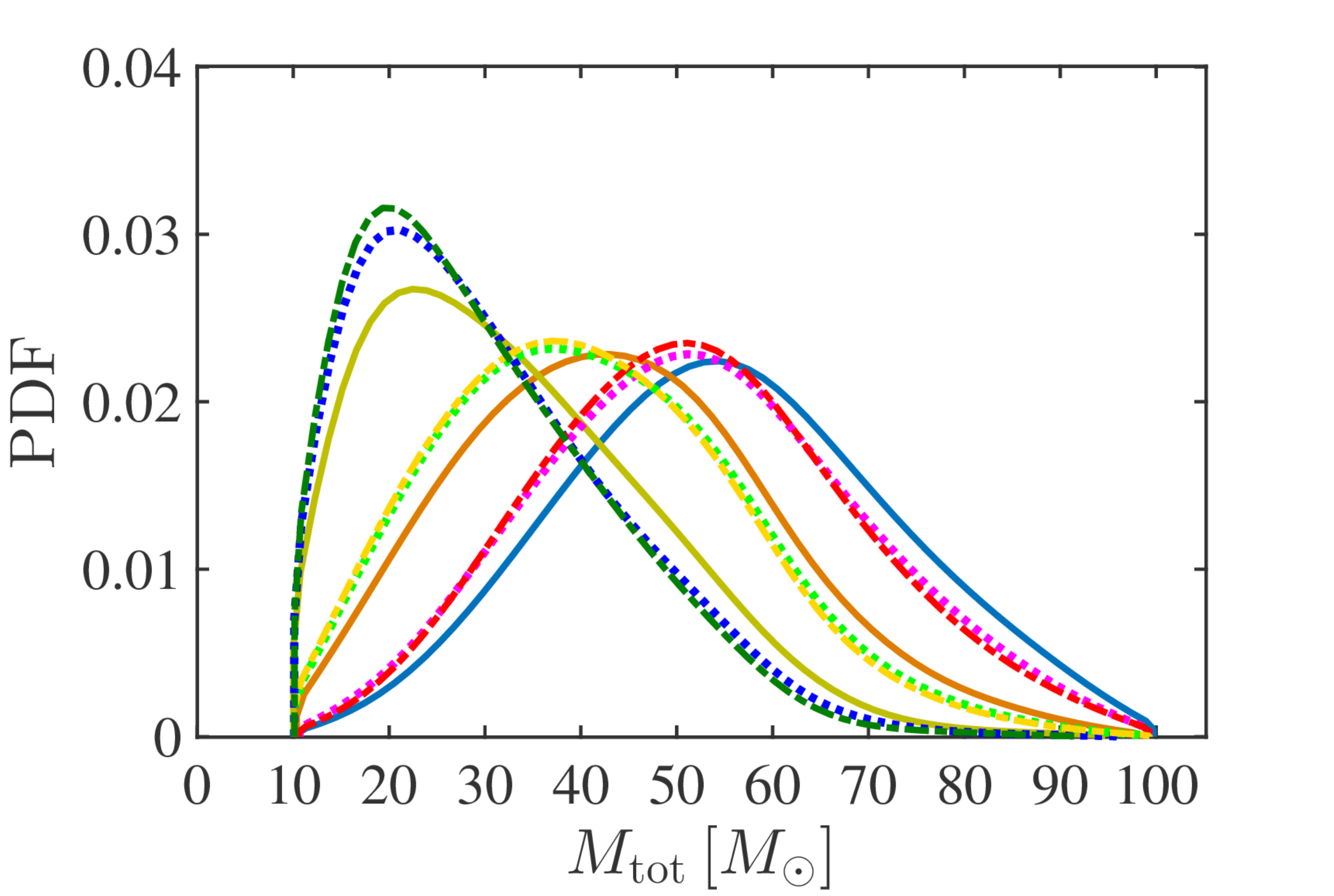}
    \includegraphics[width=80mm]{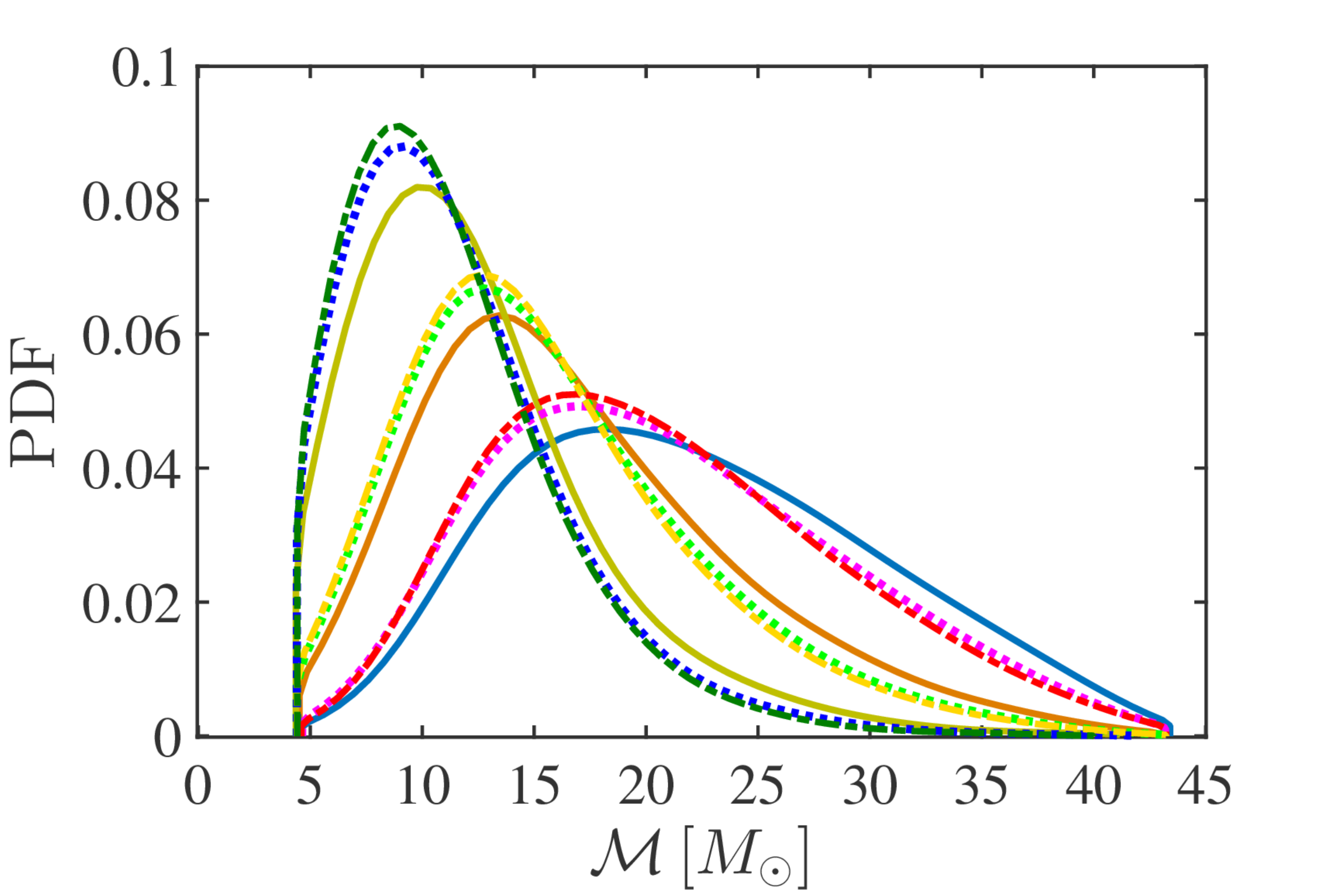}
    \\
    \includegraphics[width=80mm]{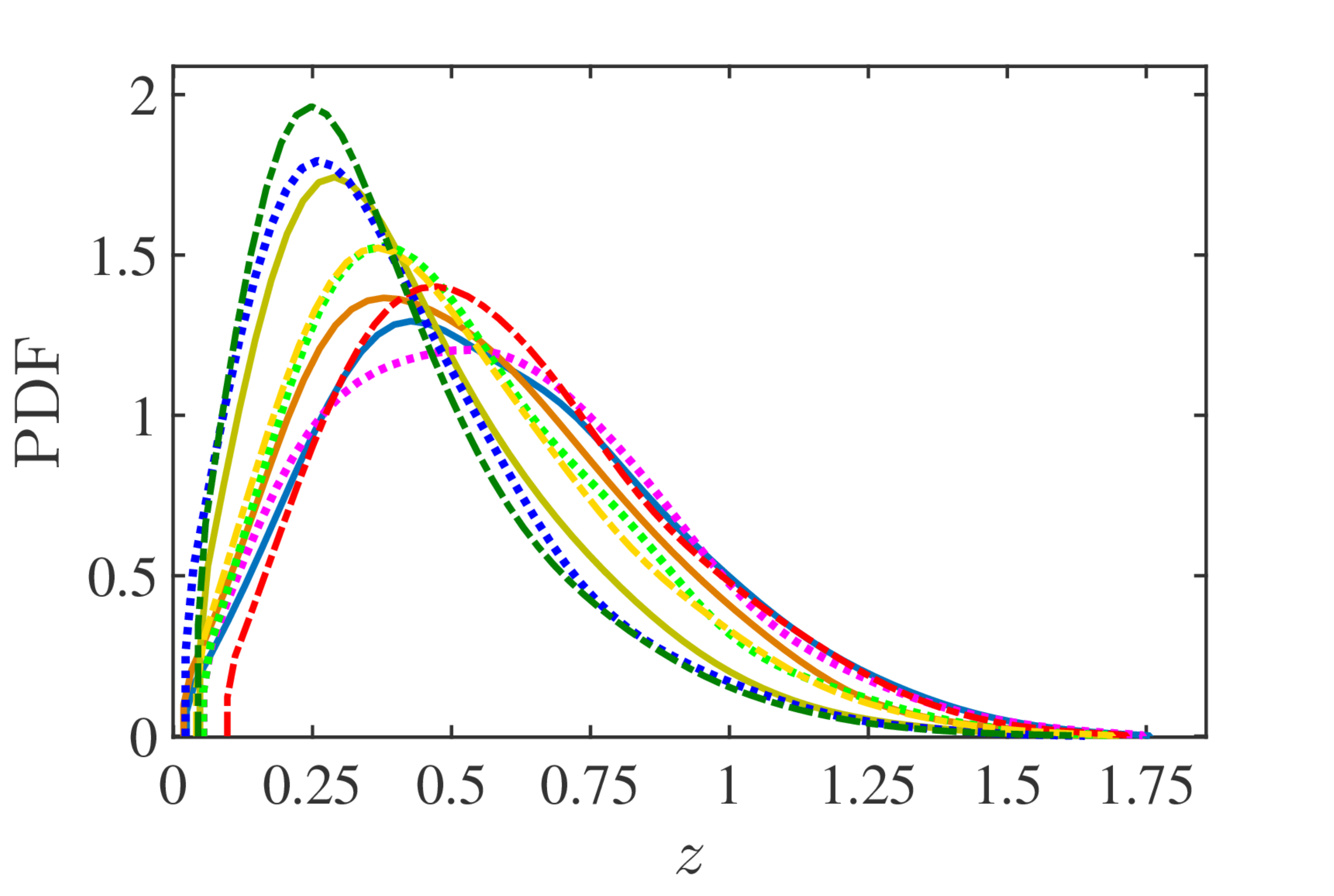}
\caption{Distributions of binary parameters with $\rm S/N > 8$ for detections with a single aLIGO detector at design sensitivity for LK-induced BBH mergers in isotropic GNs assuming $5 \, \Msun \leqslant \ m \leqslant 50 \, \Msun$ for the BBH component masses: the $1 - {\rm CDF}$ of $e_{\rm 10 Hz}$ (row 1, left), mass ratio $q$ (row 1, right), total mass $M_{\rm tot}$ (row 2, left), chirp mass $\mathcal{M}$ (row 2, right), and redshift $z$ (row 3). Results are presented for GN models with different BH mass functions ($m^{-\beta}$) and 3D number density profiles ($r^{-\alpha}$) of the BBH population as labelled in the top left panel. \label{fig:ParamDist_aLIGOVolume} } 
\end{figure*}

\begin{figure*}
    \centering
    \includegraphics[width=80mm]{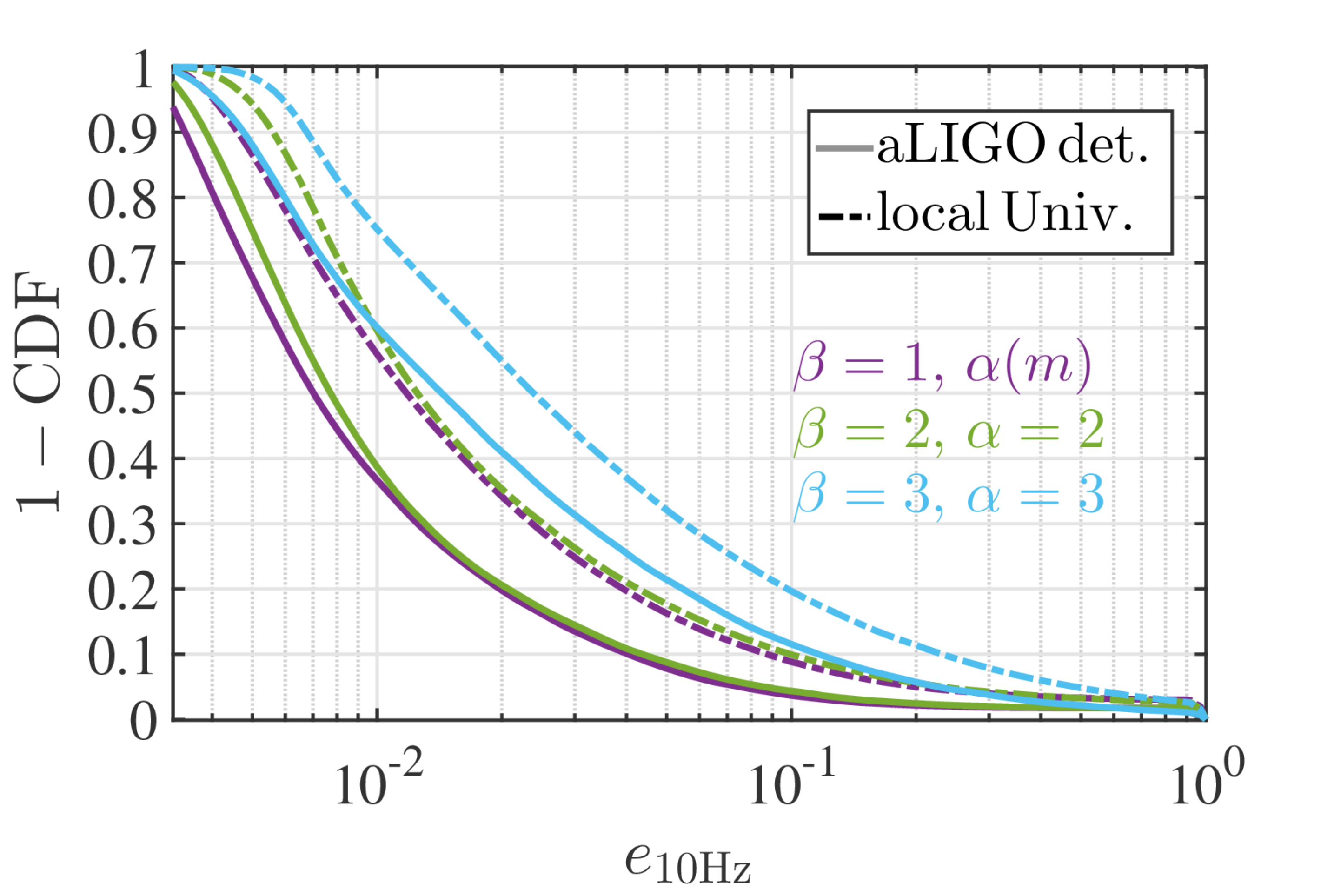}
    \includegraphics[width=80mm]{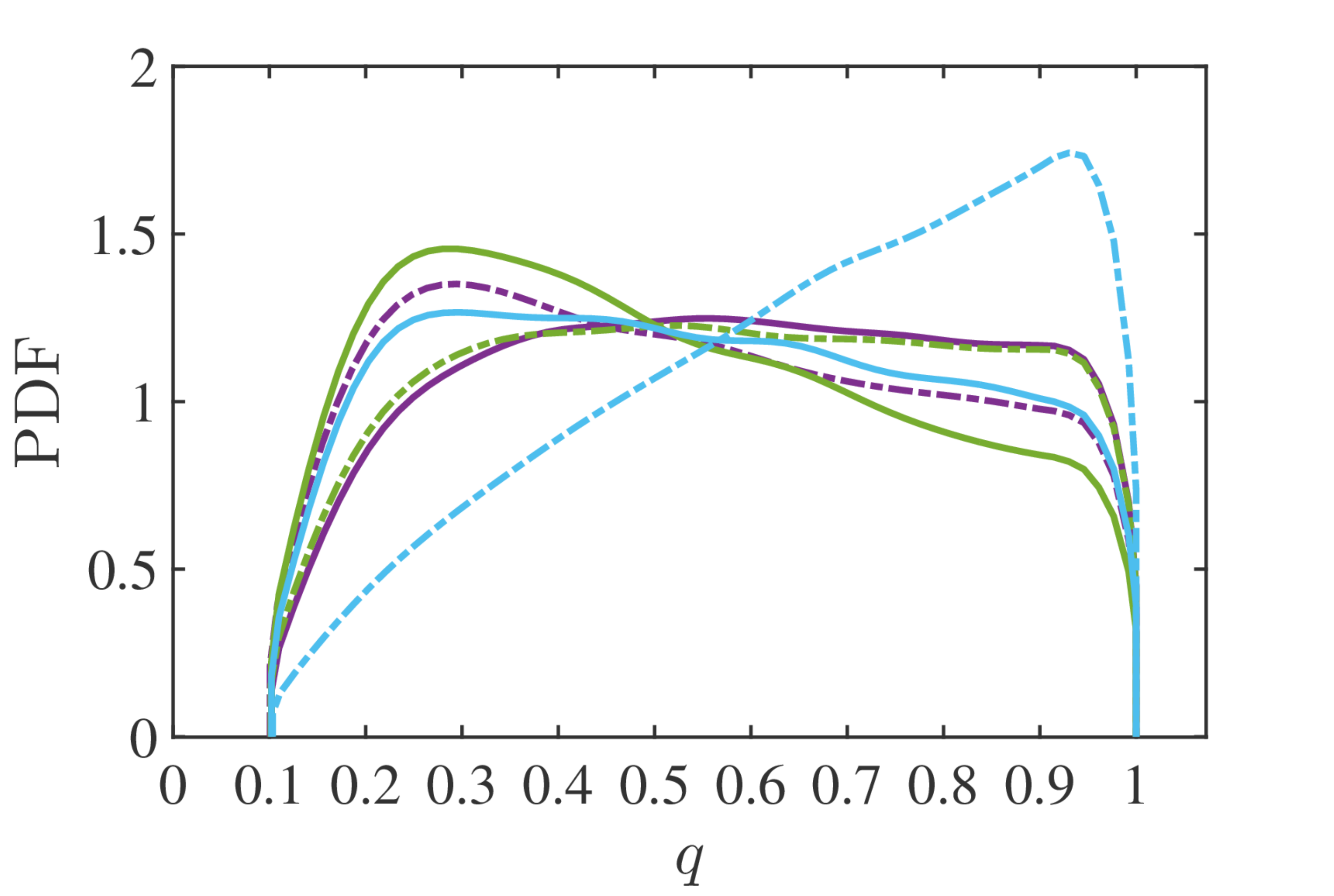}
    \\
    \includegraphics[width=80mm]{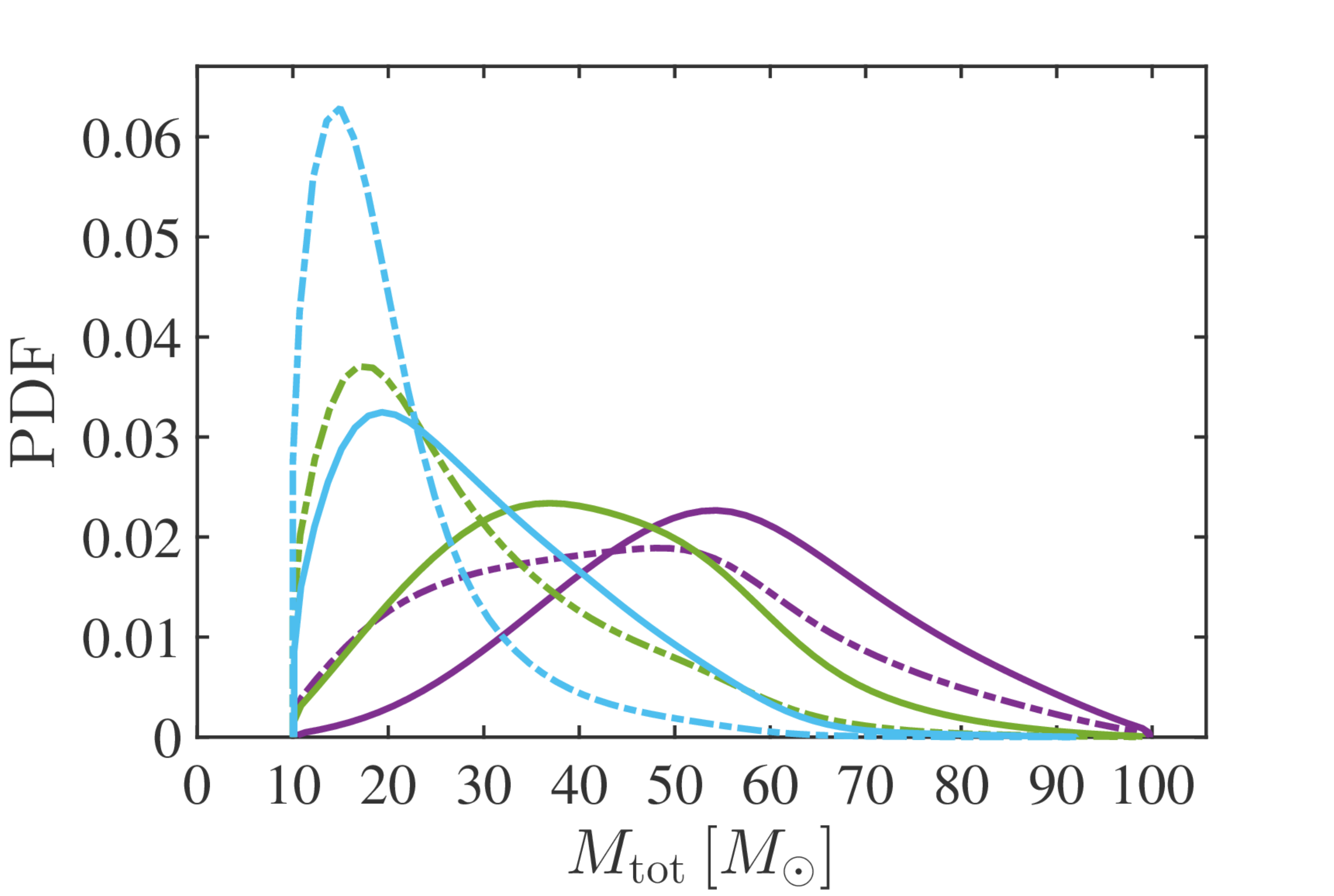}
    \includegraphics[width=80mm]{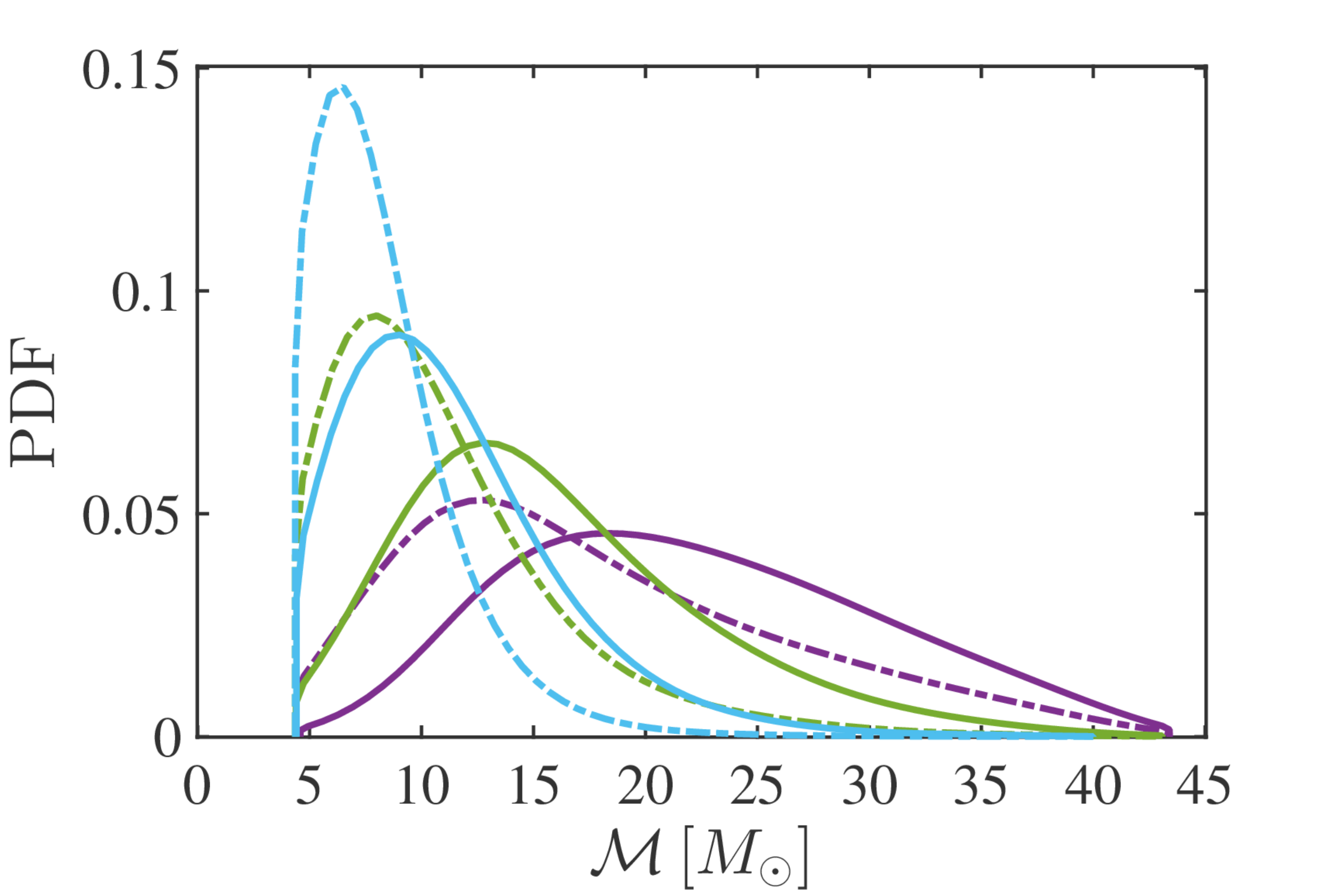}
\caption{The impact of observational bias on the distributions of binary parameters for various examples as in Figure \ref{fig:ParamDist_aLIGOVolume} (labelled in the top left panel): the $1 - {\rm CDF}$ of $e_{\rm 10 Hz}$ (row 1, left), mass ratio $q$ (row 1, right), total mass $M_{\rm tot}$ (row 2, left), and chirp mass $\mathcal{M}$ (row 2, right). Solid lines correspond to distributions of detectable mergers, and dash-dotted lines correspond to mergers per unit volume in the local Universe.  \label{fig:Impact_ObsBias_ParamDists} } 
\end{figure*}

 The distributions of considered parameters as seen by a single aLIGO detector at design sensitivity are presented in Figure \ref{fig:ParamDist_aLIGOVolume}, and the impact of observational bias on the parameter distributions are illustrated in Figure \ref{fig:Impact_ObsBias_ParamDists}. As seen, the results are significantly different from the detector-independent merger rate density in the local Universe.
\begin{itemize}
   \item The $M_{\rm tot}$ and $\mathcal{M}$ distributions $P(M_{\rm tot})$ and $P(\mathcal{M})$ (middle panels in Figure \ref{fig:ParamDist_aLIGOVolume}) for the detectable binaries significantly depend on the slope of the BH mass function $\beta$ and are weakly affected by the slope of the binary spatial distribution around the SMBH, $\alpha$. These characteristics arise from the fact that mass distributions of LK-induced BBH mergers in single GNs are governed by $\beta$, do not depend significantly on $\Msmbh$, and are weakly sensitive to $\alpha$ \citep{Fragioneetal2019}. Furthermore, a higher $\beta$ leads to systematically lower masses. Both the $M_{\rm tot}$ and $\mathcal{M}$ distributions are widened and tilted toward higher masses in comparison to the corresponding distributions of mergers per unit volume in the local Universe (bottom panels in Figure \ref{fig:Impact_ObsBias_ParamDists}) since more massive binaries with higher mass ratios are detected at larger distances (e.g. \citealt{HallEvans2019}). For instance, $P(M_{\rm tot})$ peaks at $M_{\rm tot} \sim \{ 15 \, \Msun, 18 \, \Msun, 45 \, \Msun \}$ for $\beta = \{3, 2, 1 \}$ for mergers in the local Universe, respectively, while the corresponding detected distributions peak at $M_{\rm tot} \sim \{20 \, \Msun, 40 \, \Msun, 55 \, \Msun \}$, respectively. The same trends characterise mass distributions in AdV and KAGRA detections, but the observational selection effect has less impact on them as both AdV and KAGRA have lower astrophysical reach than aLIGO. Note that the same trends characterise the distribution of reduced mass as well. As a consequence, we conclude that LK-induced BBH mergers in GNs may be detected with roughly medium masses.
   
   \item The mass ratio distribution $P(q)$ for detectable mergers (top right panel in Figure \ref{fig:ParamDist_aLIGOVolume}) is roughly uniform, and it depends on $\beta$ but is barely shaped by $\alpha$. Note that $P(q)$ is closer to a uniform distribution for $\beta = 1$, and it skews toward lower $q$ values and peaks at $q \sim 0.2 - 0.3$ for $\beta = 2, 3$. Similar to the case of $P(M_{\rm tot})$ and $P(\mathcal{M})$, these characteristics are also determined by the $\beta$ and $\alpha$ dependence of mass distributions of LK-induced BBHs in single GNs. Furthermore, observational bias generally skews $P(q)$ towards a uniform distribution (top right panel in Figure \ref{fig:Impact_ObsBias_ParamDists}), which mainly arises from the fact that observational bias widens and skews toward higher masses the mass distributions of mergers in the local Universe. Thereby a larger fraction of detectable mergers possesses mid-range $q$ values. The same trends apply for $P(q)$ in AdV and KAGRA detections, but in these cases, $p(q)$ is less uniform since the observational selection effect has less impact on the mass distribution.
   
   \item The eccentricity distribution at $10 \, \Hz$, $P(e_{\rm 10Hz})$, as seen by aLIGO (top left panel in Figure \ref{fig:ParamDist_aLIGOVolume}), depends on both $\beta$ and $\alpha$ and has a double peak at of order $10^{-3} - 10^{-2}$ and at $e_{\rm 10Hz} \sim 1$. For systems with high eccentricities, the dynamics is dominated by GW radiation within one LK cycle (Section \ref{subsec:CompPrevRes}). Moreover, $2.6 - 11.5 \%$ of binaries have $e_{\rm 10 Hz} > 0.1$ depending on $\beta$ and $\alpha$ (Table \ref{tab:Frac_e10Hz_0p1}), and $< 3.9 \%$ of them possess $e_{\rm 10 Hz} > 0.9$ (Table \ref{tab:Frac_e10Hz_0p9}). Similarly, $2.8 - 12.1 \%$ ($< 4.3 \%$) and $3.2 - 11.2 \%$ ($< 3.5 \%$) of detectable binaries have $e_{\rm 10 Hz} > 0.1$ ($e_{\rm 10 Hz} > 0.9$) as seen by AdV and KAGRA, respectively (Tables \ref{tab:Frac_e10Hz_0p1} and \ref{tab:Frac_e10Hz_0p9}). For comparison, $6.7 - 19.7 \%$ and $< 6 \% $ of mergers per unit volume in the local Universe enter the $10 \, \Hz$ frequency band with $e_{\rm 10 Hz} > 0.1$ and $e_{\rm 10 Hz} > 0.9$, respectively. As seen, $P(e_{\rm 10Hz})$ is shifted toward lower eccentricities for detections with advanced GW detectors compared to the local Universe (top left panel in Figure \ref{fig:Impact_ObsBias_ParamDists}), which can be qualitatively explained as follows. GR precession in the inner binary, being the dominant process quenching LK oscillations, preferentially selects more massive binaries with larger orbital separations via the condition $t_{\rm LK} < t_{\rm GR,inner}$ (Equation \ref{eq:Crit_Quench_GRprec}), which translates into $M_{\rm tot}^{1/2} a_{\rm in}^{-3/2} < a_{\rm in}^{5/2} M_{\rm tot}^{-3/2}$ in terms of $M_{\rm tot}$ and $a_{\rm in}$ (Section \ref{sec:Timescales}). More massive binaries with higher mass ratios (i) have larger relative weights in a detected sample compared to that for the local Universe due to their larger detection distances, (ii) possess lower residual eccentricities owing to their higher masses, larger orbital separations, and higher redshifts (Section \ref{subsec:MockSamples_GNs}). The combination of (i) and (ii) leads to an overabundance of more massive binaries with lower $e_{\rm 10Hz}$ in a detected sample compared to that for the local Universe, which shifts $P(e_{\rm 10Hz})$ toward lower eccentricities. This also explains the trend that a systematically larger fraction of binaries enter the $10\, \Hz$ frequency band for higher $\beta$ values (Tables \ref{tab:Frac_e10Hz_0p1} and \ref{tab:Frac_e10Hz_0p9}).
   
   \item We find that the $z$ distribution $P(z)$ (bottom panel in Figure \ref{fig:ParamDist_aLIGOVolume}) is also sensitive to $\beta$ and does not depend significantly on $\alpha$. As seen, $P(z)$ peaks at $\sim 0.25 - 0.28$ at fixed $\beta = 3$ depending on $\alpha$, and the peak positions expand out to $z \sim 0.34 - 0.38$ and $z \sim 0.45 - 0.55$ for $\beta = 2, 1$, respectively. The reasons for these characteristics are that larger $\beta$ values lead to systematically lighter binaries in single GNs, which possess lower detection distances. However, $\alpha$ barely shapes the mass distributions in single GNs. Finally, we note that the vast majority ($95\%$) of binaries merge with $z \lesssim 1.1$, which justifies our assumptions on the redshift independence of the SMBH mass function and the equilibrium state of GNs regarding the detectable merger population (Section \ref{sec:Props_GNs}). We find the same characteristics for $P(z)$ in AdV (KAGRA) detections. In this case, the vast majority ($95\%$) binaries merge with $z \lesssim 0.8$ ($z \lesssim 0.78$), and $P(z)$ peaks between $z \sim 0.2 - 0.4$ ($z \sim 0.15 - 0.37$), depending mainly on $\beta$ and slightly on $\alpha$.
\end{itemize}
 Note that the mild differences in results for aLIGO, AdV, and KAGRA originate from their different design sensitivities, resulting in variant detection distances in terms of binary parameters.
 
 Additionally, we investigate how different choices for (a) the radius of influence (Section \ref{subsec:GN}), (b) the outer eccentricity distribution (Section \ref{subsec:ParamDist_Outer}), and (c) the condition for binary evaporation (Section \ref{subsec:SelectCrit}) influence the distributions of binary parameters in detections. (a) We define the radius of influence to be the distance from the SMBH at which the enclosed mass of stars equals $2 \Msmbh$ and find it to be slightly lower than that obtained from Equations \eqref{eq:rmax} and \eqref{eq:Msigma}.\footnote{We parameterize the population of stars as in \citet{GondanKocsis2021}.} Since the $a_{\rm out}$ distribution yields a steep distribution with a cut-off well below $r_{\rm max}$ in single GNs (Section \ref{subsec:CompPrevRes}), a slightly lower $r_{\rm max}$ would have a negligible effect on the distributions of binary parameters in detections. (b) We run additional MC simulations with $f(e_{\rm out}) \propto e_{\rm out}^{2.6}$ for the models considered above, keeping all other parameter distributions the same. We find that the change in $f(e_{\rm out})$ mainly affects the $e_{\rm out}$ distribution and leaves the distributions of remaining orbital parameters $\{ a_{\rm in}, e_{\rm in}, e_{\rm out} \}$ and measurable binary parameters practically unchanged for the detectable binary samples. (c) Finally, we assess how the number of LK cycles allowed before evaporation influences the binary parameter distributions. As the parameter region producing LK-induced BBHs in a single GN does not sensitive to the number of LK cycles allowed before evaporation \citep{Hamersetal2018}, we repeat the simulations of this section but allow $1$ and $100$ LK cycles instead of $10$ cycles (Equation \ref{eq:Crit_Quench_GRprec}). We find that the distributions of orbital and measurable binary parameters for the detectable sample slightly depend on the number of allowed cycles.
 
 As discussed above, mainly GR precession in the inner binary shapes the relations between the parameters $\{ M_{\rm tot}, q, e_{\rm 10 Hz}\}$ since it preferentially selects more massive binaries with larger orbital separations that possess lower residual eccentricities when entering the aLIGO/AdV/KAGRA band. Accordingly, a negative correlation can be assumed between $e_{\rm 10 Hz}$ and $M_{\rm tot}$, while the correlation between $e_{\rm 10 Hz}$ and $q$ and between $M_{\rm tot}$ and $q$ may significantly depend on the actual choices for $\beta$ and $\alpha$ due to the complex nature of selection criteria (Section \ref{subsec:MockSamples_GNs}) and dynamical evolution of LK-induces systems (Section \ref{subsec:MockSamples_GNs}) in terms of orbital and mass-dependent parameters. Similarly, observational bias modifies relations between the parameters $\{ q, M_{\rm tot}, D_{\rm L} \}$ as it preferentially selects more massive binaries with higher mass ratios due to their larger detection distances, which implies a positive correlation between each 2-combination of the parameter set in question. We systematically investigate possible correlations among various binary parameters in Section \ref{subsec:CorrBBHparams} to determine how detections with advanced GW detectors shape trends between binary parameters.

\begin{table*}
\centering  
   \begin{tabular}{@{}cc|ccccccccc}
     \multicolumn{11}{c}{Spearman's Rank Correlation Coefficients} \\
     \hline
      $\beta$ & $\alpha$ & $M_{\rm tot} - q$ & $M_{\rm tot,z} - q$ & $M_{\rm tot} - D_{\rm L}$ & $M_{\rm tot,z} - D_{\rm L}$ & $q - D_{\rm L}$ & $e_{\rm 10Hz} - M_{\rm tot}$ & $e_{\rm 10Hz} - M_{\rm tot,z}$ & $e_{\rm 10Hz} - q$ & $e_{\rm 10Hz} - D_{\rm L}$ \\
     \hline\hline
     $1$ & $\alpha(m)$ & $ 0.332$ & $ 0.328$ & $0.232$ & $0.646$ & $ 0.145$ & $-0.270$ & $-0.305$ & $-0.132$ & $-0.208$  \\
     $2$ & $\alpha(m)$ & $-0.029$ & $ 0.019$ & $0.380$ & $0.683$ & $ 0.066$ & $-0.302$ & $-0.326$ & $-0.067$ & $-0.230$  \\
     $3$ & $\alpha(m)$ & $-0.454$ & $-0.409$ & $0.552$ & $0.734$ & $-0.173$ & $-0.307$ & $-0.331$ & $ 0.138$ & $-0.254$  \\
     $1$ &     $2$     & $ 0.246$ & $ 0.228$ & $0.240$ & $0.662$ & $ 0.131$ & $-0.245$ & $-0.276$ & $-0.090$ & $-0.192$  \\
     $2$ &     $2$     & $-0.130$ & $-0.076$ & $0.412$ & $0.686$ & $ 0.036$ & $-0.286$ & $-0.307$ & $-0.031$ & $-0.222$  \\
     $3$ &     $2$     & $-0.522$ & $-0.460$ & $0.577$ & $0.743$ & $-0.185$ & $-0.300$ & $-0.315$ & $ 0.101$ & $-0.242$  \\
     $1$ &     $3$     & $ 0.231$ & $ 0.209$ & $0.253$ & $0.675$ & $ 0.117$ & $-0.193$ & $-0.225$ & $-0.082$ & $-0.166$  \\
     $2$ &     $3$     & $-0.136$ & $-0.106$ & $0.439$ & $0.697$ & $ 0.019$ & $-0.243$ & $-0.268$ & $-0.007$ & $-0.209$  \\
     $3$ &     $3$     & $-0.561$ & $-0.479$ & $0.592$ & $0.755$ & $-0.211$ & $-0.277$ & $-0.290$ & $ 0.092$ & $-0.228$  \\
   \end{tabular} 
   \caption{Spearman's Rank Correlation Coefficients for GN models with different BH mass functions ($m^{-\beta}$) and 3D number density profiles ($r^{-\alpha}$) of the BBH population (cf. Figure \ref{fig:ParamDist_aLIGOVolume}) between the measurable parameters $\{ M_{\rm tot}, M_{\rm tot,z},  q, D_{\rm L}, e_{\rm 10Hz} \}$ for LK-induced BBH mergers with $\rm S/N > 8$ for detections with a single aLIGO detector at design sensitivity.  \label{tab:Correlations} }
\end{table*}

\subsection{Correlations among binary parameters}
\label{subsec:CorrBBHparams}
 
 In this section, we investigate possible correlations among various measurable parameters describing LK-induced BBH mergers in GNs, including (redshifted) mass-dependent parameters together with the luminosity distance and residual eccentricity.
 
 We use the Spearman correlation coefficient test \citep{Spearman1904} to measure the strength and direction of the possible monotonic association between binary parameters, as this test does not carry any assumptions about the distribution of the data. Spearman's rank correlation coefficient $r_{\rm S}$ is computed, and its convergence is verified as prescribed in \citet{GondanKocsis2021}. Note that the correlation between a binary parameter $X$ and the parameters $q$ and $\eta$ ($D_{\rm L}$ and $z$) are the same because there is a one-to-one correspondence between $q$ and $\eta$ ($D_{\rm L}$ and $z$). Therefore, we present results only for $q$ ($D_{\rm L}$).
 
 Table \ref{tab:Correlations} shows the correlation coefficients between various parameters for detections with a single aLIGO detector at design sensitivity. We identify the following model-independent (i.e. $\beta$ and $\alpha$ independent) correlations.
 \begin{itemize}
    \item The total mass $M_{\rm tot}$ and source distance $D_{\rm L}$ are correlated since the detection distance is larger for higher binary masses up to \mbox{$M_{\rm tot,z} \sim 160-170 \, \Msun$} \citep{HallEvans2019}. Similarly, $\mathcal{M}$ and $\mu$ are also correlated with $D_{\rm L}$ for the same reason, such as the redshifted mass parameters.
    
    \item The MC analysis shows that residual eccentricity $e_{\rm 10Hz}$ and total mass $M_{\rm tot}$ are anti-correlated, which arises mainly from the combination of two effects (Section \ref{subsec:DistaLIGOdet}): (i) massive binaries have an overabundance in a detected sample compared to that for the local Universe, (ii) massive binaries possess low residual eccentricities owing to their high masses, large orbital separations, and high redshifts. We find the same trends for other mass-dependent parameters such as $\mathcal{M}$ and $\mu$ and the corresponding redshifted masses.
    
    \item We also identify an anticorrelation between luminosity distance $D_{\rm L}$ and residual eccentricity $e_{\rm 10Hz}$, which originates directly from the correlation between $D_{\rm L}$ and $M_{\rm tot}$ and the anticorrelation between $M_{\rm tot}$ and $e_{\rm 10Hz}$.
 \end{itemize}
 We find the same model-independent correlations for detections with AdV and KAGRA.
 
 As seen in Table \ref{tab:Correlations}, model-dependent correlations occur when $r_{\rm S}$ is computed between a binary parameter $X$ and $q$, and in these cases, the parity of $r_{\rm S}$ depends only on $\beta$. Furthermore, $r_{\rm S}$ systematically increases between $q$ and $e_{\rm 10Hz}$ with $\beta$, while it systematically decreases between $q$ and the other parameters for higher $\beta$ values. We explain these findings based upon the characteristics of the $M_{\rm tot}$ distributions obtained for aLIGO detections (Section \ref{subsec:DistaLIGOdet}) as follows. Starting with $r_{\rm S}$ between $M_{\rm tot}$ and $q$, the $M_{\rm tot}$ distribution is tilted toward higher masses for lower $\beta$ values, as seen in Figure \ref{fig:ParamDist_aLIGOVolume}. Binaries near the peak produce $q \sim 1$ values, while a larger fraction of $q \lesssim 1$ values originate from binaries below the peak position leading to the obtained positive correlation between $M_{\rm tot}$ and $q$ for $\beta = 1$. As $\beta$ increases, the $M_{\rm tot}$ distribution gradually tilts toward lower masses providing a larger fraction of $q \lesssim 1$ values above the peak position, which leads to the anticorrelation between $M_{\rm tot}$ and $q$ for $\beta = 3$. Thereby, the acquired trend for $r_{\rm S}$ between $q$ and $D_{\rm L}$ ($q$ and $e_{\rm 10Hz}$) in terms of $\beta$ is straightforward since the positive (negative) correlation between $M_{\rm tot}$ and $D_{\rm L}$ ($M_{\rm tot}$ and $e_{\rm 10Hz}$) is altered by the correlation between $M_{\rm tot}$ and $q$. Moreover, the same trends apply to other mass-dependent parameters such as $\mathcal{M}$ and $\mu$ and the corresponding redshifted masses for the same reason. Finally, we note that these trends are also obtained for detections with AdV and KAGRA.

\section{Summary and Conclusions} 
\label{sec:SummAndConc}
 
 In this paper, we have extended previous studies to obtain the distribution of physical parameters describing LK-induced BBH mergers around SMBHs as seen by aLIGO/AdV/KAGRA at design sensitivity and to identify possible correlations among physical parameters for the detectable binary population. 
 
 For this purpose, we used high-precision, fully regularized N-body simulations, including PN terms up to order PN2.5, to generate MC samples of LK-induced BBH populations in isotropic GNs for various BH mass function ($m^{-\beta}$) and 3D number density distribution ($r^{-\alpha}$) models of the BBH population. We generated mock catalogues of GNs with the observed distribution of SMBHs within the astrophysical reach of aLIGO/AdV/KAGRA and associated an MC sample of LK-induced BBHs to each GN host. Waveforms were associated with BBHs using a waveform generator dedicated to simulating full inspiral-merger-ringdown waveforms of BBHs with nonzero eccentricities. To select the binary subpopulation possibly detectable by aLIGO/AdV/KAGRA, we calculated the SNR value for each binary by utilizing the generated waveform and taking into account the source direction, inclination, and polarization angles. Finally, we used a rank correlation test to measure correlations among source parameters for the detectable binaries.
 
 Our main results for the detectable LK-induced BBH populations merging in GNs can be summarized as follows.
\begin{itemize}
   \item We find that $\sim 3 - 12 \%$ of binaries have residual eccentricity $e_{\rm 10Hz}$ beyond $0.1$ when entering the aLIGO/AdV/KAGRA band and $\lesssim 4 \%$ of them have $e_{\rm 10Hz} > 0.9$, depending significantly on both $\beta$ and $\alpha$ and slightly on the selected advanced GW detector (Tables \ref{tab:Frac_e10Hz_0p1} and \ref{tab:Frac_e10Hz_0p9}). Furthermore, we find that larger $\beta$ values result in systematically higher binary fractions. For comparison, $\sim 7 - 20 \%$ and $\lesssim 6 \% $ of mergers per unit volume in the local Universe enter the $10 \, \Hz$ frequency band with $e_{\rm 10 Hz} > 0.1$ and $e_{\rm 10 Hz} > 0.9$, respectively (Tables \ref{tab:Frac_e10Hz_0p1} and \ref{tab:Frac_e10Hz_0p9}). Binary fractions in the aLIGO/AdV/KAGRA band are somewhat lower for the detected sample, which is mainly due to the combination of observational bias and the quenching of the LK mechanism by GR precession in the inner binary (Section \ref{subsec:DistaLIGOdet}). The $e_{\rm 10Hz}$ distribution of detectable mergers has a double peak, one at of order $\sim 10^{-3} - 10^{-2}$ and one at $e_{\rm 10Hz} \sim 1$, where the peak at high eccentricities corresponds to binaries for which the dynamics are dominated by GW radiation within one LK cycle (Section \ref{subsec:DistaLIGOdet}).
   
   \item Observational bias shifts the mass distributions toward higher masses (Figure \ref{fig:Impact_ObsBias_ParamDists}). These distributions are weakly sensitive to $\alpha$ and are mainly shaped by $\beta$, where larger $\beta$ values result in systematically lighter BBH mergers (Figure \ref{fig:ParamDist_aLIGOVolume}). For instance, the $M_{\rm tot}$ distributions peak at $M_{\rm tot} \sim \{ 15 \, \Msun, 18 \, \Msun, 45 \, \Msun \}$ for $\beta = \{3, 2, 1 \}$ for mergers in the local Universe, respectively, while the corresponding distributions, as seen by aLIGO, peak at $M_{\rm tot} \sim \{20 \, \Msun, 40 \, \Msun, 55 \, \Msun \}$, respectively (Figure \ref{fig:Impact_ObsBias_ParamDists}). The peaks are located at mildly lower $M_{\rm tot}$ values for detections with AdV and KAGRA due to the detectors' lower astrophysical reach. The $\beta$ and $\alpha$ dependent characteristics of mass distributions originate from the fact that these distributions in single GNs are governed by $\beta$, do not depend significantly on $\Msmbh$, and are barely affected by $\alpha$ (Section \ref{subsec:DistaLIGOdet}). Binaries typically merge with $\sim 1/2 \times {\rm max}(M_{\rm tot})$ for $\beta = 1$ and for aLIGO detections. The typical masses are even lower for detections with AdV/KAGRA or larger $\beta$. Consequently, LK-induced BBH mergers around SMBHs may be detected with roughly medium masses.
   
   \item $P(q)$ is roughly uniform, and it is closer to a uniform distribution for $\beta = 1$ (Figure \ref{fig:ParamDist_aLIGOVolume}). We find that $P(q)$ is tilted toward lower values as $\beta$ increases and peaks at $q \sim 0.2 - 0.3$. Furthermore, $P(q)$ is weakly sensitive to $\alpha$. Finally, observational bias tilts $P(q)$ towards a uniform distribution (Figure \ref{fig:Impact_ObsBias_ParamDists}).
   
   \item We find that the redshift distribution $P(z)$ peaks between $z \sim 0.15 - 0.55$ depending mainly on the considered advanced GW detector and $\beta$, and it is mildly affected by $\alpha$ (Figure \ref{fig:ParamDist_aLIGOVolume}). Here, larger $\beta$ values lead to systematically lower peak positions as the fraction of lighter binaries with lower detection distances increases with $\beta$. Finally, we find that the vast majority of binaries merge with $z \lesssim 1.1$ for detections with single advanced GW detectors (Section \ref{subsec:DistaLIGOdet}).
   
   \item We identified model-independent (i.e. $\beta$ and $\alpha$ independent) correlations between various source parameters as follows (Table \ref{tab:Correlations}). First, we found a significant negative correlation between residual eccentricity $e_{\rm 10Hz}$ and the mass parameters $\{ M_{\rm tot}, \mathcal{M}, \mu \}$ together with the corresponding redshifted masses. These findings arise mainly from the combination of the observational selection effect and the quenching of the LK mechanism by GR precession in the inner binary (Section \ref{subsec:CorrBBHparams}). Furthermore, we found a significant positive correlation between luminosity distance $D_{\rm L}$ and the (redshifted) mass parameters due to observational bias. Finally, a negative correlation between luminosity distance $D_{\rm L}$ and residual eccentricity $e_{\rm 10Hz}$ is a direct consequence of the correlation between $\{e_{\rm 10Hz}, D_{\rm L} \}$ and $M_{\rm tot}$.
\end{itemize}
 
 We worked under the assumption that GNs in the detection volume is (i) dynamically relaxed, (ii) spherically symmetric (non-rotating), and (iii) host an SMBH in their centre. Note, however, that these assumptions may be violated to some degree. (i) Nuclear star clusters around SMBHs of mass \mbox{$\Msmbh \lesssim 10^7 \Msun$} reach an equilibrium distribution within a Hubble time (e.g. \citealt{Merritt2013,BarOr2013,Gondanetal2018b}); otherwise, the 3D number density profiles of stellar populations are shallower than that in relaxed GNs (e.g. \citealt{EmamiLoeb2020}). Since we found in Section \ref{subsec:DistaLIGOdet} that the distributions of measurable binary parameters do not depend significantly on the 3D number density of the BBH population, we expect similar distributions to those presented in Figure \ref{fig:ParamDist_aLIGOVolume} when taking into account dynamical relaxation. Accordingly, the directions of model-independent correlations obtained in Section \ref{subsec:CorrBBHparams} are also expected to remain unchanged when the dynamical relaxation of GNs is considered. (ii) A substantial fraction of GNs possesses non-spherical (rotating) nuclear star clusters in their centre (e.g. \citealt{GeorgievBoker2014,Chatzopoulosetal2015,Feldmeieretal2017}) in which $P(a_{\rm out})$ and $P(a_{\rm in})$ are expected to be shifted toward larger and lower values, respectively, compared to spherical GNs, depending on the level of axisymmetry \citep{PetrovichAntonini2017,BubPetrovich2020}. Accordingly, larger $a_{\rm out}$ values indicate lighter BBH mergers in non-spherical GNs, as the gravitational pull of the SMBH tends to break more massive binaries at smaller distances from the SMBH (Section \ref{subsec:SelectCrit}). Similarly, lower $a_{\rm in}$ values also indicate lighter BBH mergers mainly due to the selection effect of GR precession in the inner orbit (Section \ref{subsec:DistaLIGOdet}). Furthermore, lighter BBHs and lower orbital separations lead to larger $e_{\rm 10Hz}$ values in non-spherical GNs (Section \ref{subsec:MockSamples_GNs}) than in spherical ones. Consequently, the consideration of non-spherical GNs may shift the observable mass distributions toward lower masses and the distributions of residual eccentricity toward higher values in Figure \ref{fig:ParamDist_aLIGOVolume}. Finally, lighter BBHs imply lower detection distances (Section \ref{subsec:DistaLIGOdet}), which shifts the redshift distributions in Figure \ref{fig:ParamDist_aLIGOVolume} toward lower values. Moreover, the directions of correlations between mass-dependent parameters and $D_{\rm L}$ ($e_{\rm 10Hz}$) remain unchanged when considering non-spherical GNs as the observational selection effect (formation mechanism) determines them. Thereby, the direction of correlation between $D_{\rm L}$ and $e_{\rm 10Hz}$ also does not change. (iii) The absence of an SMBH in the GN's core has the opposite effect on $P(a_{\rm out})$ as the presence of rotation, i.e. it is shifted toward lower values (e.g. \citealt{BubPetrovich2020}). This implies more massive BBH mergers and larger $a_{\in}$ compared to those in spherical GNs. Accordingly, the observable mass distributions in Figure \ref{fig:ParamDist_aLIGOVolume} are expected to be shifted toward larger masses, the $e_{\rm 10Hz}$ distributions toward lower values, and the redshift distributions toward larger values. Finally, the directions of model-independent correlations are expected to be unchanged, similar to the case of non-spherical GNs.
 
 In conclusion, our findings for the binary parameter distributions, together with the redshift distribution and model-independent correlations among binary parameters, may be useful to statistically disentangle this merger channel from others in the BBH merger sample detected by aLIGO-AdV-KAGRA. The identification of individual BBH mergers in this formation channel may also be possible by detecting the eccentricity oscillation with LISA in the local Universe up to a few ${\rm Mpc}$, with observation periods shorter than the mission lifetime \citep{Hoangetal2019,RandallXianyu2019}.
 
 If detected, population properties of LK-induced BBHs merging in GNs with an SMBH in their centre may serve to test theories describing the formation and evolution of stellar binaries and the dynamics of stellar populations in the central regions of GNs. Furthermore, these triple systems are an outstanding probe of the BH nature of compact objects and of strong-field physics (e.g. \citealt{Cardosoetal2021}). Finally, we note that detections with LISA may be used to probe the spin parameter of the SMBHs and the SMBH spin effects \citep{FangChen2019,FangHuang2020}.

\section*{Data Availability}
 
 The data underlying this article will be shared on reasonable request to the corresponding author.

\section*{Acknowledgment}

 We thank the anonymous referee for constructive comments which helped improve the quality of the paper. L\'aszl\'o Gond\'an is supported by the \mbox{{\'U}NKP-21-4} New National Excellence Programmes of the Ministry for Innovation and Technology from the source of the National Research, Development and Innovation Fund.
 

 \bibliographystyle{mnras}
 \bibliography{refs}

\begin{thebibliography}{}
\makeatletter
\relax
\def\mn@urlcharsother{\let\do\@makeother \do\$\do\&\do\#\do\^\do\_\do\%\do\~}
\def\mn@doi{\begingroup\mn@urlcharsother \@ifnextchar [ {\mn@doi@}
  {\mn@doi@[]}}
\def\mn@doi@[#1]#2{\def\@tempa{#1}\ifx\@tempa\@empty \href
  {http://dx.doi.org/#2} {doi:#2}\else \href {http://dx.doi.org/#2} {#1}\fi
  \endgroup}
\def\mn@eprint#1#2{\mn@eprint@#1:#2::\@nil}
\def\mn@eprint@arXiv#1{\href {http://arxiv.org/abs/#1} {{\tt arXiv:#1}}}
\def\mn@eprint@dblp#1{\href {http://dblp.uni-trier.de/rec/bibtex/#1.xml}
  {dblp:#1}}
\def\mn@eprint@#1:#2:#3:#4\@nil{\def\@tempa {#1}\def\@tempb {#2}\def\@tempc
  {#3}\ifx \@tempc \@empty \let \@tempc \@tempb \let \@tempb \@tempa \fi \ifx
  \@tempb \@empty \def\@tempb {arXiv}\fi \@ifundefined
  {mn@eprint@\@tempb}{\@tempb:\@tempc}{\expandafter \expandafter \csname
  mn@eprint@\@tempb\endcsname \expandafter{\@tempc}}}

\bibitem[\protect\citeauthoryear{{Aarseth}}{{Aarseth}}{2012}]{Aarseth2012}
{Aarseth} S.~J.,  2012, \mn@doi [\mnras] {10.1111/j.1365-2966.2012.20666.x},
  \href {http://adsabs.harvard.edu/abs/2012MNRAS.422..841A} {422, 841}

\bibitem[\protect\citeauthoryear{{Aasi} et~al.,}{{Aasi}
  et~al.}{2015}]{Aasietal2015}
{Aasi} J.,  et~al., 2015, \mn@doi [Classical and Quantum Gravity]
  {10.1088/0264-9381/32/11/115012}, \href
  {http://adsabs.harvard.edu/abs/2015CQGra..32k5012A} {32, 115012}

\bibitem[\protect\citeauthoryear{{Abadie} et~al.,}{{Abadie}
  et~al.}{2010}]{Abadieetal2010}
{Abadie} J.,  et~al., 2010, \mn@doi [Classical and Quantum Gravity]
  {10.1088/0264-9381/27/17/173001}, \href
  {https://ui.adsabs.harvard.edu/abs/2010CQGra..27q3001A} {27, 173001}

\bibitem[\protect\citeauthoryear{{Abbott} et~al.,}{{Abbott}
  et~al.}{2016}]{Abbottetal2016}
{Abbott} B.~P.,  et~al., 2016, \mn@doi [Physical Review Letters]
  {10.1103/PhysRevLett.116.061102}, \href
  {http://adsabs.harvard.edu/abs/2016PhRvL.116f1102A} {116, 061102}

\bibitem[\protect\citeauthoryear{{Abbott} et~al.,}{{Abbott}
  et~al.}{2017}]{Abbottetal2017}
{Abbott} B.~P.,  et~al., 2017, \mn@doi [\prl] {10.1103/PhysRevLett.119.161101},
  \href {https://ui.adsabs.harvard.edu/abs/2017PhRvL.119p1101A} {119, 161101}

\bibitem[\protect\citeauthoryear{{Abbott} et~al.,}{{Abbott}
  et~al.}{2018}]{Abbottetal2018}
{Abbott} B.~P.,  et~al., 2018, \mn@doi [Living Reviews in Relativity]
  {10.1007/s41114-018-0012-9}, \href
  {https://ui.adsabs.harvard.edu/abs/2018LRR....21....3A} {21, 3}

\bibitem[\protect\citeauthoryear{{Abbott} et~al.,}{{Abbott}
  et~al.}{2019a}]{Abbotetal2019b}
{Abbott} B.~P.,  et~al., 2019a, \mn@doi [Physical Review X]
  {10.1103/PhysRevX.9.031040}, \href
  {https://ui.adsabs.harvard.edu/abs/2019PhRvX...9c1040A} {9, 031040}

\bibitem[\protect\citeauthoryear{{Abbott} et~al.,}{{Abbott}
  et~al.}{2019b}]{Abbottetal2019}
{Abbott} B.~P.,  et~al., 2019b, \mn@doi [\apjl] {10.3847/2041-8213/ab3800},
  \href {https://ui.adsabs.harvard.edu/abs/2019ApJ...882L..24A} {882, L24}

\bibitem[\protect\citeauthoryear{{Abbott} et~al.,}{{Abbott}
  et~al.}{2019c}]{Abbottetal2019c}
{Abbott} B.~P.,  et~al., 2019c, \mn@doi [\apj] {10.3847/1538-4357/ab3c2d},
  \href {https://ui.adsabs.harvard.edu/abs/2019ApJ...883..149A} {883, 149}

\bibitem[\protect\citeauthoryear{{Abbott} et~al.,}{{Abbott}
  et~al.}{2020}]{Abbotetal2020a}
{Abbott} B.~P.,  et~al., 2020, \mn@doi [Living Reviews in Relativity]
  {10.1007/s41114-020-00026-9}, \href
  {https://ui.adsabs.harvard.edu/abs/2020LRR....23....3A} {23, 3}

\bibitem[\protect\citeauthoryear{{Abbott} et~al.,}{{Abbott}
  et~al.}{2021a}]{Abbotetal2021d}
{Abbott} R.,  et~al., 2021a, \mn@doi [Physical Review X]
  {10.1103/PhysRevX.11.021053}, \href
  {https://ui.adsabs.harvard.edu/abs/2021PhRvX..11b1053A} {11, 021053}

\bibitem[\protect\citeauthoryear{{Abbott} et~al.,}{{Abbott}
  et~al.}{2021b}]{Abbotetal2021a}
{Abbott} R.,  et~al., 2021b, \mn@doi [\apjl] {10.3847/2041-8213/abe949}, \href
  {https://ui.adsabs.harvard.edu/abs/2021ApJ...913L...7A} {913, L7}

\bibitem[\protect\citeauthoryear{{Acernese} et~al.,}{{Acernese}
  et~al.}{2015}]{Acerneseetal2015}
{Acernese} F.,  et~al., 2015, \mn@doi [Classical and Quantum Gravity]
  {10.1088/0264-9381/32/2/024001}, \href
  {http://adsabs.harvard.edu/abs/2015CQGra..32b4001A} {32, 024001}

\bibitem[\protect\citeauthoryear{{Aharon} \& {Perets}}{{Aharon} \&
  {Perets}}{2016}]{AharonPerets2016}
{Aharon} D.,  {Perets} H.~B.,  2016, \mn@doi [\apjl]
  {10.3847/2041-8205/830/1/L1}, \href
  {http://adsabs.harvard.edu/abs/2016ApJ...830L...1A} {830, L1}

\bibitem[\protect\citeauthoryear{{Alexander}}{{Alexander}}{2017}]{Alexander2017}
{Alexander} T.,  2017, \mn@doi [\araa] {10.1146/annurev-astro-091916-055306},
  \href {http://adsabs.harvard.edu/abs/2017ARA26A..55...17A} {55, 17}

\bibitem[\protect\citeauthoryear{{Alexander} \& {Hopman}}{{Alexander} \&
  {Hopman}}{2009}]{AlexanderHopman2009}
{Alexander} T.,  {Hopman} C.,  2009, \mn@doi [\apj]
  {10.1088/0004-637X/697/2/1861}, \href
  {http://adsabs.harvard.edu/abs/2009ApJ...697.1861A} {697, 1861}

\bibitem[\protect\citeauthoryear{{Amaro-Seoane}, {Freitag}  \&
  {Spurzem}}{{Amaro-Seoane} et~al.}{2004}]{AmaroSeoaneetal2004}
{Amaro-Seoane} P.,  {Freitag} M.,   {Spurzem} R.,  2004, \mn@doi [\mnras]
  {10.1111/j.1365-2966.2004.07956.x}, \href
  {https://ui.adsabs.harvard.edu/abs/2004MNRAS.352..655A} {352, 655}

\bibitem[\protect\citeauthoryear{{Antognini}}{{Antognini}}{2015}]{Antognini2015MNRAS}
{Antognini} J.~M.~O.,  2015, \mn@doi [\mnras] {10.1093/mnras/stv1552}, \href
  {https://ui.adsabs.harvard.edu/abs/2015MNRAS.452.3610A} {452, 3610}

\bibitem[\protect\citeauthoryear{{Antonini} \& {Perets}}{{Antonini} \&
  {Perets}}{2012}]{AntoniniPerets2012}
{Antonini} F.,  {Perets} H.~B.,  2012, \mn@doi [\apj]
  {10.1088/0004-637X/757/1/27}, \href
  {http://adsabs.harvard.edu/abs/2012ApJ...757...27A} {757, 27}

\bibitem[\protect\citeauthoryear{{Antonini} \& {Rasio}}{{Antonini} \&
  {Rasio}}{2016}]{AntoniniRasiol2016}
{Antonini} F.,  {Rasio} F.~A.,  2016, \mn@doi [\apj]
  {10.3847/0004-637X/831/2/187}, \href
  {https://ui.adsabs.harvard.edu/abs/2016ApJ...831..187A} {831, 187}

\bibitem[\protect\citeauthoryear{{Antonini}, {Murray}  \& {Mikkola}}{{Antonini}
  et~al.}{2014}]{Antoninietal2014}
{Antonini} F.,  {Murray} N.,   {Mikkola} S.,  2014, \mn@doi [\apj]
  {10.1088/0004-637X/781/1/45}, \href
  {http://adsabs.harvard.edu/abs/2014ApJ...781...45A} {781, 45}

\bibitem[\protect\citeauthoryear{{Antonini}, {Chatterjee}, {Rodriguez},
  {Morscher}  \& et al.}{{Antonini} et~al.}{2016}]{Antoninietal2016}
{Antonini} F.,  {Chatterjee} S.,  {Rodriguez} C.~L.,  {Morscher} M.,   et al.
  2016, \mn@doi [\apj] {10.3847/0004-637X/816/2/65}, \href
  {http://adsabs.harvard.edu/abs/2016ApJ...816...65A} {816, 65}

\bibitem[\protect\citeauthoryear{{Antonini}, {Toonen}  \& {Hamers}}{{Antonini}
  et~al.}{2017}]{Antoninietal2017}
{Antonini} F.,  {Toonen} S.,   {Hamers} A.~S.,  2017, \mn@doi [\apj]
  {10.3847/1538-4357/aa6f5e}, \href
  {http://adsabs.harvard.edu/abs/2017ApJ...841...77A} {841, 77}

\bibitem[\protect\citeauthoryear{{Arca Sedda}}{{Arca
  Sedda}}{2020}]{ArcaSedda2020}
{Arca Sedda} M.,  2020, \mn@doi [\apj] {10.3847/1538-4357/ab723b}, \href
  {https://ui.adsabs.harvard.edu/abs/2020ApJ...891...47A} {891, 47}

\bibitem[\protect\citeauthoryear{{Arca-Sedda} \&
  {Capuzzo-Dolcetta}}{{Arca-Sedda} \&
  {Capuzzo-Dolcetta}}{2019}]{ArcaSeddaCapuzzo2019}
{Arca-Sedda} M.,  {Capuzzo-Dolcetta} R.,  2019, \mn@doi [\mnras]
  {10.1093/mnras/sty3096}, \href
  {https://ui.adsabs.harvard.edu/abs/2019MNRAS.483..152A} {483, 152}

\bibitem[\protect\citeauthoryear{{Arca-Sedda} \& {Gualandris}}{{Arca-Sedda} \&
  {Gualandris}}{2018}]{ArcaSeddaGualandris2018}
{Arca-Sedda} M.,  {Gualandris} A.,  2018, \mn@doi [\mnras]
  {10.1093/mnras/sty922}, \href
  {https://ui.adsabs.harvard.edu/abs/2018MNRAS.477.4423A} {477, 4423}

\bibitem[\protect\citeauthoryear{{Arca Sedda}, {Li}  \& {Kocsis}}{{Arca Sedda}
  et~al.}{2021}]{ArcaSeddaetal2021}
{Arca Sedda} M.,  {Li} G.,   {Kocsis} B.,  2021, \mn@doi [\aap]
  {10.1051/0004-6361/202038795}, \href
  {https://ui.adsabs.harvard.edu/abs/2021A&A...650A.189A} {650, A189}

\bibitem[\protect\citeauthoryear{{Bahcall} \& {Wolf}}{{Bahcall} \&
  {Wolf}}{1977}]{BahcallWolf1977}
{Bahcall} J.~N.,  {Wolf} R.~A.,  1977, \mn@doi [\apj] {10.1086/155534}, \href
  {http://adsabs.harvard.edu/abs/1977ApJ...216..883B} {216, 883}

\bibitem[\protect\citeauthoryear{{Baibhav}, {Berti}, {Gerosa}, {Mapelli},
  {Giacobbo}, {Bouffanais}  \& {Di Carlo}}{{Baibhav}
  et~al.}{2019}]{Baibhavetal2019}
{Baibhav} V.,  {Berti} E.,  {Gerosa} D.,  {Mapelli} M.,  {Giacobbo} N.,
  {Bouffanais} Y.,   {Di Carlo} U.~N.,  2019, \mn@doi [\prd]
  {10.1103/PhysRevD.100.064060}, \href
  {https://ui.adsabs.harvard.edu/abs/2019PhRvD.100f4060B} {100, 064060}

\bibitem[\protect\citeauthoryear{{Bailyn}, {Jain}, {Coppi}  \&
  {Orosz}}{{Bailyn} et~al.}{1998}]{Bailynetal1998}
{Bailyn} C.~D.,  {Jain} R.~K.,  {Coppi} P.,   {Orosz} J.~A.,  1998, \mn@doi
  [\apj] {10.1086/305614}, \href
  {http://adsabs.harvard.edu/abs/1998ApJ...499..367B} {499, 367}

\bibitem[\protect\citeauthoryear{{Baldassare}, {Reines}, {Gallo}  \&
  {Greene}}{{Baldassare} et~al.}{2015}]{Baldassareetal2015}
{Baldassare} V.~F.,  {Reines} A.~E.,  {Gallo} E.,   {Greene} J.~E.,  2015,
  \mn@doi [\apjl] {10.1088/2041-8205/809/1/L14}, \href
  {https://ui.adsabs.harvard.edu/abs/2015ApJ...809L..14B} {809, L14}

\bibitem[\protect\citeauthoryear{{Bar-Or}, {Kupi}  \& {Alexander}}{{Bar-Or}
  et~al.}{2013}]{BarOr2013}
{Bar-Or} B.,  {Kupi} G.,   {Alexander} T.,  2013, \mn@doi [\apj]
  {10.1088/0004-637X/764/1/52}, \href
  {http://adsabs.harvard.edu/abs/2013ApJ...764...52B} {764, 52}

\bibitem[\protect\citeauthoryear{{Barack} et~al.,}{{Barack}
  et~al.}{2019}]{Baracketal2019}
{Barack} L.,  et~al., 2019, \mn@doi [Classical and Quantum Gravity]
  {10.1088/1361-6382/ab0587}, \href
  {https://ui.adsabs.harvard.edu/abs/2019CQGra..36n3001B} {36, 143001}

\bibitem[\protect\citeauthoryear{{Barth}, {Greene}  \& {Ho}}{{Barth}
  et~al.}{2005}]{Barthetal2005}
{Barth} A.~J.,  {Greene} J.~E.,   {Ho} L.~C.,  2005, \mn@doi [\apjl]
  {10.1086/428365}, \href {http://adsabs.harvard.edu/abs/2005ApJ...619L.151B}
  {619, L151}

\bibitem[\protect\citeauthoryear{{Bartko} et~al.,}{{Bartko}
  et~al.}{2009}]{Bartkoetal2009}
{Bartko} H.,  et~al., 2009, \mn@doi [\apj] {10.1088/0004-637X/697/2/1741},
  \href {http://adsabs.harvard.edu/abs/2009ApJ...697.1741B} {697, 1741}

\bibitem[\protect\citeauthoryear{{Baumgardt}, {Makino}  \&
  {Ebisuzaki}}{{Baumgardt} et~al.}{2004}]{Baumgardtetal2004}
{Baumgardt} H.,  {Makino} J.,   {Ebisuzaki} T.,  2004, \mn@doi [\apj]
  {10.1086/423299}, \href
  {https://ui.adsabs.harvard.edu/abs/2004ApJ...613.1143B} {613, 1143}

\bibitem[\protect\citeauthoryear{{Baumgardt}, {Amaro-Seoane}  \&
  {Sch{\"o}del}}{{Baumgardt} et~al.}{2018}]{Baumgardtetal2018}
{Baumgardt} H.,  {Amaro-Seoane} P.,   {Sch{\"o}del} R.,  2018, \mn@doi [\aap]
  {10.1051/0004-6361/201730462}, \href
  {http://adsabs.harvard.edu/abs/2018A26A...609A..28B} {609, A28}

\bibitem[\protect\citeauthoryear{{Belczynski}, {Wiktorowicz}, {Fryer}, {Holz}
  \& {Kalogera}}{{Belczynski} et~al.}{2012}]{Belczynskietal2012}
{Belczynski} K.,  {Wiktorowicz} G.,  {Fryer} C.~L.,  {Holz} D.~E.,   {Kalogera}
  V.,  2012, \mn@doi [\apj] {10.1088/0004-637X/757/1/91}, \href
  {http://adsabs.harvard.edu/abs/2012ApJ...757...91B} {757, 91}

\bibitem[\protect\citeauthoryear{{Belczynski}, {Holz}, {Bulik}  \&
  {O'Shaughnessy}}{{Belczynski} et~al.}{2016a}]{Belczynskietal2016}
{Belczynski} K.,  {Holz} D.~E.,  {Bulik} T.,   {O'Shaughnessy} R.,  2016a,
  \mn@doi [\nat] {10.1038/nature18322}, \href
  {https://ui.adsabs.harvard.edu/abs/2016Natur.534..512B} {534, 512}

\bibitem[\protect\citeauthoryear{{Belczynski}, {Repetto}, {Holz},
  {O'Shaughnessy}, {Bulik}, {Berti}, {Fryer}  \& {Dominik}}{{Belczynski}
  et~al.}{2016b}]{Belczynskietal2016b}
{Belczynski} K.,  {Repetto} S.,  {Holz} D.~E.,  {O'Shaughnessy} R.,  {Bulik}
  T.,  {Berti} E.,  {Fryer} C.,   {Dominik} M.,  2016b, \mn@doi [\apj]
  {10.3847/0004-637X/819/2/108}, \href
  {https://ui.adsabs.harvard.edu/abs/2016ApJ...819..108B} {819, 108}

\bibitem[\protect\citeauthoryear{{Belczynski} et~al.,}{{Belczynski}
  et~al.}{2020}]{Belczynskietal2020}
{Belczynski} K.,  et~al., 2020, \mn@doi [\apj] {10.3847/1538-4357/ab6d77},
  \href {https://ui.adsabs.harvard.edu/abs/2020ApJ...890..113B} {890, 113}

\bibitem[\protect\citeauthoryear{{Binney} \& {Tremaine}}{{Binney} \&
  {Tremaine}}{1987}]{BinneyTremaine1987}
{Binney} J.,  {Tremaine} S.,  1987, {Galactic dynamics}

\bibitem[\protect\citeauthoryear{{Breivik}, {Rodriguez}, {Larson}, {Kalogera}
  \& {Rasio}}{{Breivik} et~al.}{2016}]{Breiviketal2016}
{Breivik} K.,  {Rodriguez} C.~L.,  {Larson} S.~L.,  {Kalogera} V.,   {Rasio}
  F.~A.,  2016, \mn@doi [\apjl] {10.3847/2041-8205/830/1/L18}, \href
  {https://ui.adsabs.harvard.edu/abs/2016ApJ...830L..18B} {830, L18}

\bibitem[\protect\citeauthoryear{{Brown} \& {Zimmerman}}{{Brown} \&
  {Zimmerman}}{2010}]{BrownZimmerman2010}
{Brown} D.~A.,  {Zimmerman} P.~J.,  2010, \mn@doi [\prd]
  {10.1103/PhysRevD.81.024007}, \href
  {https://ui.adsabs.harvard.edu/abs/2010PhRvD..81b4007B} {81, 024007}

\bibitem[\protect\citeauthoryear{{Bub} \& {Petrovich}}{{Bub} \&
  {Petrovich}}{2020}]{BubPetrovich2020}
{Bub} M.~W.,  {Petrovich} C.,  2020, \mn@doi [\apj] {10.3847/1538-4357/ab8461},
  \href {https://ui.adsabs.harvard.edu/abs/2020ApJ...894...15B} {894, 15}

\bibitem[\protect\citeauthoryear{{Cardoso}, {Duque}  \& {Khanna}}{{Cardoso}
  et~al.}{2021}]{Cardosoetal2021}
{Cardoso} V.,  {Duque} F.,   {Khanna} G.,  2021, \mn@doi [\prd]
  {10.1103/PhysRevD.103.L081501}, \href
  {https://ui.adsabs.harvard.edu/abs/2021PhRvD.103h1501C} {103, L081501}

\bibitem[\protect\citeauthoryear{{Chamberlain}, {Moore}, {Gerosa}  \&
  {Yunes}}{{Chamberlain} et~al.}{2019}]{Chamberlainetal2019}
{Chamberlain} K.,  {Moore} C.~J.,  {Gerosa} D.,   {Yunes} N.,  2019, \mn@doi
  [\prd] {10.1103/PhysRevD.99.024025}, \href
  {https://ui.adsabs.harvard.edu/abs/2019PhRvD..99b4025C} {99, 024025}

\bibitem[\protect\citeauthoryear{{Chatzopoulos}, {Fritz}, {Gerhard},
  {Gillessen}, {Wegg}, {Genzel}  \& {Pfuhl}}{{Chatzopoulos}
  et~al.}{2015}]{Chatzopoulosetal2015}
{Chatzopoulos} S.,  {Fritz} T.~K.,  {Gerhard} O.,  {Gillessen} S.,  {Wegg} C.,
  {Genzel} R.,   {Pfuhl} O.,  2015, \mn@doi [\mnras] {10.1093/mnras/stu2452},
  \href {https://ui.adsabs.harvard.edu/abs/2015MNRAS.447..948C} {447, 948}

\bibitem[\protect\citeauthoryear{{Cutler} \& {Flanagan}}{{Cutler} \&
  {Flanagan}}{1994}]{CutlerFlanagan1994}
{Cutler} C.,  {Flanagan} {\'E}.~E.,  1994, \mn@doi [\prd]
  {10.1103/PhysRevD.49.2658}, \href
  {http://adsabs.harvard.edu/abs/1994PhRvD..49.2658C} {49, 2658}

\bibitem[\protect\citeauthoryear{{Do}, {Lu}, {Ghez}, {Morris}, {Yelda},
  {Martinez}, {Wright}  \& {Matthews}}{{Do} et~al.}{2013}]{Doetal2013}
{Do} T.,  {Lu} J.~R.,  {Ghez} A.~M.,  {Morris} M.~R.,  {Yelda} S.,  {Martinez}
  G.~D.,  {Wright} S.~A.,   {Matthews} K.,  2013, \mn@doi [\apj]
  {10.1088/0004-637X/764/2/154}, \href
  {https://ui.adsabs.harvard.edu/abs/2013ApJ...764..154D} {764, 154}

\bibitem[\protect\citeauthoryear{{Dominik}, {Belczynski}, {Fryer}, {Holz},
  {Berti}, {Bulik}, {Mandel}  \& {O'Shaughnessy}}{{Dominik}
  et~al.}{2012}]{Dominiketal2012}
{Dominik} M.,  {Belczynski} K.,  {Fryer} C.,  {Holz} D.~E.,  {Berti} E.,
  {Bulik} T.,  {Mandel} I.,   {O'Shaughnessy} R.,  2012, \mn@doi [\apj]
  {10.1088/0004-637X/759/1/52}, \href
  {https://ui.adsabs.harvard.edu/abs/2012ApJ...759...52D} {759, 52}

\bibitem[\protect\citeauthoryear{{Dominik} et~al.,}{{Dominik}
  et~al.}{2015}]{Dominiketal2015}
{Dominik} M.,  et~al., 2015, \mn@doi [\apj] {10.1088/0004-637X/806/2/263},
  \href {https://ui.adsabs.harvard.edu/abs/2015ApJ...806..263D} {806, 263}

\bibitem[\protect\citeauthoryear{{East}, {McWilliams}, {Levin}  \&
  {Pretorius}}{{East} et~al.}{2013}]{Eastetal2013}
{East} W.~E.,  {McWilliams} S.~T.,  {Levin} J.,   {Pretorius} F.,  2013,
  \mn@doi [\prd] {10.1103/PhysRevD.87.043004}, \href
  {https://ui.adsabs.harvard.edu/abs/2013PhRvD..87d3004E} {87, 043004}

\bibitem[\protect\citeauthoryear{{Eldridge} \& {Stanway}}{{Eldridge} \&
  {Stanway}}{2016}]{EldridgeStanway2016}
{Eldridge} J.~J.,  {Stanway} E.~R.,  2016, \mn@doi [\mnras]
  {10.1093/mnras/stw1772}, \href
  {https://ui.adsabs.harvard.edu/abs/2016MNRAS.462.3302E} {462, 3302}

\bibitem[\protect\citeauthoryear{{Emami} \& {Loeb}}{{Emami} \&
  {Loeb}}{2020}]{EmamiLoeb2020}
{Emami} R.,  {Loeb} A.,  2020, \mn@doi [\jcap] {10.1088/1475-7516/2020/02/021},
  \href {https://ui.adsabs.harvard.edu/abs/2020JCAP...02..021E} {2020, 021}

\bibitem[\protect\citeauthoryear{{Fang} \& {Huang}}{{Fang} \&
  {Huang}}{2020}]{FangHuang2020}
{Fang} Y.,  {Huang} Q.-G.,  2020, \mn@doi [\prd] {10.1103/PhysRevD.102.104002},
  \href {https://ui.adsabs.harvard.edu/abs/2020PhRvD.102j4002F} {102, 104002}

\bibitem[\protect\citeauthoryear{{Fang}, {Chen}  \& {Huang}}{{Fang}
  et~al.}{2019}]{FangChen2019}
{Fang} Y.,  {Chen} X.,   {Huang} Q.-G.,  2019, \mn@doi [\apj]
  {10.3847/1538-4357/ab510e}, \href
  {https://ui.adsabs.harvard.edu/abs/2019ApJ...887..210F} {887, 210}

\bibitem[\protect\citeauthoryear{{Farmer}, {Renzo}, {de Mink}, {Marchant}  \&
  {Justham}}{{Farmer} et~al.}{2019}]{Farmeretal2019}
{Farmer} R.,  {Renzo} M.,  {de Mink} S.~E.,  {Marchant} P.,   {Justham} S.,
  2019, \mn@doi [\apj] {10.3847/1538-4357/ab518b}, \href
  {https://ui.adsabs.harvard.edu/abs/2019ApJ...887...53F} {887, 53}

\bibitem[\protect\citeauthoryear{{Farr}, {Sravan}, {Cantrell}, {Kreidberg},
  {Bailyn}, {Mandel}  \& {Kalogera}}{{Farr} et~al.}{2011}]{Farretal2011}
{Farr} W.~M.,  {Sravan} N.,  {Cantrell} A.,  {Kreidberg} L.,  {Bailyn} C.~D.,
  {Mandel} I.,   {Kalogera} V.,  2011, \mn@doi [\apj]
  {10.1088/0004-637X/741/2/103}, \href
  {http://adsabs.harvard.edu/abs/2011ApJ...741..103F} {741, 103}

\bibitem[\protect\citeauthoryear{{Feldmeier-Krause}, {Zhu}, {Neumayer}, {van de
  Ven}, {de Zeeuw}  \& {Sch{\"o}del}}{{Feldmeier-Krause}
  et~al.}{2017}]{Feldmeieretal2017}
{Feldmeier-Krause} A.,  {Zhu} L.,  {Neumayer} N.,  {van de Ven} G.,  {de Zeeuw}
  P.~T.,   {Sch{\"o}del} R.,  2017, \mn@doi [\mnras] {10.1093/mnras/stw3377},
  \href {https://ui.adsabs.harvard.edu/abs/2017MNRAS.466.4040F} {466, 4040}

\bibitem[\protect\citeauthoryear{{Feldmeier} et~al.,}{{Feldmeier}
  et~al.}{2014}]{Feldmeieretal2014}
{Feldmeier} A.,  et~al., 2014, \mn@doi [\aap] {10.1051/0004-6361/201423777},
  \href {https://ui.adsabs.harvard.edu/abs/2014A&A...570A...2F} {570, A2}

\bibitem[\protect\citeauthoryear{{Fishbach} \& {Holz}}{{Fishbach} \&
  {Holz}}{2017}]{FishbachHolz2017}
{Fishbach} M.,  {Holz} D.~E.,  2017, \mn@doi [\apjl]
  {10.3847/2041-8213/aa9bf6}, \href
  {https://ui.adsabs.harvard.edu/abs/2017ApJ...851L..25F} {851, L25}

\bibitem[\protect\citeauthoryear{{Fishbach}, {Holz}  \& {Farr}}{{Fishbach}
  et~al.}{2017}]{Fishbachetal2017}
{Fishbach} M.,  {Holz} D.~E.,   {Farr} B.,  2017, \mn@doi [The Astrophysical
  Journal] {10.3847/2041-8213/aa7045}, \href
  {https://ui.adsabs.harvard.edu/abs/2017ApJ...840L..24F} {840, L24}

\bibitem[\protect\citeauthoryear{{Fragione} \& {Bromberg}}{{Fragione} \&
  {Bromberg}}{2019}]{FragioneBromberg2019}
{Fragione} G.,  {Bromberg} O.,  2019, \mn@doi [\mnras] {10.1093/mnras/stz2024},
  \href {https://ui.adsabs.harvard.edu/abs/2019MNRAS.488.4370F} {488, 4370}

\bibitem[\protect\citeauthoryear{{Fragione} \& {Kocsis}}{{Fragione} \&
  {Kocsis}}{2019}]{FragioneKocsis2019}
{Fragione} G.,  {Kocsis} B.,  2019, \mn@doi [\mnras] {10.1093/mnras/stz1175},
  \href {https://ui.adsabs.harvard.edu/abs/2019MNRAS.486.4781F} {486, 4781}

\bibitem[\protect\citeauthoryear{{Fragione} \& {Kocsis}}{{Fragione} \&
  {Kocsis}}{2020}]{FragioneKocsis2020}
{Fragione} G.,  {Kocsis} B.,  2020, \mn@doi [\mnras] {10.1093/mnras/staa443},
  \href {https://ui.adsabs.harvard.edu/abs/2020MNRAS.493.3920F} {493, 3920}

\bibitem[\protect\citeauthoryear{{Fragione} \& {Sari}}{{Fragione} \&
  {Sari}}{2018}]{FragioneSari2018}
{Fragione} G.,  {Sari} R.,  2018, \mn@doi [\apj] {10.3847/1538-4357/aaa0d7},
  \href {http://adsabs.harvard.edu/abs/2018ApJ...852...51F} {852, 51}

\bibitem[\protect\citeauthoryear{{Fragione}, {Grishin}, {Leigh}, {Perets}  \&
  {Perna}}{{Fragione} et~al.}{2019a}]{Fragioneetal2019}
{Fragione} G.,  {Grishin} E.,  {Leigh} N. W.~C.,  {Perets} H.~B.,   {Perna} R.,
   2019a, \mn@doi [\mnras] {10.1093/mnras/stz1651}, \href
  {https://ui.adsabs.harvard.edu/abs/2019MNRAS.488...47F} {488, 47}

\bibitem[\protect\citeauthoryear{{Fragione}, {Leigh}  \& {Perna}}{{Fragione}
  et~al.}{2019b}]{Fragioneetal2019b}
{Fragione} G.,  {Leigh} N. W.~C.,   {Perna} R.,  2019b, \mn@doi [\mnras]
  {10.1093/mnras/stz1803}, \href
  {https://ui.adsabs.harvard.edu/abs/2019MNRAS.488.2825F} {488, 2825}

\bibitem[\protect\citeauthoryear{{Freitag}, {Amaro-Seoane}  \&
  {Kalogera}}{{Freitag} et~al.}{2006}]{Freitagetal2006}
{Freitag} M.,  {Amaro-Seoane} P.,   {Kalogera} V.,  2006, \mn@doi [\apj]
  {10.1086/506193}, \href {http://adsabs.harvard.edu/abs/2006ApJ...649...91F}
  {649, 91}

\bibitem[\protect\citeauthoryear{{Gallego-Cano}, {Sch{\"o}del}, {Dong},
  {Nogueras-Lara}, {Gallego-Calvente}, {Amaro-Seoane}  \&
  {Baumgardt}}{{Gallego-Cano} et~al.}{2018}]{GallegoCanoetal2018}
{Gallego-Cano} E.,  {Sch{\"o}del} R.,  {Dong} H.,  {Nogueras-Lara} F.,
  {Gallego-Calvente} A.~T.,  {Amaro-Seoane} P.,   {Baumgardt} H.,  2018,
  \mn@doi [\aap] {10.1051/0004-6361/201730451}, \href
  {https://ui.adsabs.harvard.edu/abs/2018A&A...609A..26G} {609, A26}

\bibitem[\protect\citeauthoryear{{Gayathri} et~al.,}{{Gayathri}
  et~al.}{2022}]{Gayathrietal2022}
{Gayathri} V.,  et~al., 2022, \mn@doi [Nature Astronomy]
  {10.1038/s41550-021-01568-w}, \href
  {https://ui.adsabs.harvard.edu/abs/2022NatAs...6..344G} {6, 344}

\bibitem[\protect\citeauthoryear{{Geller}, {Leigh}, {Giersz}, {Kremer}  \&
  {Rasio}}{{Geller} et~al.}{2019}]{Gelleretal2019}
{Geller} A.~M.,  {Leigh} N. W.~C.,  {Giersz} M.,  {Kremer} K.,   {Rasio} F.~A.,
   2019, \mn@doi [\apj] {10.3847/1538-4357/ab0214}, \href
  {https://ui.adsabs.harvard.edu/abs/2019ApJ...872..165G} {872, 165}

\bibitem[\protect\citeauthoryear{{Georgiev} \& {B{\"o}ker}}{{Georgiev} \&
  {B{\"o}ker}}{2014}]{GeorgievBoker2014}
{Georgiev} I.~Y.,  {B{\"o}ker} T.,  2014, \mn@doi [\mnras]
  {10.1093/mnras/stu797}, \href
  {https://ui.adsabs.harvard.edu/abs/2014MNRAS.441.3570G} {441, 3570}

\bibitem[\protect\citeauthoryear{{Gerosa} \& {Berti}}{{Gerosa} \&
  {Berti}}{2019}]{GerosaBerti2019}
{Gerosa} D.,  {Berti} E.,  2019, \mn@doi [\prd] {10.1103/PhysRevD.100.041301},
  \href {https://ui.adsabs.harvard.edu/abs/2019PhRvD.100d1301G} {100, 041301}

\bibitem[\protect\citeauthoryear{{Gerosa} \& {Fishbach}}{{Gerosa} \&
  {Fishbach}}{2021}]{GerosaFishbach2021}
{Gerosa} D.,  {Fishbach} M.,  2021, \mn@doi [Nature Astronomy]
  {10.1038/s41550-021-01398-w}, \href
  {https://ui.adsabs.harvard.edu/abs/2021NatAs...5..749G} {5, 749}

\bibitem[\protect\citeauthoryear{{Giacobbo}, {Mapelli}  \& {Spera}}{{Giacobbo}
  et~al.}{2018}]{Giacobboetal2018}
{Giacobbo} N.,  {Mapelli} M.,   {Spera} M.,  2018, \mn@doi [\mnras]
  {10.1093/mnras/stx2933}, \href
  {https://ui.adsabs.harvard.edu/abs/2018MNRAS.474.2959G} {474, 2959}

\bibitem[\protect\citeauthoryear{{Gillessen}, {Eisenhauer}, {Trippe},
  {Alexander}, {Genzel}, {Martins}  \& {Ott}}{{Gillessen}
  et~al.}{2009}]{Gillessenetal2009}
{Gillessen} S.,  {Eisenhauer} F.,  {Trippe} S.,  {Alexander} T.,  {Genzel} R.,
  {Martins} F.,   {Ott} T.,  2009, \mn@doi [\apj]
  {10.1088/0004-637X/692/2/1075}, \href
  {https://ui.adsabs.harvard.edu/abs/2009ApJ...692.1075G} {692, 1075}

\bibitem[\protect\citeauthoryear{{Gillessen} et~al.,}{{Gillessen}
  et~al.}{2017}]{Gillessenetal2017}
{Gillessen} S.,  et~al., 2017, \mn@doi [\apj] {10.3847/1538-4357/aa5c41}, \href
  {https://ui.adsabs.harvard.edu/abs/2017ApJ...837...30G} {837, 30}

\bibitem[\protect\citeauthoryear{{Gond{\'a}n} \& {Kocsis}}{{Gond{\'a}n} \&
  {Kocsis}}{2019}]{GondanKocsis2019}
{Gond{\'a}n} L.,  {Kocsis} B.,  2019, \mn@doi [\apj]
  {10.3847/1538-4357/aaf893}, \href
  {https://ui.adsabs.harvard.edu/abs/2019ApJ...871..178G} {871, 178}

\bibitem[\protect\citeauthoryear{{Gond{\'a}n} \& {Kocsis}}{{Gond{\'a}n} \&
  {Kocsis}}{2021}]{GondanKocsis2021}
{Gond{\'a}n} L.,  {Kocsis} B.,  2021, \mn@doi [\mnras]
  {10.1093/mnras/stab1722}, \href
  {https://ui.adsabs.harvard.edu/abs/2021MNRAS.506.1665G} {506, 1665}

\bibitem[\protect\citeauthoryear{{Gond{\'a}n} \& {Kocsis}}{{Gond{\'a}n} \&
  {Kocsis}}{2022}]{GondanKocsis2022}
{Gond{\'a}n} L.,  {Kocsis} B.,  2022, \mn@doi [\mnras]
  {10.1093/mnras/stac1985}, \href
  {https://ui.adsabs.harvard.edu/abs/2022MNRAS.515.3299G} {515, 3299}

\bibitem[\protect\citeauthoryear{{Gond{\'a}n}, {Kocsis}, {Raffai}  \&
  {Frei}}{{Gond{\'a}n} et~al.}{2018a}]{Gondanetal2018a}
{Gond{\'a}n} L.,  {Kocsis} B.,  {Raffai} P.,   {Frei} Z.,  2018a, \mn@doi
  [\apj] {10.3847/1538-4357/aaad0e}, \href
  {https://ui.adsabs.harvard.edu/abs/2018ApJ...855...34G} {855, 34}

\bibitem[\protect\citeauthoryear{{Gond{\'a}n}, {Kocsis}, {Raffai}  \&
  {Frei}}{{Gond{\'a}n} et~al.}{2018b}]{Gondanetal2018b}
{Gond{\'a}n} L.,  {Kocsis} B.,  {Raffai} P.,   {Frei} Z.,  2018b, \mn@doi
  [\apj] {10.3847/1538-4357/aabfee}, \href
  {https://ui.adsabs.harvard.edu/abs/2018ApJ...860....5G} {860, 5}

\bibitem[\protect\citeauthoryear{{Graham}, {Driver}, {Allen}  \&
  {Liske}}{{Graham} et~al.}{2007}]{Grahametal2007}
{Graham} A.~W.,  {Driver} S.~P.,  {Allen} P.~D.,   {Liske} J.,  2007, \mn@doi
  [\mnras] {10.1111/j.1365-2966.2007.11770.x}, \href
  {http://adsabs.harvard.edu/abs/2007MNRAS.378..198G} {378, 198}

\bibitem[\protect\citeauthoryear{{Greene} \& {Ho}}{{Greene} \&
  {Ho}}{2006}]{GreeneHo2006}
{Greene} J.~E.,  {Ho} L.~C.,  2006, \mn@doi [\apj] {10.1086/500353}, \href
  {http://adsabs.harvard.edu/abs/2006ApJ...641..117G} {641, 117}

\bibitem[\protect\citeauthoryear{{Gregory} \& {Thompson}}{{Gregory} \&
  {Thompson}}{1978}]{GregoryThompson1978}
{Gregory} S.~A.,  {Thompson} L.~A.,  1978, \mn@doi [\apj] {10.1086/156198},
  \href {https://ui.adsabs.harvard.edu/abs/1978ApJ...222..784G} {222, 784}

\bibitem[\protect\citeauthoryear{{Gruzinov}, {Levin}  \& {Zhu}}{{Gruzinov}
  et~al.}{2020}]{Gruzinovetal2020}
{Gruzinov} A.,  {Levin} Y.,   {Zhu} J.,  2020, \mn@doi [\apj]
  {10.3847/1538-4357/abbfaa}, \href
  {https://ui.adsabs.harvard.edu/abs/2020ApJ...905...11G} {905, 11}

\bibitem[\protect\citeauthoryear{{Hall} \& {Evans}}{{Hall} \&
  {Evans}}{2019}]{HallEvans2019}
{Hall} E.~D.,  {Evans} M.,  2019, \mn@doi [Classical and Quantum Gravity]
  {10.1088/1361-6382/ab41d6}, \href
  {https://ui.adsabs.harvard.edu/abs/2019CQGra..36v5002H} {36, 225002}

\bibitem[\protect\citeauthoryear{{Hamers} \& {Safarzadeh}}{{Hamers} \&
  {Safarzadeh}}{2020}]{HamersSafarzadeh2020}
{Hamers} A.~S.,  {Safarzadeh} M.,  2020, \mn@doi [\apj]
  {10.3847/1538-4357/ab9b27}, \href
  {https://ui.adsabs.harvard.edu/abs/2020ApJ...898...99H} {898, 99}

\bibitem[\protect\citeauthoryear{{Hamers}, {Bar-Or}, {Petrovich}  \&
  {Antonini}}{{Hamers} et~al.}{2018}]{Hamersetal2018}
{Hamers} A.~S.,  {Bar-Or} B.,  {Petrovich} C.,   {Antonini} F.,  2018, \mn@doi
  [\apj] {10.3847/1538-4357/aadae2}, \href
  {http://cdsads.u-strasbg.fr/abs/2018ApJ...865....2H} {865, 2}

\bibitem[\protect\citeauthoryear{{Hamers}, {Fragione}, {Neunteufel}  \&
  {Kocsis}}{{Hamers} et~al.}{2021}]{Hamersetal2021}
{Hamers} A.~S.,  {Fragione} G.,  {Neunteufel} P.,   {Kocsis} B.,  2021, \mn@doi
  [\mnras] {10.1093/mnras/stab2136}, \href
  {https://ui.adsabs.harvard.edu/abs/2021MNRAS.506.5345H} {506, 5345}

\bibitem[\protect\citeauthoryear{Hill}{Hill}{1878}]{Hill1878}
Hill G.~W.,  1878, American Journal of Mathematics, 1, 5

\bibitem[\protect\citeauthoryear{{Hills}}{{Hills}}{1988}]{Hills1988}
{Hills} J.~G.,  1988, \mn@doi [\nat] {10.1038/331687a0}, \href
  {https://ui.adsabs.harvard.edu/abs/1988Natur.331..687H} {331, 687}

\bibitem[\protect\citeauthoryear{{Hoang}, {Naoz}, {Kocsis}, {Rasio}  \& et
  al.}{{Hoang} et~al.}{2018}]{Hoangetal2018}
{Hoang} B.-M.,  {Naoz} S.,  {Kocsis} B.,  {Rasio} F.~A.,   et al. 2018, \mn@doi
  [\apj] {10.3847/1538-4357/aaafce}, \href
  {http://cdsads.u-strasbg.fr/abs/2018ApJ...856..140H} {856, 140}

\bibitem[\protect\citeauthoryear{{Hoang}, {Naoz}, {Kocsis}, {Farr}  \&
  {McIver}}{{Hoang} et~al.}{2019}]{Hoangetal2019}
{Hoang} B.-M.,  {Naoz} S.,  {Kocsis} B.,  {Farr} W.~M.,   {McIver} J.,  2019,
  \mn@doi [\apjl] {10.3847/2041-8213/ab14f7}, \href
  {https://ui.adsabs.harvard.edu/abs/2019ApJ...875L..31H} {875, L31}

\bibitem[\protect\citeauthoryear{{Hogg}}{{Hogg}}{1999}]{Hogg1999}
{Hogg} D.~W.,  1999, arXiv e-prints, \href
  {https://ui.adsabs.harvard.edu/abs/1999astro.ph..5116H} {pp
  astro--ph/9905116}

\bibitem[\protect\citeauthoryear{{Hong} \& {Lee}}{{Hong} \&
  {Lee}}{2015}]{HongLee2015}
{Hong} J.,  {Lee} H.~M.,  2015, \mn@doi [\mnras] {10.1093/mnras/stv035}, \href
  {http://cdsads.u-strasbg.fr/abs/2015MNRAS.448..754H} {448, 754}

\bibitem[\protect\citeauthoryear{{Hopman} \& {Alexander}}{{Hopman} \&
  {Alexander}}{2006}]{HopmanAlexander2006}
{Hopman} C.,  {Alexander} T.,  2006, \mn@doi [\apjl] {10.1086/506273}, \href
  {http://adsabs.harvard.edu/abs/2006ApJ...645L.133H} {645, L133}

\bibitem[\protect\citeauthoryear{{Huerta} et~al.,}{{Huerta}
  et~al.}{2017}]{Huertaetal2017}
{Huerta} E.~A.,  et~al., 2017, \mn@doi [\prd] {10.1103/PhysRevD.95.024038},
  \href {https://ui.adsabs.harvard.edu/abs/2017PhRvD..95b4038H} {95, 024038}

\bibitem[\protect\citeauthoryear{{Huerta} et~al.,}{{Huerta}
  et~al.}{2018}]{Huertaetal2018}
{Huerta} E.~A.,  et~al., 2018, \mn@doi [\prd] {10.1103/PhysRevD.97.024031},
  \href {https://ui.adsabs.harvard.edu/abs/2018PhRvD..97b4031H} {97, 024031}

\bibitem[\protect\citeauthoryear{{Inayoshi}, {Tamanini}, {Caprini}  \&
  {Haiman}}{{Inayoshi} et~al.}{2017}]{Inayoshietal2017}
{Inayoshi} K.,  {Tamanini} N.,  {Caprini} C.,   {Haiman} Z.,  2017, \mn@doi
  [\prd] {10.1103/PhysRevD.96.063014}, \href
  {https://ui.adsabs.harvard.edu/abs/2017PhRvD..96f3014I} {96, 063014}

\bibitem[\protect\citeauthoryear{{Jeans}}{{Jeans}}{1919}]{Jeans1919}
{Jeans} J.~H.,  1919, \mn@doi [\mnras] {10.1093/mnras/79.6.408}, \href
  {https://ui.adsabs.harvard.edu/abs/1919MNRAS..79..408J} {79, 408}

\bibitem[\protect\citeauthoryear{{Kagra Collaboration}, {Akutsu}, {Ando},
  {Arai}, {Arai}, {Araki}, {Araya}  \& {Aritomi}}{{Kagra Collaboration}
  et~al.}{2019}]{KagraColletal2018}
{Kagra Collaboration} {Akutsu} T.,  {Ando} M.,  {Arai} K.,  {Arai} Y.,  {Araki}
  S.,  {Araya} A.,   {Aritomi} N.,  2019, \mn@doi [Nature Astronomy]
  {10.1038/s41550-018-0658-y}, \href
  {https://ui.adsabs.harvard.edu/abs/2019NatAs...3...35K} {3, 35}

\bibitem[\protect\citeauthoryear{{Katz}, {Dong}  \& {Malhotra}}{{Katz}
  et~al.}{2011}]{Katzetal2011}
{Katz} B.,  {Dong} S.,   {Malhotra} R.,  2011, \mn@doi [\prl]
  {10.1103/PhysRevLett.107.181101}, \href
  {https://ui.adsabs.harvard.edu/abs/2011PhRvL.107r1101K} {107, 181101}

\bibitem[\protect\citeauthoryear{{Keshet}, {Hopman}  \& {Alexander}}{{Keshet}
  et~al.}{2009}]{Keshetetal2009}
{Keshet} U.,  {Hopman} C.,   {Alexander} T.,  2009, \mn@doi [\apjl]
  {10.1088/0004-637X/698/1/L64}, \href
  {http://adsabs.harvard.edu/abs/2009ApJ...698L..64K} {698, L64}

\bibitem[\protect\citeauthoryear{{Kimpson}, {Spera}, {Mapelli}  \&
  {Ziosi}}{{Kimpson} et~al.}{2016}]{Kimpsonetal2016}
{Kimpson} T.~O.,  {Spera} M.,  {Mapelli} M.,   {Ziosi} B.~M.,  2016, \mn@doi
  [\mnras] {10.1093/mnras/stw2085}, \href
  {http://cdsads.u-strasbg.fr/abs/2016MNRAS.463.2443K} {463, 2443}

\bibitem[\protect\citeauthoryear{{Kocsis}}{{Kocsis}}{2013}]{Kocsis2013}
{Kocsis} B.,  2013, \mn@doi [\apj] {10.1088/0004-637X/763/2/122}, \href
  {https://ui.adsabs.harvard.edu/abs/2013ApJ...763..122K} {763, 122}

\bibitem[\protect\citeauthoryear{{Kocsis} \& {Levin}}{{Kocsis} \&
  {Levin}}{2012}]{KocsisLevin2012}
{Kocsis} B.,  {Levin} J.,  2012, \mn@doi [\prd] {10.1103/PhysRevD.85.123005},
  \href {http://adsabs.harvard.edu/abs/2012PhRvD..85l3005K} {85, 123005}

\bibitem[\protect\citeauthoryear{{Kocsis} \& {Tremaine}}{{Kocsis} \&
  {Tremaine}}{2011}]{KocsisTremaine2011}
{Kocsis} B.,  {Tremaine} S.,  2011, \mn@doi [\mnras]
  {10.1111/j.1365-2966.2010.17897.x}, \href
  {https://ui.adsabs.harvard.edu/abs/2011MNRAS.412..187K} {412, 187}

\bibitem[\protect\citeauthoryear{{Kormendy} \& {Ho}}{{Kormendy} \&
  {Ho}}{2013}]{KormendyHo2013}
{Kormendy} J.,  {Ho} L.~C.,  2013, \mn@doi [\araa]
  {10.1146/annurev-astro-082708-101811}, \href
  {https://ui.adsabs.harvard.edu/abs/2013ARA&A..51..511K} {51, 511}

\bibitem[\protect\citeauthoryear{{Kozai}}{{Kozai}}{1962}]{Kozai1962}
{Kozai} Y.,  1962, \mn@doi [\aj] {10.1086/108790}, \href
  {http://cdsads.u-strasbg.fr/abs/1962AJ.....67..591K} {67, 591}

\bibitem[\protect\citeauthoryear{{Leigh}, {Antonini}, {Stone}, {Shara}  \&
  {Merritt}}{{Leigh} et~al.}{2016}]{Leighetal2016}
{Leigh} N. W.~C.,  {Antonini} F.,  {Stone} N.~C.,  {Shara} M.~M.,   {Merritt}
  D.,  2016, \mn@doi [\mnras] {10.1093/mnras/stw2018}, \href
  {https://ui.adsabs.harvard.edu/abs/2016MNRAS.463.1605L} {463, 1605}

\bibitem[\protect\citeauthoryear{{Leigh} et~al.,}{{Leigh}
  et~al.}{2018}]{Leighetal2018}
{Leigh} N.~W.~C.,  et~al., 2018, \mn@doi [\mnras] {10.1093/mnras/stx3134},
  \href {https://ui.adsabs.harvard.edu/abs/2018MNRAS.474.5672L} {474, 5672}

\bibitem[\protect\citeauthoryear{{Lenon}, {Nitz}  \& {Brown}}{{Lenon}
  et~al.}{2020}]{Lenonetal2020}
{Lenon} A.~K.,  {Nitz} A.~H.,   {Brown} D.~A.,  2020, \mn@doi [\mnras]
  {10.1093/mnras/staa2120}, \href
  {https://ui.adsabs.harvard.edu/abs/2020MNRAS.497.1966L} {497, 1966}

\bibitem[\protect\citeauthoryear{{Li}, {Naoz}, {Holman}  \& {Loeb}}{{Li}
  et~al.}{2014}]{Lieetal2011}
{Li} G.,  {Naoz} S.,  {Holman} M.,   {Loeb} A.,  2014, \mn@doi [\apj]
  {10.1088/0004-637X/791/2/86}, \href
  {https://ui.adsabs.harvard.edu/abs/2014ApJ...791...86L} {791, 86}

\bibitem[\protect\citeauthoryear{{Lidov}}{{Lidov}}{1962}]{Lidov1962}
{Lidov} M.~L.,  1962, \mn@doi [\planss] {10.1016/0032-0633(62)90129-0}, \href
  {http://adsabs.harvard.edu/abs/1962P26SS....9..719L} {9, 719}

\bibitem[\protect\citeauthoryear{{Lidov} \& {Ziglin}}{{Lidov} \&
  {Ziglin}}{1976}]{LidovZiglin1976}
{Lidov} M.~L.,  {Ziglin} S.~L.,  1976, \mn@doi [Celestial Mechanics]
  {10.1007/BF01229100}, \href
  {http://cdsads.u-strasbg.fr/abs/1976CeMec..13..471L} {13, 471}

\bibitem[\protect\citeauthoryear{{Limongi} \& {Chieffi}}{{Limongi} \&
  {Chieffi}}{2018}]{Limongietal2018}
{Limongi} M.,  {Chieffi} A.,  2018, \mn@doi [\apjs] {10.3847/1538-4365/aacb24},
  \href {https://ui.adsabs.harvard.edu/abs/2018ApJS..237...13L} {237, 13}

\bibitem[\protect\citeauthoryear{{Lithwick} \& {Naoz}}{{Lithwick} \&
  {Naoz}}{2011}]{LithwickNaoz2011}
{Lithwick} Y.,  {Naoz} S.,  2011, \mn@doi [\apj] {10.1088/0004-637X/742/2/94},
  \href {https://ui.adsabs.harvard.edu/abs/2011ApJ...742...94L} {742, 94}

\bibitem[\protect\citeauthoryear{{Liu} \& {Lai}}{{Liu} \&
  {Lai}}{2017}]{LiuLai2017}
{Liu} B.,  {Lai} D.,  2017, \mn@doi [\apjl] {10.3847/2041-8213/aa8727}, \href
  {http://adsabs.harvard.edu/abs/2017ApJ...846L..11L} {846, L11}

\bibitem[\protect\citeauthoryear{{Liu} \& {Lai}}{{Liu} \&
  {Lai}}{2018}]{LiuLai2018}
{Liu} B.,  {Lai} D.,  2018, \mn@doi [\apj] {10.3847/1538-4357/aad09f}, \href
  {http://adsabs.harvard.edu/abs/2018ApJ...863...68L} {863, 68}

\bibitem[\protect\citeauthoryear{{Liu} \& {Lai}}{{Liu} \&
  {Lai}}{2019}]{LiuLai2019}
{Liu} B.,  {Lai} D.,  2019, \mn@doi [\mnras] {10.1093/mnras/sty3432}, \href
  {https://ui.adsabs.harvard.edu/abs/2019MNRAS.483.4060L} {483, 4060}

\bibitem[\protect\citeauthoryear{{Liu}, {Cao}  \& {Shao}}{{Liu}
  et~al.}{2020}]{Liuetal2020}
{Liu} X.,  {Cao} Z.,   {Shao} L.,  2020, \mn@doi [\prd]
  {10.1103/PhysRevD.101.044049}, \href
  {https://ui.adsabs.harvard.edu/abs/2020PhRvD.101d4049L} {101, 044049}

\bibitem[\protect\citeauthoryear{{Lockhart}, {Lu}, {Peiris}, {Rich}, {Bouchez}
  \& {Ghez}}{{Lockhart} et~al.}{2018}]{Lockhartetal2018}
{Lockhart} K.~E.,  {Lu} J.~R.,  {Peiris} H.~V.,  {Rich} R.~M.,  {Bouchez} A.,
  {Ghez} A.~M.,  2018, \mn@doi [\apj] {10.3847/1538-4357/aaaa71}, \href
  {https://ui.adsabs.harvard.edu/abs/2018ApJ...854..121L} {854, 121}

\bibitem[\protect\citeauthoryear{{Lower}, {Thrane}, {Lasky}  \&
  {Smith}}{{Lower} et~al.}{2018}]{Loweretal2018}
{Lower} M.~E.,  {Thrane} E.,  {Lasky} P.~D.,   {Smith} R.,  2018, \mn@doi
  [\prd] {10.1103/PhysRevD.98.083028}, \href
  {https://ui.adsabs.harvard.edu/abs/2018PhRvD..98h3028L} {98, 083028}

\bibitem[\protect\citeauthoryear{{Mandel} \& {Broekgaarden}}{{Mandel} \&
  {Broekgaarden}}{2022}]{MandelBroekgaarden2022}
{Mandel} I.,  {Broekgaarden} F.~S.,  2022, \mn@doi [Living Reviews in
  Relativity] {10.1007/s41114-021-00034-3}, \href
  {https://ui.adsabs.harvard.edu/abs/2022LRR....25....1M} {25, 1}

\bibitem[\protect\citeauthoryear{{Mao} et~al.,}{{Mao} et~al.}{2017}]{Mao2017}
{Mao} Q.,  et~al., 2017, \mn@doi [\apj] {10.3847/1538-4357/835/2/161}, \href
  {https://ui.adsabs.harvard.edu/abs/2017ApJ...835..161M} {835, 161}

\bibitem[\protect\citeauthoryear{{Mapelli}, {Spera}, {Montanari}, {Limongi},
  {Chieffi}, {Giacobbo}, {Bressan}  \& {Bouffanais}}{{Mapelli}
  et~al.}{2020}]{Mapellietal2020}
{Mapelli} M.,  {Spera} M.,  {Montanari} E.,  {Limongi} M.,  {Chieffi} A.,
  {Giacobbo} N.,  {Bressan} A.,   {Bouffanais} Y.,  2020, \mn@doi [\apj]
  {10.3847/1538-4357/ab584d}, \href
  {https://ui.adsabs.harvard.edu/abs/2020ApJ...888...76M} {888, 76}

\bibitem[\protect\citeauthoryear{{Marchant}, {Renzo}, {Farmer}, {Pappas},
  {Taam}, {de Mink}  \& {Kalogera}}{{Marchant} et~al.}{2019}]{Marchantetal2018}
{Marchant} P.,  {Renzo} M.,  {Farmer} R.,  {Pappas} K. M.~W.,  {Taam} R.~E.,
  {de Mink} S.~E.,   {Kalogera} V.,  2019, \mn@doi [\apj]
  {10.3847/1538-4357/ab3426}, \href
  {https://ui.adsabs.harvard.edu/abs/2019ApJ...882...36M} {882, 36}

\bibitem[\protect\citeauthoryear{{Martinez} et~al.,}{{Martinez}
  et~al.}{2020}]{Martinezetal2020}
{Martinez} M. A.~S.,  et~al., 2020, \mn@doi [\apj] {10.3847/1538-4357/abba25},
  \href {https://ui.adsabs.harvard.edu/abs/2020ApJ...903...67M} {903, 67}

\bibitem[\protect\citeauthoryear{{Mastrobuono-Battisti}, {Perets}  \&
  {Loeb}}{{Mastrobuono-Battisti} et~al.}{2014}]{Mastrobuono-Battisti2014}
{Mastrobuono-Battisti} A.,  {Perets} H.~B.,   {Loeb} A.,  2014, \mn@doi [\apj]
  {10.1088/0004-637X/796/1/40}, \href
  {https://ui.adsabs.harvard.edu/abs/2014ApJ...796...40M} {796, 40}

\bibitem[\protect\citeauthoryear{{M{\'a}th{\'e}}, {Sz{\"o}lgy{\'e}n}  \&
  {Kocsis}}{{M{\'a}th{\'e}} et~al.}{2022}]{Matheetal2022}
{M{\'a}th{\'e}} G.,  {Sz{\"o}lgy{\'e}n} {\'A}.,   {Kocsis} B.,  2022, arXiv
  e-prints, \href {https://ui.adsabs.harvard.edu/abs/2022arXiv220207665M} {p.
  arXiv:2202.07665}

\bibitem[\protect\citeauthoryear{{McKernan} et~al.,}{{McKernan}
  et~al.}{2018}]{McKernanetal2018}
{McKernan} B.,  et~al., 2018, \mn@doi [The Astrophysical Journal]
  {10.3847/1538-4357/aadae5}, \href
  {https://ui.adsabs.harvard.edu/abs/2018ApJ...866...66M} {866, 66}

\bibitem[\protect\citeauthoryear{{Meiron}, {Kocsis}  \& {Loeb}}{{Meiron}
  et~al.}{2017}]{Meironetal2017}
{Meiron} Y.,  {Kocsis} B.,   {Loeb} A.,  2017, \mn@doi [\apj]
  {10.3847/1538-4357/834/2/200}, \href
  {https://ui.adsabs.harvard.edu/abs/2017ApJ...834..200M} {834, 200}

\bibitem[\protect\citeauthoryear{{Merritt}}{{Merritt}}{2004}]{Merritt2004}
{Merritt} D.,  2004, in {Ho} L.~C.,  ed., Coevolution of Black Holes and
  Galaxies. p.~263 (\mn@eprint {arXiv} {astro-ph/0301257})

\bibitem[\protect\citeauthoryear{{Merritt}}{{Merritt}}{2013}]{Merritt2013}
{Merritt} D.,  2013, {Dynamics and Evolution of Galactic Nuclei}

\bibitem[\protect\citeauthoryear{{Michaely} \& {Perets}}{{Michaely} \&
  {Perets}}{2020}]{MichaelyPerets2020}
{Michaely} E.,  {Perets} H.~B.,  2020, \mn@doi [\mnras]
  {10.1093/mnras/staa2720}, \href
  {https://ui.adsabs.harvard.edu/abs/2020MNRAS.tmp.2577M} {}

\bibitem[\protect\citeauthoryear{{Michimura} et~al.,}{{Michimura}
  et~al.}{2020}]{Michimuraetal2020}
{Michimura} Y.,  et~al., 2020, \mn@doi [\prd] {10.1103/PhysRevD.102.022008},
  \href {https://ui.adsabs.harvard.edu/abs/2020PhRvD.102b2008M} {102, 022008}

\bibitem[\protect\citeauthoryear{{Mikkola} \& {Merritt}}{{Mikkola} \&
  {Merritt}}{2006}]{MikkolaMerritt2006}
{Mikkola} S.,  {Merritt} D.,  2006, \mn@doi [\mnras]
  {10.1111/j.1365-2966.2006.10854.x}, \href
  {https://ui.adsabs.harvard.edu/abs/2006MNRAS.372..219M} {372, 219}

\bibitem[\protect\citeauthoryear{{Mikkola} \& {Merritt}}{{Mikkola} \&
  {Merritt}}{2008}]{MikkolaMerritt2008}
{Mikkola} S.,  {Merritt} D.,  2008, \mn@doi [\aj]
  {10.1088/0004-6256/135/6/2398}, \href
  {https://ui.adsabs.harvard.edu/abs/2008AJ....135.2398M} {135, 2398}

\bibitem[\protect\citeauthoryear{{Miller} \& {Lauburg}}{{Miller} \&
  {Lauburg}}{2009}]{MillerLauburg2009}
{Miller} M.~C.,  {Lauburg} V.~M.,  2009, \mn@doi [The Astrophysical Journal]
  {10.1088/0004-637X/692/1/917}, \href
  {https://ui.adsabs.harvard.edu/abs/2009ApJ...692..917M} {692, 917}

\bibitem[\protect\citeauthoryear{{Naoz}}{{Naoz}}{2016}]{Naoz2016}
{Naoz} S.,  2016, \mn@doi [\araa] {10.1146/annurev-astro-081915-023315}, \href
  {https://ui.adsabs.harvard.edu/abs/2016ARA&A..54..441N} {54, 441}

\bibitem[\protect\citeauthoryear{{Naoz} \& {Silk}}{{Naoz} \&
  {Silk}}{2014}]{NaozSilk2014}
{Naoz} S.,  {Silk} J.,  2014, \mn@doi [\apj] {10.1088/0004-637X/795/2/102},
  \href {https://ui.adsabs.harvard.edu/abs/2014ApJ...795..102N} {795, 102}

\bibitem[\protect\citeauthoryear{{Naoz}, {Kocsis}, {Loeb}  \& {Yunes}}{{Naoz}
  et~al.}{2013}]{Naozetal2013}
{Naoz} S.,  {Kocsis} B.,  {Loeb} A.,   {Yunes} N.,  2013, \mn@doi [\apj]
  {10.1088/0004-637X/773/2/187}, \href
  {https://ui.adsabs.harvard.edu/abs/2013ApJ...773..187N} {773, 187}

\bibitem[\protect\citeauthoryear{{Neumayer}, {Seth}  \& {B{\"o}ker}}{{Neumayer}
  et~al.}{2020}]{Neumayeretal2020}
{Neumayer} N.,  {Seth} A.,   {B{\"o}ker} T.,  2020, \mn@doi [\aapr]
  {10.1007/s00159-020-00125-0}, \href
  {https://ui.adsabs.harvard.edu/abs/2020A&ARv..28....4N} {28, 4}

\bibitem[\protect\citeauthoryear{{Nitz}, {Kumar}, {Wang}, {Kastha}, {Wu},
  {Sch{\"a}fer}, {Dhurkunde}  \& {Capano}}{{Nitz} et~al.}{2021}]{Nitzetal2021}
{Nitz} A.~H.,  {Kumar} S.,  {Wang} Y.-F.,  {Kastha} S.,  {Wu} S.,
  {Sch{\"a}fer} M.,  {Dhurkunde} R.,   {Capano} C.~D.,  2021, arXiv e-prints,
  \href {https://ui.adsabs.harvard.edu/abs/2021arXiv211206878N} {p.
  arXiv:2112.06878}

\bibitem[\protect\citeauthoryear{{O'Leary}, {Kocsis}  \& {Loeb}}{{O'Leary}
  et~al.}{2009}]{OLearyetal2009}
{O'Leary} R.~M.,  {Kocsis} B.,   {Loeb} A.,  2009, \mn@doi [\mnras]
  {10.1111/j.1365-2966.2009.14653.x}, \href
  {http://adsabs.harvard.edu/abs/2009MNRAS.395.2127O} {395, 2127}

\bibitem[\protect\citeauthoryear{{O'Shaughnessy}, {Kalogera}  \&
  {Belczynski}}{{O'Shaughnessy} et~al.}{2010}]{OShaughnessyetal2010}
{O'Shaughnessy} R.,  {Kalogera} V.,   {Belczynski} K.,  2010, \mn@doi [\apj]
  {10.1088/0004-637X/716/1/615}, \href
  {https://ui.adsabs.harvard.edu/abs/2010ApJ...716..615O} {716, 615}

\bibitem[\protect\citeauthoryear{{{\"O}pik}}{{{\"O}pik}}{1924}]{Opik1924}
{{\"O}pik} E.,  1924, Publications of the Tartu Astrofizica Observatory, \href
  {https://ui.adsabs.harvard.edu/abs/1924PTarO..25f...1O} {25, 1}

\bibitem[\protect\citeauthoryear{{{\"O}zel}, {Psaltis}, {Narayan}  \&
  {McClintock}}{{{\"O}zel} et~al.}{2010}]{Ozeletal2010}
{{\"O}zel} F.,  {Psaltis} D.,  {Narayan} R.,   {McClintock} J.~E.,  2010,
  \mn@doi [\apj] {10.1088/0004-637X/725/2/1918}, \href
  {http://adsabs.harvard.edu/abs/2010ApJ...725.1918O} {725, 1918}

\bibitem[\protect\citeauthoryear{{Panamarev}, {Just}, {Spurzem}, {Berczik},
  {Wang}  \& {Arca Sedda}}{{Panamarev} et~al.}{2019}]{Panamarevetal2019}
{Panamarev} T.,  {Just} A.,  {Spurzem} R.,  {Berczik} P.,  {Wang} L.,   {Arca
  Sedda} M.,  2019, \mn@doi [\mnras] {10.1093/mnras/stz208}, \href
  {https://ui.adsabs.harvard.edu/abs/2019MNRAS.484.3279P} {484, 3279}

\bibitem[\protect\citeauthoryear{{Paumard} et~al.,}{{Paumard}
  et~al.}{2006}]{Paumardetal2006}
{Paumard} T.,  et~al., 2006, \mn@doi [\apj] {10.1086/503273}, \href
  {https://ui.adsabs.harvard.edu/abs/2006ApJ...643.1011P} {643, 1011}

\bibitem[\protect\citeauthoryear{{Peebles}}{{Peebles}}{1972}]{Peebles1972}
{Peebles} P.~J.~E.,  1972, \mn@doi [\apj] {10.1086/151797}, \href
  {https://ui.adsabs.harvard.edu/abs/1972ApJ...178..371P} {178, 371}

\bibitem[\protect\citeauthoryear{{Peters}}{{Peters}}{1964}]{Peters1964}
{Peters} P.~C.,  1964, \mn@doi [Physical Review] {10.1103/PhysRev.136.B1224},
  \href {http://adsabs.harvard.edu/abs/1964PhRv..136.1224P} {136, 1224}

\bibitem[\protect\citeauthoryear{{Petrovich} \& {Antonini}}{{Petrovich} \&
  {Antonini}}{2017}]{PetrovichAntonini2017}
{Petrovich} C.,  {Antonini} F.,  2017, \mn@doi [\apj]
  {10.3847/1538-4357/aa8628}, \href
  {https://ui.adsabs.harvard.edu/abs/2017ApJ...846..146P} {846, 146}

\bibitem[\protect\citeauthoryear{{Planck Collaboration} et~al.,}{{Planck
  Collaboration} et~al.}{2020}]{Planck2018}
{Planck Collaboration} et~al., 2020, \mn@doi [\aap]
  {10.1051/0004-6361/201833910}, \href
  {https://ui.adsabs.harvard.edu/abs/2020A&A...641A...6P} {641, A6}

\bibitem[\protect\citeauthoryear{{Poisson} \& {Will}}{{Poisson} \&
  {Will}}{1995}]{PoissonWill1995}
{Poisson} E.,  {Will} C.~M.,  1995, \mn@doi [\prd] {10.1103/PhysRevD.52.848},
  \href {https://ui.adsabs.harvard.edu/abs/1995PhRvD..52..848P} {52, 848}

\bibitem[\protect\citeauthoryear{{Preto} \& {Amaro-Seoane}}{{Preto} \&
  {Amaro-Seoane}}{2010}]{PretoAmaroSeoane2010}
{Preto} M.,  {Amaro-Seoane} P.,  2010, \mn@doi [\apjl]
  {10.1088/2041-8205/708/1/L42}, \href
  {https://ui.adsabs.harvard.edu/abs/2010ApJ...708L..42P} {708, L42}

\bibitem[\protect\citeauthoryear{{Raghavan} et~al.,}{{Raghavan}
  et~al.}{2010}]{Raghavanetal2010}
{Raghavan} D.,  et~al., 2010, \mn@doi [\apjs] {10.1088/0067-0049/190/1/1},
  \href {https://ui.adsabs.harvard.edu/abs/2010ApJS..190....1R} {190, 1}

\bibitem[\protect\citeauthoryear{{Randall} \& {Xianyu}}{{Randall} \&
  {Xianyu}}{2018a}]{RandallXianyu2018a}
{Randall} L.,  {Xianyu} Z.-Z.,  2018a, \mn@doi [\apj]
  {10.3847/1538-4357/aaa1a2}, \href
  {https://ui.adsabs.harvard.edu/abs/2018ApJ...853...93R} {853, 93}

\bibitem[\protect\citeauthoryear{{Randall} \& {Xianyu}}{{Randall} \&
  {Xianyu}}{2018b}]{RandallXianyu2018b}
{Randall} L.,  {Xianyu} Z.-Z.,  2018b, \mn@doi [\apj]
  {10.3847/1538-4357/aad7fe}, \href
  {http://adsabs.harvard.edu/abs/2018ApJ...864..134R} {864, 134}

\bibitem[\protect\citeauthoryear{{Randall} \& {Xianyu}}{{Randall} \&
  {Xianyu}}{2019}]{RandallXianyu2019}
{Randall} L.,  {Xianyu} Z.-Z.,  2019, arXiv e-prints, \href
  {https://ui.adsabs.harvard.edu/abs/2019arXiv190208604R} {p. arXiv:1902.08604}

\bibitem[\protect\citeauthoryear{{Rauch} \& {Tremaine}}{{Rauch} \&
  {Tremaine}}{1996}]{RauchTremaine1996}
{Rauch} K.~P.,  {Tremaine} S.,  1996, \mn@doi [\na]
  {10.1016/S1384-1076(96)00012-7}, \href
  {https://ui.adsabs.harvard.edu/abs/1996NewA....1..149R} {1, 149}

\bibitem[\protect\citeauthoryear{{Rodriguez} \& {Antonini}}{{Rodriguez} \&
  {Antonini}}{2018}]{RodriguezAntonini2018}
{Rodriguez} C.~L.,  {Antonini} F.,  2018, \mn@doi [\apj]
  {10.3847/1538-4357/aacea4}, \href
  {https://ui.adsabs.harvard.edu/abs/2018ApJ...863....7R} {863, 7}

\bibitem[\protect\citeauthoryear{{Rodriguez}, {Amaro-Seoane}, {Chatterjee},
  {Kremer}, {Rasio}, {Samsing}, {Ye}  \& {Zevin}}{{Rodriguez}
  et~al.}{2018a}]{Rodriguezetal2018b}
{Rodriguez} C.~L.,  {Amaro-Seoane} P.,  {Chatterjee} S.,  {Kremer} K.,  {Rasio}
  F.~A.,  {Samsing} J.,  {Ye} C.~S.,   {Zevin} M.,  2018a, \mn@doi [\prd]
  {10.1103/PhysRevD.98.123005}, \href
  {https://ui.adsabs.harvard.edu/abs/2018PhRvD..98l3005R} {98, 123005}

\bibitem[\protect\citeauthoryear{{Rodriguez}, {Amaro-Seoane}, {Chatterjee},
  {Kremer}, {Rasio}, {Samsing}, {Ye}  \& {Zevin}}{{Rodriguez}
  et~al.}{2018b}]{Rodriguezetal2018}
{Rodriguez} C.~L.,  {Amaro-Seoane} P.,  {Chatterjee} S.,  {Kremer} K.,  {Rasio}
  F.~A.,  {Samsing} J.,  {Ye} C.~S.,   {Zevin} M.,  2018b, \mn@doi [\prd]
  {10.1103/PhysRevD.98.123005}, \href
  {https://ui.adsabs.harvard.edu/abs/2018PhRvD..98l3005R} {98, 123005}

\bibitem[\protect\citeauthoryear{{Romero-Shaw}, {Lasky}  \&
  {Thrane}}{{Romero-Shaw} et~al.}{2019}]{RomeroShaw2019}
{Romero-Shaw} I.~M.,  {Lasky} P.~D.,   {Thrane} E.,  2019, \mn@doi [\mnras]
  {10.1093/mnras/stz2996}, \href
  {https://ui.adsabs.harvard.edu/abs/2019MNRAS.490.5210R} {490, 5210}

\bibitem[\protect\citeauthoryear{{Romero-Shaw}, {Lasky}, {Thrane}  \&
  {Calder{\'o}n Bustillo}}{{Romero-Shaw} et~al.}{2020}]{RomeroShawetal2020}
{Romero-Shaw} I.,  {Lasky} P.~D.,  {Thrane} E.,   {Calder{\'o}n Bustillo} J.,
  2020, \mn@doi [\apjl] {10.3847/2041-8213/abbe26}, \href
  {https://ui.adsabs.harvard.edu/abs/2020ApJ...903L...5R} {903, L5}

\bibitem[\protect\citeauthoryear{{Romero-Shaw}, {Lasky}  \&
  {Thrane}}{{Romero-Shaw} et~al.}{2021}]{RomeroShawetal2021}
{Romero-Shaw} I.,  {Lasky} P.~D.,   {Thrane} E.,  2021, \mn@doi [\apjl]
  {10.3847/2041-8213/ac3138}, \href
  {https://ui.adsabs.harvard.edu/abs/2021ApJ...921L..31R} {921, L31}

\bibitem[\protect\citeauthoryear{{Romero-Shaw}, {Lasky}  \&
  {Thrane}}{{Romero-Shaw} et~al.}{2022}]{RomeroShawetal2022}
{Romero-Shaw} I.~M.,  {Lasky} P.~D.,   {Thrane} E.,  2022, arXiv e-prints,
  \href {https://ui.adsabs.harvard.edu/abs/2022arXiv220614695R} {p.
  arXiv:2206.14695}

\bibitem[\protect\citeauthoryear{{Samsing}}{{Samsing}}{2018}]{Samsing2018}
{Samsing} J.,  2018, \mn@doi [\prd] {10.1103/PhysRevD.97.103014}, \href
  {http://cdsads.u-strasbg.fr/abs/2018PhRvD..97j3014S} {97, 103014}

\bibitem[\protect\citeauthoryear{{Samsing} \& {Ramirez-Ruiz}}{{Samsing} \&
  {Ramirez-Ruiz}}{2017}]{SamsingRamirezRuiz2017}
{Samsing} J.,  {Ramirez-Ruiz} E.,  2017, \mn@doi [\apjl]
  {10.3847/2041-8213/aa6f0b}, \href
  {http://adsabs.harvard.edu/abs/2017ApJ...840L..14S} {840, L14}

\bibitem[\protect\citeauthoryear{{Samsing}, {MacLeod}  \&
  {Ramirez-Ruiz}}{{Samsing} et~al.}{2014}]{Samsingetal2014}
{Samsing} J.,  {MacLeod} M.,   {Ramirez-Ruiz} E.,  2014, \mn@doi [\apj]
  {10.1088/0004-637X/784/1/71}, \href
  {http://adsabs.harvard.edu/abs/2014ApJ...784...71S} {784, 71}

\bibitem[\protect\citeauthoryear{{Samsing}, {D'Orazio}, {Askar}  \&
  {Giersz}}{{Samsing} et~al.}{2018}]{Samsingetal2018}
{Samsing} J.,  {D'Orazio} D.~J.,  {Askar} A.,   {Giersz} M.,  2018, arXiv
  e-prints, \href {https://ui.adsabs.harvard.edu/abs/2018arXiv180208654S} {p.
  arXiv:1802.08654}

\bibitem[\protect\citeauthoryear{{Samsing}, {D'Orazio}, {Kremer}, {Rodriguez}
  \& {Askar}}{{Samsing} et~al.}{2020}]{Samsingetal2020}
{Samsing} J.,  {D'Orazio} D.~J.,  {Kremer} K.,  {Rodriguez} C.~L.,   {Askar}
  A.,  2020, \mn@doi [\prd] {10.1103/PhysRevD.101.123010}, \href
  {https://ui.adsabs.harvard.edu/abs/2020PhRvD.101l3010S} {101, 123010}

\bibitem[\protect\citeauthoryear{{Samsing} et~al.,}{{Samsing}
  et~al.}{2022}]{Samsingetal2022}
{Samsing} J.,  et~al., 2022, \mn@doi [\nat] {10.1038/s41586-021-04333-1}, \href
  {https://ui.adsabs.harvard.edu/abs/2022Natur.603..237S} {603, 237}

\bibitem[\protect\citeauthoryear{{Sana} et~al.,}{{Sana}
  et~al.}{2012}]{Sanaetal2012}
{Sana} H.,  et~al., 2012, \mn@doi [Science] {10.1126/science.1223344}, \href
  {https://ui.adsabs.harvard.edu/abs/2012Sci...337..444S} {337, 444}

\bibitem[\protect\citeauthoryear{{Sch{\"o}del}, {Ott}, {Genzel}, {Eckart},
  {Mouawad}  \& {Alexander}}{{Sch{\"o}del} et~al.}{2003}]{Schodeletal2003}
{Sch{\"o}del} R.,  {Ott} T.,  {Genzel} R.,  {Eckart} A.,  {Mouawad} N.,
  {Alexander} T.,  2003, \mn@doi [\apj] {10.1086/378122}, \href
  {https://ui.adsabs.harvard.edu/abs/2003ApJ...596.1015S} {596, 1015}

\bibitem[\protect\citeauthoryear{{Sch{\"o}del} et~al.,}{{Sch{\"o}del}
  et~al.}{2007}]{Schodeletal2007}
{Sch{\"o}del} R.,  et~al., 2007, \mn@doi [\aap] {10.1051/0004-6361:20065089},
  \href {https://ui.adsabs.harvard.edu/abs/2007A26A...469..125S} {469, 125}

\bibitem[\protect\citeauthoryear{{Sch{\"o}del}, {Gallego-Cano}, {Dong},
  {Nogueras-Lara}, {Gallego-Calvente}, {Amaro-Seoane}  \&
  {Baumgardt}}{{Sch{\"o}del} et~al.}{2018}]{Schodeletal2018}
{Sch{\"o}del} R.,  {Gallego-Cano} E.,  {Dong} H.,  {Nogueras-Lara} F.,
  {Gallego-Calvente} A.~T.,  {Amaro-Seoane} P.,   {Baumgardt} H.,  2018,
  \mn@doi [\aap] {10.1051/0004-6361/201730452}, \href
  {https://ui.adsabs.harvard.edu/abs/2018A&A...609A..27S} {609, A27}

\bibitem[\protect\citeauthoryear{{Seth}, {Blum}, {Bastian}, {Caldwell}  \&
  {Debattista}}{{Seth} et~al.}{2008}]{Sethetal2008}
{Seth} A.~C.,  {Blum} R.~D.,  {Bastian} N.,  {Caldwell} N.,   {Debattista}
  V.~P.,  2008, \mn@doi [\apj] {10.1086/591935}, \href
  {https://ui.adsabs.harvard.edu/abs/2008ApJ...687..997S} {687, 997}

\bibitem[\protect\citeauthoryear{{Shankar}}{{Shankar}}{2013}]{Shankar2013}
{Shankar} F.,  2013, \mn@doi [Classical and Quantum Gravity]
  {10.1088/0264-9381/30/24/244001}, \href
  {https://ui.adsabs.harvard.edu/abs/2013CQGra..30x4001S} {30, 244001}

\bibitem[\protect\citeauthoryear{{Shankar}, {Salucci}, {Granato}, {De Zotti}
  \& {Danese}}{{Shankar} et~al.}{2004}]{Shankaretal2004}
{Shankar} F.,  {Salucci} P.,  {Granato} G.~L.,  {De Zotti} G.,   {Danese} L.,
  2004, \mn@doi [\mnras] {10.1111/j.1365-2966.2004.08261.x}, \href
  {http://adsabs.harvard.edu/abs/2004MNRAS.354.1020S} {354, 1020}

\bibitem[\protect\citeauthoryear{{Shankar}, {Weinberg}  \&
  {Miralda-Escud{\'e}}}{{Shankar} et~al.}{2009}]{Shankaretal2009}
{Shankar} F.,  {Weinberg} D.~H.,   {Miralda-Escud{\'e}} J.,  2009, \mn@doi
  [\apj] {10.1088/0004-637X/690/1/20}, \href
  {http://adsabs.harvard.edu/abs/2009ApJ...690...20S} {690, 20}

\bibitem[\protect\citeauthoryear{{Shemmer}, {Netzer}, {Maiolino}, {Oliva},
  {Croom}, {Corbett}  \& {di Fabrizio}}{{Shemmer}
  et~al.}{2004}]{Shemmeretal2004}
{Shemmer} O.,  {Netzer} H.,  {Maiolino} R.,  {Oliva} E.,  {Croom} S.,
  {Corbett} E.,   {di Fabrizio} L.,  2004, \mn@doi [\apj] {10.1086/423607},
  \href {https://ui.adsabs.harvard.edu/abs/2004ApJ...614..547S} {614, 547}

\bibitem[\protect\citeauthoryear{{Sijacki}, {Vogelsberger}, {Genel},
  {Springel}, {Torrey}, {Snyder}, {Nelson}  \& {Hernquist}}{{Sijacki}
  et~al.}{2015}]{Sijackietal2015}
{Sijacki} D.,  {Vogelsberger} M.,  {Genel} S.,  {Springel} V.,  {Torrey} P.,
  {Snyder} G.~F.,  {Nelson} D.,   {Hernquist} L.,  2015, \mn@doi [\mnras]
  {10.1093/mnras/stv1340}, \href
  {http://adsabs.harvard.edu/abs/2015MNRAS.452..575S} {452, 575}

\bibitem[\protect\citeauthoryear{{Silsbee} \& {Tremaine}}{{Silsbee} \&
  {Tremaine}}{2017}]{SilsbeeTremaine2017}
{Silsbee} K.,  {Tremaine} S.,  2017, \mn@doi [\apj] {10.3847/1538-4357/aa5729},
  \href {http://adsabs.harvard.edu/abs/2017ApJ...836...39S} {836, 39}

\bibitem[\protect\citeauthoryear{Spearman}{Spearman}{1904}]{Spearman1904}
Spearman C.,  1904, The American Journal of Psychology, 15, 72

\bibitem[\protect\citeauthoryear{{Spera}, {Trani}  \& {Mencagli}}{{Spera}
  et~al.}{2022}]{Speraetal2022}
{Spera} M.,  {Trani} A.~A.,   {Mencagli} M.,  2022, \mn@doi [Galaxies]
  {10.3390/galaxies10040076}, \href
  {https://ui.adsabs.harvard.edu/abs/2022Galax..10...76S} {10, 76}

\bibitem[\protect\citeauthoryear{{Spitzer}}{{Spitzer}}{1987}]{Spitzer1987}
{Spitzer} L.,  1987, Dynamical Evolution of Globular Clusters.
Princeton Univ. Press, Princeton, NJ

\bibitem[\protect\citeauthoryear{{Sz{\"o}lgy{\'e}n} \&
  {Kocsis}}{{Sz{\"o}lgy{\'e}n} \& {Kocsis}}{2018}]{SzolgyeneKocsis2018}
{Sz{\"o}lgy{\'e}n} {\'A}.,  {Kocsis} B.,  2018, \mn@doi [Physical Review
  Letters] {10.1103/PhysRevLett.121.101101}, \href
  {http://adsabs.harvard.edu/abs/2018PhRvL.121j1101S} {121, 101101}

\bibitem[\protect\citeauthoryear{{Tagawa}, {Kocsis}, {Haiman}, {Bartos},
  {Omukai}  \& {Samsing}}{{Tagawa} et~al.}{2021}]{Tagawaetal2020}
{Tagawa} H.,  {Kocsis} B.,  {Haiman} Z.,  {Bartos} I.,  {Omukai} K.,
  {Samsing} J.,  2021, \mn@doi [\apjl] {10.3847/2041-8213/abd4d3}, \href
  {https://ui.adsabs.harvard.edu/abs/2021ApJ...907L..20T} {907, L20}

\bibitem[\protect\citeauthoryear{{Tak{\'a}tsy}, {B{\'e}csy}  \&
  {Raffai}}{{Tak{\'a}tsy} et~al.}{2019}]{Takatsyetal2019}
{Tak{\'a}tsy} J.,  {B{\'e}csy} B.,   {Raffai} P.,  2019, \mn@doi [\mnras]
  {10.1093/mnras/stz820}, \href
  {https://ui.adsabs.harvard.edu/abs/2019MNRAS.486..570T} {486, 570}

\bibitem[\protect\citeauthoryear{{Thanjavur}, {Simard}, {Bluck}  \&
  {Mendel}}{{Thanjavur} et~al.}{2016}]{Thanjavuretal2016}
{Thanjavur} K.,  {Simard} L.,  {Bluck} A.~F.~L.,   {Mendel} T.,  2016, \mn@doi
  [\mnras] {10.1093/mnras/stw495}, \href
  {https://ui.adsabs.harvard.edu/abs/2016MNRAS.459...44T} {459, 44}

\bibitem[\protect\citeauthoryear{{The LIGO Scientific Collaboration}
  et~al.,}{{The LIGO Scientific Collaboration} et~al.}{2021a}]{Abbotetal2021c}
{The LIGO Scientific Collaboration} et~al., 2021a, arXiv e-prints, \href
  {https://ui.adsabs.harvard.edu/abs/2021arXiv211103606T} {p. arXiv:2111.03606}

\bibitem[\protect\citeauthoryear{{The LIGO Scientific Collaboration}
  et~al.,}{{The LIGO Scientific Collaboration} et~al.}{2021b}]{Abbotetal2021b}
{The LIGO Scientific Collaboration} et~al., 2021b, arXiv e-prints, \href
  {https://ui.adsabs.harvard.edu/abs/2021arXiv211103634T} {p. arXiv:2111.03634}

\bibitem[\protect\citeauthoryear{{Thorne}}{{Thorne}}{1987}]{Thorne1987}
{Thorne} K.~S.,  1987, 300 Years of Gravitation.
Cambridge Univ. Press, Cambridge

\bibitem[\protect\citeauthoryear{{Trani}, {Rastello}, {Di Carlo},
  {Santoliquido}, {Tanikawa}  \& {Mapelli}}{{Trani}
  et~al.}{2022}]{Tranietal2022}
{Trani} A.~A.,  {Rastello} S.,  {Di Carlo} U.~N.,  {Santoliquido} F.,
  {Tanikawa} A.,   {Mapelli} M.,  2022, \mn@doi [\mnras]
  {10.1093/mnras/stac122}, \href
  {https://ui.adsabs.harvard.edu/abs/2022MNRAS.511.1362T} {511, 1362}

\bibitem[\protect\citeauthoryear{{Trippe} et~al.,}{{Trippe}
  et~al.}{2008}]{Trippeetal2008}
{Trippe} S.,  et~al., 2008, \mn@doi [\aap] {10.1051/0004-6361:200810191}, \href
  {https://ui.adsabs.harvard.edu/abs/2008A&A...492..419T} {492, 419}

\bibitem[\protect\citeauthoryear{{Ueda}, {Akiyama}, {Hasinger}, {Miyaji}  \&
  {Watson}}{{Ueda} et~al.}{2014}]{Uedaetal2014}
{Ueda} Y.,  {Akiyama} M.,  {Hasinger} G.,  {Miyaji} T.,   {Watson} M.~G.,
  2014, \mn@doi [\apj] {10.1088/0004-637X/786/2/104}, \href
  {https://ui.adsabs.harvard.edu/abs/2014ApJ...786..104U} {786, 104}

\bibitem[\protect\citeauthoryear{{Unnikrishnan}}{{Unnikrishnan}}{2013}]{Unnikrishnan2013}
{Unnikrishnan} C.~S.,  2013, \mn@doi [International Journal of Modern Physics
  D] {10.1142/S0218271813410101}, \href
  {https://ui.adsabs.harvard.edu/abs/2013IJMPD..2241010U} {22, 1341010}

\bibitem[\protect\citeauthoryear{{VanLandingham}, {Miller}, {Hamilton}  \&
  {Richardson}}{{VanLandingham} et~al.}{2016}]{VanLandinghametal2016}
{VanLandingham} J.~H.,  {Miller} M.~C.,  {Hamilton} D.~P.,   {Richardson}
  D.~C.,  2016, \mn@doi [\apj] {10.3847/0004-637X/828/2/77}, \href
  {https://ui.adsabs.harvard.edu/abs/2016ApJ...828...77V} {828, 77}

\bibitem[\protect\citeauthoryear{{Vasiliev}}{{Vasiliev}}{2017}]{Vasiliev2017}
{Vasiliev} E.,  2017, \mn@doi [\apj] {10.3847/1538-4357/aa8cc8}, \href
  {https://ui.adsabs.harvard.edu/abs/2017ApJ...848...10V} {848, 10}

\bibitem[\protect\citeauthoryear{{Venumadhav}, {Zackay}, {Roulet}, {Dai}  \&
  {Zaldarriaga}}{{Venumadhav} et~al.}{2019}]{Venumadhavetal2019a}
{Venumadhav} T.,  {Zackay} B.,  {Roulet} J.,  {Dai} L.,   {Zaldarriaga} M.,
  2019, \mn@doi [\prd] {10.1103/PhysRevD.100.023011}, \href
  {https://ui.adsabs.harvard.edu/abs/2019PhRvD.100b3011V} {100, 023011}

\bibitem[\protect\citeauthoryear{{Venumadhav}, {Zackay}, {Roulet}, {Dai}  \&
  {Zaldarriaga}}{{Venumadhav} et~al.}{2020}]{Venumadhavetal2019b}
{Venumadhav} T.,  {Zackay} B.,  {Roulet} J.,  {Dai} L.,   {Zaldarriaga} M.,
  2020, \mn@doi [\prd] {10.1103/PhysRevD.101.083030}, \href
  {https://ui.adsabs.harvard.edu/abs/2020PhRvD.101h3030V} {101, 083030}

\bibitem[\protect\citeauthoryear{{Weinberg}}{{Weinberg}}{1972}]{Weinberg1972}
{Weinberg} S.,  1972, {Gravitation and Cosmology: Principles and Applications
  of the General Theory of Relativity}

\bibitem[\protect\citeauthoryear{{Wen}}{{Wen}}{2003}]{Wen2003}
{Wen} L.,  2003, \mn@doi [\apj] {10.1086/378794}, \href
  {http://adsabs.harvard.edu/abs/2003ApJ...598..419W} {598, 419}

\bibitem[\protect\citeauthoryear{{Woosley}}{{Woosley}}{2017}]{Woosley2017}
{Woosley} S.~E.,  2017, \mn@doi [\apj] {10.3847/1538-4357/836/2/244}, \href
  {https://ui.adsabs.harvard.edu/abs/2017ApJ...836..244W} {836, 244}

\bibitem[\protect\citeauthoryear{{Woosley}}{{Woosley}}{2019}]{Woosley2019}
{Woosley} S.~E.,  2019, \mn@doi [\apj] {10.3847/1538-4357/ab1b41}, \href
  {https://ui.adsabs.harvard.edu/abs/2019ApJ...878...49W} {878, 49}

\bibitem[\protect\citeauthoryear{{Wu}, {Cao}  \& {Zhu}}{{Wu}
  et~al.}{2020}]{Wuetal2020}
{Wu} S.,  {Cao} Z.,   {Zhu} Z.-H.,  2020, \mn@doi [\mnras]
  {10.1093/mnras/staa1176}, \href
  {https://ui.adsabs.harvard.edu/abs/2020MNRAS.495..466W} {495, 466}

\bibitem[\protect\citeauthoryear{{Yelda}, {Ghez}, {Lu}, {Do}, {Meyer}, {Morris}
   \& {Matthews}}{{Yelda} et~al.}{2014}]{Yeldaetal2014}
{Yelda} S.,  {Ghez} A.~M.,  {Lu} J.~R.,  {Do} T.,  {Meyer} L.,  {Morris} M.~R.,
    {Matthews} K.,  2014, \mn@doi [\apj] {10.1088/0004-637X/783/2/131}, \href
  {http://adsabs.harvard.edu/abs/2014ApJ...783..131Y} {783, 131}

\bibitem[\protect\citeauthoryear{{Yu}, {Ma}, {Giesler}  \& {Chen}}{{Yu}
  et~al.}{2020}]{Yuetal2020}
{Yu} H.,  {Ma} S.,  {Giesler} M.,   {Chen} Y.,  2020, \mn@doi [\prd]
  {10.1103/PhysRevD.102.123009}, \href
  {https://ui.adsabs.harvard.edu/abs/2020PhRvD.102l3009Y} {102, 123009}

\bibitem[\protect\citeauthoryear{{Yusef-Zadeh}, {Bushouse}  \&
  {Wardle}}{{Yusef-Zadeh} et~al.}{2012}]{YusefZadehetal2012}
{Yusef-Zadeh} F.,  {Bushouse} H.,   {Wardle} M.,  2012, \mn@doi [\apj]
  {10.1088/0004-637X/744/1/24}, \href
  {https://ui.adsabs.harvard.edu/abs/2012ApJ...744...24Y} {744, 24}

\bibitem[\protect\citeauthoryear{{Zackay}, {Dai}, {Venumadhav}, {Roulet}  \&
  {Zaldarriaga}}{{Zackay} et~al.}{2019a}]{Zackayetal2019a}
{Zackay} B.,  {Dai} L.,  {Venumadhav} T.,  {Roulet} J.,   {Zaldarriaga} M.,
  2019a, arXiv e-prints, \href
  {https://ui.adsabs.harvard.edu/abs/2019arXiv191009528Z} {p. arXiv:1910.09528}

\bibitem[\protect\citeauthoryear{{Zackay}, {Venumadhav}, {Dai}, {Roulet}  \&
  {Zaldarriaga}}{{Zackay} et~al.}{2019b}]{Zackayetal2019b}
{Zackay} B.,  {Venumadhav} T.,  {Dai} L.,  {Roulet} J.,   {Zaldarriaga} M.,
  2019b, \mn@doi [\prd] {10.1103/PhysRevD.100.023007}, \href
  {https://ui.adsabs.harvard.edu/abs/2019PhRvD.100b3007Z} {100, 023007}

\bibitem[\protect\citeauthoryear{{Zevin}, {Samsing}, {Rodriguez}, {Haster}  \&
  {Ramirez-Ruiz}}{{Zevin} et~al.}{2019}]{Zevinetal2018}
{Zevin} M.,  {Samsing} J.,  {Rodriguez} C.,  {Haster} C.-J.,   {Ramirez-Ruiz}
  E.,  2019, \mn@doi [\apj] {10.3847/1538-4357/aaf6ec}, \href
  {https://ui.adsabs.harvard.edu/abs/2019ApJ...871...91Z} {871, 91}

\bibitem[\protect\citeauthoryear{{Zhang}, {Shao}  \& {Zhu}}{{Zhang}
  et~al.}{2019}]{Zhangetal2019}
{Zhang} F.,  {Shao} L.,   {Zhu} W.,  2019, \mn@doi [\apj]
  {10.3847/1538-4357/ab1b28}, \href
  {https://ui.adsabs.harvard.edu/abs/2019ApJ...877...87Z} {877, 87}

\makeatother
\end{thebibliography}

\bsp
\label{lastpage}
\end{document}